\documentclass[dvipsnames,svgnames]{svproc}
\usepackage{makeidx}  %
\makeindex
\usepackage{url}

\usepackage{graphicx}
\usepackage{amsfonts,amsmath,amssymb}
\usepackage[english]{babel}
\usepackage[a4paper,margin=1.4in]{geometry}
\usepackage[some]{background}
\usepackage{framed}
\usepackage{uniinput}

\usepackage{mathtools,mathrsfs,wasysym}
\usepackage{textcomp, gensymb}
\usepackage[sectionbib]{chapterbib}
\usepackage[colorlinks,urlcolor=blue,citecolor=blue,linkcolor=blue]{hyperref}
\usepackage{cleveref}
\usepackage{microtype}
\usepackage{enumitem, subfigure}
\usepackage{changes}
\usepackage{acronym}
\usepackage{xcolor}
\usepackage{tcolorbox}
\definecolor{lightgray}{gray}{0.75}
\usepackage{afterpage}

\usepackage{jdefs} %
\usepackage{olivares_defs} %
\usepackage[sort&compress,numbers,square]{natbib} %

\begin{document}
\pagenumbering{gobble}
\begin{titlepage}
\afterpage{%
\newgeometry{left=1in, right=1in, top=1.25in}
\pagecolor{black}
\color{white}

	\begin{tikzpicture}[remember picture,overlay]
	\node[opacity=1] at (current page.center) {\includegraphics[width=\paperwidth,height=\paperheight]{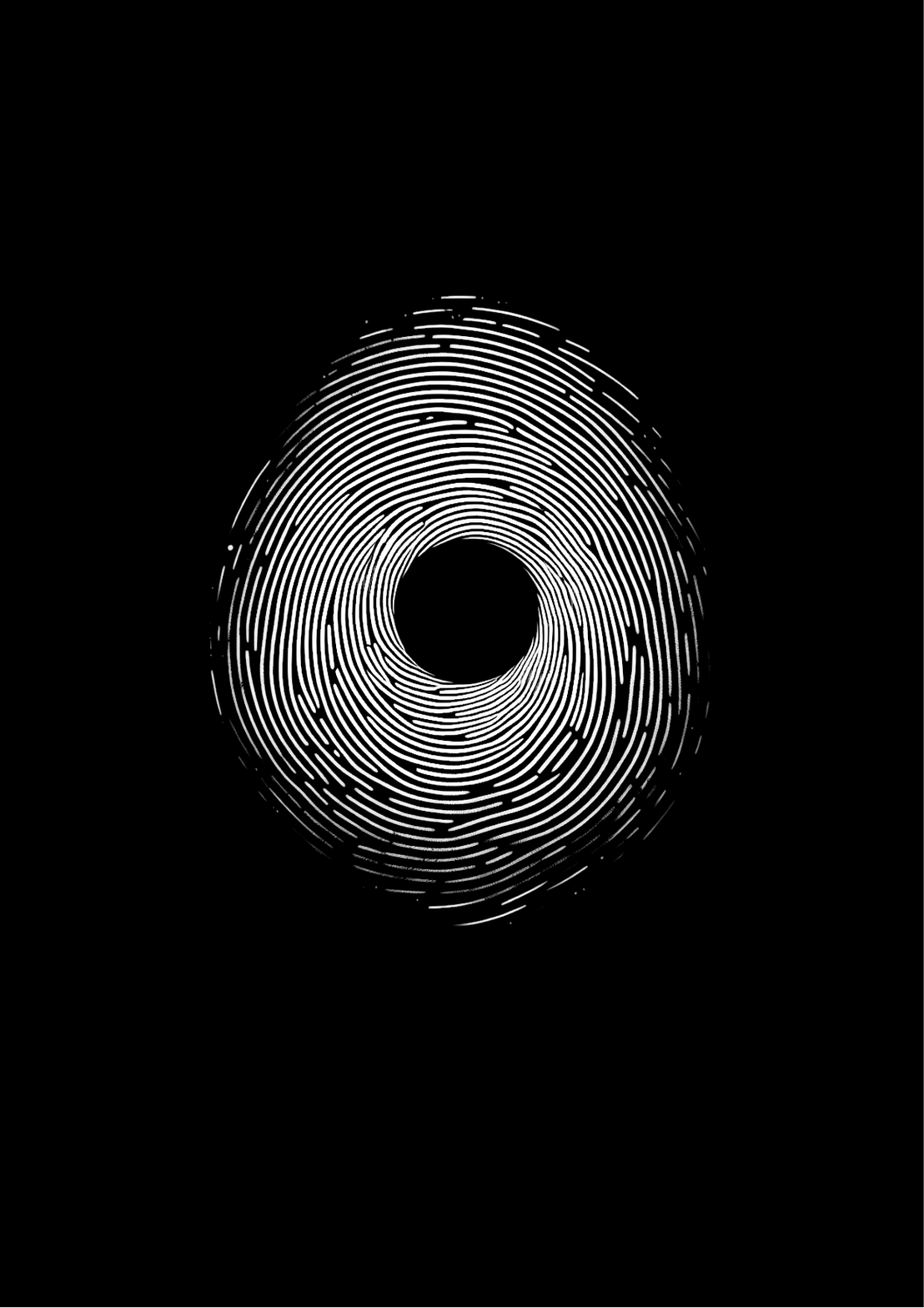}};
\end{tikzpicture}%
\vspace*{-1cm}
\begin{center}
	\color{white}
	{\Huge\bfseries
 Black hole mimickers:\\[0.3cm] from theory to observation}
\end{center}\vspace{17.5cm}

\begin{minipage}{0.24\linewidth}
\large{
	\textbf{\large \textcolor{orange}{Editors:}}\\[0.2cm]
		Suvendu Giri,~\textsuperscript{20, 7}\\
		Luis Lehner,~\textsuperscript{20}\\
		Nils Siemonsen,~\textsuperscript{6}\\
		George N.~Wong~\textsuperscript{6, 21}
        }
\end{minipage}
\hfill
\begin{minipage}{0.02\linewidth}
	\rule{1pt}{85pt}
\end{minipage}
\hfill
\begin{minipage}{0.65\linewidth}
\large{
		Cosimo Bambi,~\textsuperscript{1,2}  \hspace{1pt}
		Ramy Brustein,~\textsuperscript{3}  \hspace{1pt}
		Vitor Cardoso,~\textsuperscript{4,5}  \hspace{1pt}
		Andrew Chael,~\textsuperscript{6}  \hspace{1pt}
		Ulf Danielsson,~\textsuperscript{7}  \hspace{1pt}
		Anuradha Gupta,~\textsuperscript{8}  \hspace{1pt}
		Pierre Heidmann,~\textsuperscript{9}  \hspace{1pt}
		Steven Liebling,~\textsuperscript{10}  \hspace{1pt}
        Elisa Maggio,~\textsuperscript{11}  \hspace{1pt}
		Andrea Maselli,~\textsuperscript{12, 13}  \hspace{1pt}
		Samir Mathur,~\textsuperscript{14}  \hspace{1pt}
		Lia Medeiros,~\textsuperscript{15}  \hspace{1pt}\\
		Alex B. Nielsen,~\textsuperscript{16} \hspace{1pt}
		Héctor R. Olivares-Sánchez,~\textsuperscript{17} \hspace{1pt} \\
		Paolo Pani~\textsuperscript{18,19}
        }
\end{minipage}
\pagecolor{white}
\color{black}
\restoregeometry
}
\end{titlepage}

\afterpage{%
\newgeometry{left=1in, right=1in, top=1.25in}
\noindent
\begin{center}
$^{1}$ \textit{Center for Astronomy and Astrophysics, Center for Field Theory and Particle Physics, and Department of Physics, Fudan University, Shanghai 200438, China}\smallskip \\
$^{2}$ \textit{School of Natural Sciences and Humanities, New Uzbekistan University, Tashkent 100007, Uzbekistan}\smallskip \\
$^3$ \textit{Department of Physics, Ben-Gurion University, Beer-Sheva 84105, Israel}\smallskip \\
$^4$ \textit{Center of Gravity, Niels Bohr Institute, Blegdamsvej 17, 2100 Copenhagen, Denmark}\smallskip \\
$^5$ \textit{CENTRA, Departamento de F\'{\i}sica, Instituto Superior T\'ecnico -- IST, Universidade de Lisboa -- UL, Avenida Rovisco Pais 1, 1049-001 Lisboa, Portugal}\smallskip \\
$^6$ \textit{Princeton Gravity Initiative, Princeton University, Princeton NJ, 08544}\smallskip \\
$^7$ \textit{Institutionen för fysik och astronomi, Uppsala Universitet, Box 803, SE-751 08 Uppsala, Sweden}\smallskip \\
$^8$ \textit{Department of Physics and Astronomy, The University of Mississippi,\\ University, Mississippi 38677, USA} \smallskip \\ %
$^9$ \textit{Department of Physics and Center for Cosmology and AstroParticle Physics (CCAPP),\\ The Ohio State University, Columbus, OH 43210, USA}\smallskip \\
$^{10}$ \textit{Long Island University, Brookville, NY 11548}\smallskip \\
$^{11}$ \textit{Max Planck Institute for Gravitational Physics (Albert Einstein Institute), \\Am Mühlenberg
1, Potsdam 14476, Germany}\smallskip \\
$^{12}$  \textit{Gran Sasso Science Institute (GSSI), I-67100 L’Aquila, Italy}\smallskip \\
$^{13}$ \textit{INFN, Laboratori Nazionali del Gran Sasso, I-67100 Assergi, Italy}\smallskip \\
$^{14}$ \textit{Department of Physics, The Ohio State University, Columbus, OH 43210, USA} \smallskip \\
$^{15}$ \textit{Center for Gravitation, Cosmology and Astrophysics, Department of Physics,\\ University of Wisconsin–Milwaukee, P.O. Box 413, Milwaukee, WI 53201, USA}\smallskip \\
$^{16}$ \textit{Institute of Mathematics and Physics, University of Stavanger, NO-4036, Stavanger, Norway}\smallskip \\
$^{17}$ \textit{
		Departamento de Matem\'{a}tica da Universidade de Aveiro and Centre for Research and Development in Mathematics and Applications (CIDMA), Campus de Santiago, 3810-193 Aveiro, Portugal}\smallskip \\
$^{18}$ \textit{Dipartimento di Fisica, Sapienza Università di Roma}\smallskip \\
$^{19}$ \textit{INFN, Sezione di Roma, Piazzale Aldo Moro 5, 00185, Roma, Italy}\smallskip \\
$^{20}$ \textit{Perimeter Institute for Theoretical Physics, Canada}\smallskip \\
$^{21}$ \textit{Institute for Advanced Study, Princeton NJ, USA}
\end{center}
\restoregeometry
}

\frontmatter          %
\pagestyle{headings}  %
\chapter*{A vision for black hole mimickers}
    \vspace{-15pt}
    \textit{\footnotesize The black hole paradigm, while remarkably successful, raises fundamental questions—both classical and quantum—about the nature of spacetime, horizons, and singularities. Black hole mimickers, horizonless ultra-compact objects, have emerged as potential alternatives that seek to resolve some of these puzzles while remaining consistent with current observational constraints. Recent breakthroughs in gravitational-wave astronomy and horizon-scale electromagnetic imaging have opened new avenues to test this paradigm—making this an opportune moment to systematically investigate such alternatives.}

    \textit{\footnotesize This vision document presents a snapshot of the field as discussed at the \emph{Black Hole Mimickers: From Theory to Observation workshop}, where experts from gravitational wave astronomy, very long baseline interferometry, numerical and mathematical relativity, and high-energy physics converged to assess the current frontiers. By highlighting key open questions and proposing concrete pathways forward, this document aims to guide future efforts to probe the nature of compact objects. As the field stands at the crossroads of theoretical innovation and observational breakthroughs, we outline strategies to harness upcoming observational capabilities to fundamentally test the black hole paradigm.}

    \section*{Motivation and Context}

    The black hole paradigm has exhibited extraordinary success both as an exact mathematical solution and as a framework for interpreting observational data.\footnote{As well as its key role in holographic and quantum information theory efforts.} %
    But it also comes with fundamental conceptual puzzles: the classical black hole hypothesis is practically non-falsifiable, black holes harbor classical spacetime singularities, and event horizons raise deep and longstanding questions in the context of a quantum theory. A satisfactory microscopic description of quantum effects that resolve these issues in astrophysically relevant black holes (non-supersymmetric, asymptotically flat, and in four spacetime dimensions) is still lacking.
    These shortcomings highlight fundamental limitations in our current understanding of black holes, particularly at the interface with quantum physics.

    On the other hand, the observational landscape of black hole physics has transformed in recent years. The detection of gravitational waves from the merger of binary black holes by the LIGO-Virgo-KAGRA (LVK) detector network has opened an unprecedented window into the nonlinear, 
    highly dynamical regime of strong field gravity. At the same time, very long baseline interferometry, through the Event Horizon Telescope (EHT), has produced the first images of horizon-scale structure surrounding putative supermassive black holes. Together, these breakthroughs compel a timely and pressing question: can we---and if so, how---test the classical black hole paradigm with these new observational tools?

    Black hole mimickers---regular, ultra compact, and horizonless objects---challenge the paradigm by trying to address or solve at least some of the theoretical issues alluded above and to explain (some or all) observations. They emerge in a variety of theoretical frameworks, from top-down constructions rooted in string theory (such as fuzzballs, frozen stars, topological stars, and AdS black shells) to bottom-up phenomenological models (like boson stars, gravastars, and fluid stars). For these objects to truly challenge the black hole paradigm, there remain two primary challenges: first, to develop these proposals into physically consistent and well-motivated models amenable to observational scrutiny (particularly in dynamical settings); and second, to identify concrete observational (ideally ``smoking-gun'') signatures that reliably distinguish a mimicker from a classical black hole.

    These broad challenges divide into three large domains, classifying the open problems: (i) the motivation and construction of self-consistent models of black hole mimickers, (ii) the gravitational properties and dynamical behavior leading to gravitational waves, and (iii) the electromagnetic characteristics and plasma interactions impacting very long baseline interferometric observations.

    \section*{Theoretical model building}

    While the set of proposed black hole mimicking objects (both bottom-up and top-down) is vast, few may be elevated to the status of a well-motivated model, and fewer are true black hole mimickers. To qualify as such, proposed models must satisfy stringent criteria. A high-energy physics origin (be it particle physics or models of quantum gravity) is desirable particularly as a framework to understand ``plausible'' formation mechanisms. Theoretical puzzles associated with classical horizons and singularities are solved by the proposed models only if co-existence with classical black holes can be ruled out. More concretely, for a given proposal the following questions arise:

    \begin{itemize}
        \item \textit{Motivation:} 
        Does the proposal descend from a ``well-motivated'' high-energy physics origin?
    
        \item \emph{Formation mechanisms:} Do the proposed mimickers possess a well-established formation mechanism?

         \item \emph{Completeness:} Can these objects plausibly replace \emph{all} classical black holes, thereby resolving all  theoretical issues associated with black holes?
    
        \item \emph{Longevity:} Are these objects sufficiently long-lived in realistic astrophysical environments?
    \end{itemize}
    A self-consistent framework, descending from their high-energy origin, is required to approach these issues and further questions about their macroscopic properties and dynamics. As discussed in further detail below, these include gravitational characteristics, from concerns about stability, to dynamics of binary mimicker systems; but also electromagnetic properties, ranging from interactions with accreting plasma to jet launching. While practical in nature, such framework is crucial to ultimately subject any given mimicker to observational scrutiny. Therefore, finding such a framework, for a given mimicker, is a major open problem. Notably, there is an absence of proposals passing 
    {\em all} of the above. Of course, a successful model should also be consistent with the growing, continuously improving, body of observations so far consistently addressed by the black hole paradigm.

    \section*{Gravitational wave observations}

    Gravitational wave astronomy with the LVK detector network and future observatories offers a powerful window into the strong field regime of gravity and enables us to probe the nature of compact objects. Recent work suggests that black hole mimickers may exhibit unique and distinctive signatures associated with their stable light rings and ergoregions, allowing for tests of the black hole mimicker hypothesis. These include instability processes as well as modified quasi-normal mode spectra relevant both during the inspiral and ringdown phases of a binary mimicker. Yet major challenges remain. Simplifying assumptions had been made to investigate many of these properties, leaving fundamental questions unanswered: 
    \begin{itemize}        
        \item \emph{Instabilities:} What is the role of linear instabilities? How do these alter the objects in the nonlinear regime?

        \item \emph{Light ring dynamics:} What are the nonlinear dynamical properties of stable light rings?

        \item \emph{Merger dynamics:} Are the merger and ringdown phases of (particularly equal-mass) mimicker binaries truly analogous to those of classical black holes?

        \item \emph{Environmental effects:} To what extent could the astrophysical environment obscure tests of the black hole (mimicker) paradigm?

        \item \emph{Universality: }Are these properties universal across different proposed mimickers and scales?
    \end{itemize}
    Addressing these issues is essential to assess the theoretical viability of the black hole mimicker paradigm and use gravitational wave observations to reliably and convincingly test the mimicker hypothesis.

    \section*{Electromagnetic observations}

    Event-horizon-scale radio observations from the EHT have yielded the first images of horizon scale structure around the putative supermassive black holes at the centers of our own galaxy and the nearby elliptical galaxy M87.
    The EHT's images of M87$^*$ and Sgr\,A$^*$ offer insights not only into the surrounding accretion flow and emission processes, but also into the underlying spacetime geometry and nature of the central compact object itself. These constraints arise from calibrated measurements of the ring size, as well as thermodynamic considerations, such as comparisons between expected blackbody emission from hypothetical surfaces and the observed brightness depression at the center of the image. However, the interpretation of electromagnetic observations is necessarily complicated by uncertainties in accretion physics, emission models, and plasma dynamics. In this context, black hole mimickers raise new questions: 
    \begin{itemize}
        \item \emph{Accretion dynamics:} How do mimickers interact with infalling matter? Is the plasma fully absorbed or does a reflecting surface modify the accretion process and jet structure?
    
        \item \emph{Photon ring and shadow:} Can the geometry and brightness of the photon ring or shadow region provide model-independent evidence against the presence of an event horizon?

        \item \emph{Emission:} How do surface emission and photon orbit structures differ from those of classical black holes, and how can they be used to distinguish mimickers?
    \end{itemize}
    Understanding these features is vital for interpreting current and future data and guiding the design and observational strategy of next-generation instruments, which promise significantly enhanced resolution and dynamic range.

    \vspace{1em}

    This vision document compiles the key results, insights, and open questions that emerged over the course of the workshop. In addition to summarizing the talks and discussions of the workshop, we have sought to capture a snapshot of the field as it stands today---highlighting both its current frontiers and the conceptual and technical challenges that remain.
    Our hope is that this document serves not only as a record of the workshop, but also as a springboard for future research. By bringing together diverse communities working on different aspects of black hole mimickers from theoretical modeling to observational strategies, we aim to foster continued collaboration and accelerate progress toward a deeper understanding of compact objects and the nature of spacetime.

    \vspace{1em}

    We are grateful to all of the participants who made this workshop such a successful, intellectually stimulating experience. Special thanks are owed to the Princeton Center for Theoretical Science and the Princeton Gravity Initiative as well as Charlene Borsack and Sarah Siddall for their support. 
	
	\vspace{1cm}
	\begin{flushright}\noindent
		April 2025\hfill Suvendu Giri\\ Luis Lehner\\ Nils Siemonsen\\ George N.~Wong
	\end{flushright}
	\clearpage
	\begin{center}
		\includegraphics[width=\linewidth]{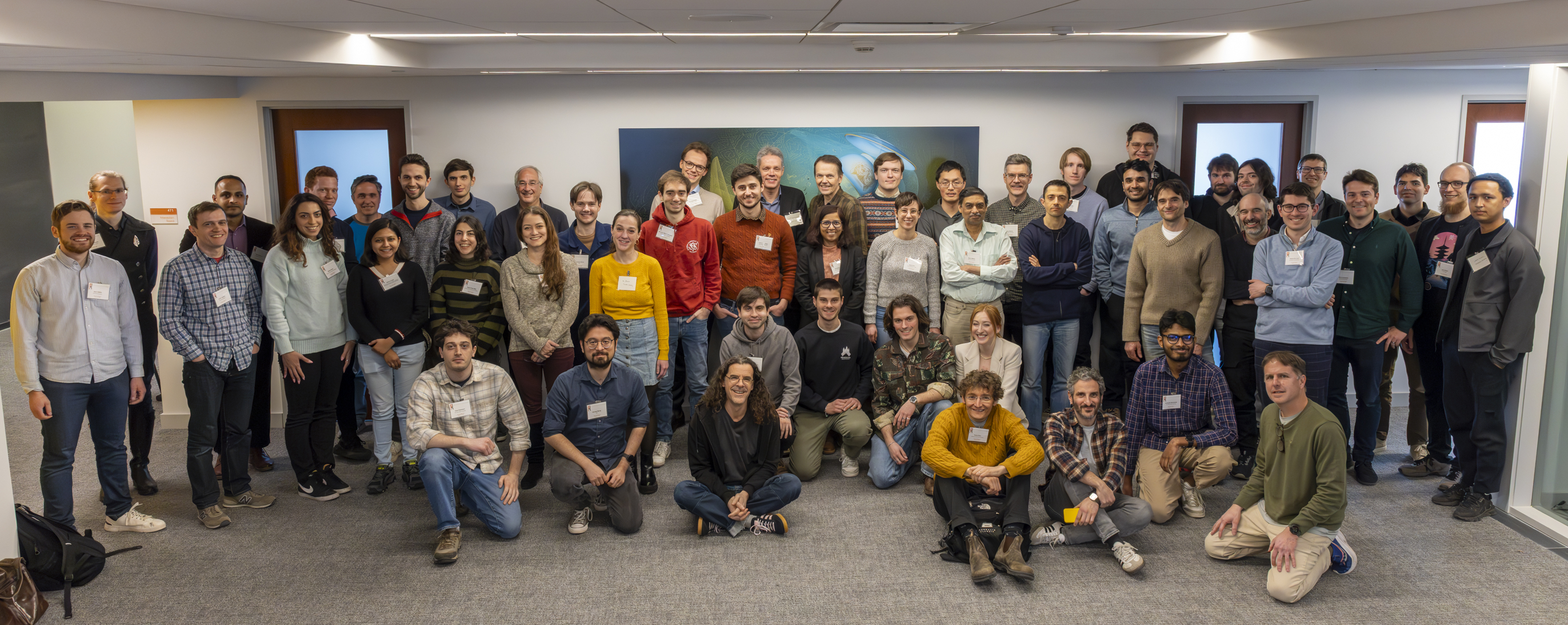}
        \emph{Demonstrating strong-field attraction: workshop participants bound together by shared curiosity about compact objects.}
	\end{center}
    \medskip
	\begin{center}\large{\textbf{Organization}}\end{center}
	\emph{Black Hole Mimickers: From Theory to Observation} was organized at the Princeton Center for Theoretical Sciences (PCTS), Princeton University, NJ, USA from March 3--5 2025. Workshop details, including the schedule and speaker list, can be found on the official website: \url{https://pcts.princeton.edu/events/2025/black-hole-mimickers-theory-observation}. 
    \\
    
    \noindent 
    Some of the recorded talks are available \href{https://www.kaltura.com/index.php/extwidget/preview/partner_id/1449362/uiconf_id/14292362/entry_id/1_ucqiokmv/embed/dynamic}{here}, and the associated Hamilton Colloquium can be viewed \href{https://www.kaltura.com/index.php/extwidget/preview/partner_id/1449362/uiconf_id/14292362/entry_id/1_ucqiokmv/embed/dynamic}{here}.
	\subsection*{Organizing committee}
	   Suvendu Giri \\
		 Luis Lehner \\
		 Nils Siemonsen \\
	   George N.~Wong \\
		
	\subsection*{Administrative support}
		Charlene Borsack \\
		Sarah Siddall \\
		
	\begin{multicols}{3}[\subsection*{Invited Speakers}]
		Cosimo Bambi \\
		Ramy Brustein \\
		Vitor Cardoso \\
		Andrew Chael \\
		Ulf Danielsson \\
		Anuradha Gupta \\
		Pierre Heidmann \\
		Steven Liebling \\
		Alex Lupsasca \\
		Andrea Maselli \\
		Elisa Maggio \\
		Samir Mathur \\
		Lia Medeiros \\
		Alex B.~Nielsen \\
		H\'{e}ctor R.~Olivares-S\'{a}nchez \\
		Paolo Pani \\
		Frans Pretorius \\
	\end{multicols}
	\subsection*{Sponsoring Institutions}
	Princeton Center for Theoretical Sciences (PCTS)\\
	Princeton Gravity Initiative (PGI)
	\tableofcontents
	\mainmatter              %
\title{Resolution of the information paradox -- the fuzzball paradigm}
\author{Samir D. Mathur}
\institute{\textit{Department of Physics,
 The Ohio State University,
  Columbus, OH 43210, USA}
}

\maketitle 

\begin{abstract}
The black hole information paradox has played a central role in the development of quantum gravity ideas, for the past 50 years. 
In this period the paradox has been sharpened to a precise contradiction through the small corrections theorem, and the contradiction appears to be resolved in string theory through a radical change in the picture of the black hole interior.  This change, encoded in the `fuzzball paradigm', establishes a new mode of failure of the semiclassical approximation: Einstein's equations break down if space stretches too {\it fast}. This observation has a direct implication for cosmology:  it predicts an extra energy at the scale of the cosmological horizon which has the right order to agree with dark energy. 
\end{abstract}

In 1975 Hawking noted a remarkable fact: the gravitational field of a black hole destabilized the quantum field theory vacuum around the horizon \cite{Hawking:1975vcx}. Particle pairs are spontaneously created out of this vacuum. One member of the pair (we will call it $b$) escapes to infinity as ‘Hawking radiation’. The other member (which we will call $c$) has negative energy and falls into the hole to lower its mass. The two members of the pair are in an entangled state which we can schematically write as 
\begin{equation}
|\psi\rangle_{pair}=\frac{1}{\sqrt{2}}\left ( |0\rangle_b|0\rangle_c+|1\rangle_b|1\rangle_c\right)
\label{one}
\end{equation}
Thus the entanglement of the radiation with the remaining hole keeps going up monotonically, leading to a sharp conflict at the endpoint of evaporation.  This is known as the black hole information paradox.

Since the Hawking evaporation process involves a large number of radiated quanta, it was hoped by some that the puzzle could be resolved by delicate correlations among the radiated quanta, effects  that were missed in Hawking's leading order computation. But this hope was dashed by the ‘small corrections theorem’ \cite{Mathur:2009hf}. Suppose the state of the emitted pair at evaporation step $i$ had a small correction
\begin{equation}
|\psi\rangle_{pair}=\frac{1}{\sqrt{2}}\left ( |0\rangle_{b_i}|0\rangle_{c_i}+|1\rangle_{b_i}|1\rangle_{c_i}\right)+|\delta\psi_i\rangle, ~~~~\Big | |\delta \psi_i\rangle \Big |<\epsilon
\label{threeq}
\end{equation}
Suppose we further assume that once the radiated quanta recede sufficiently far from the hole, then their state does not change in any significant way (no long-distance nonlocality). Then the entanglement at step $N$ must keep on growing monotonically as
\begin{equation}
S_{ent}(N+1)>S_{ent}(N)+\log 2 -2\epsilon
\label{four}
\end{equation}
In other words, small corrections to the usual semiclassical horizon dynamics cannot resolve the information paradox; we need order {\it unity} corrections.

String theory provides a well-defined theory of quantum gravity which does not suffer from loop divergences. In this theory one can count the degeneracy of states for extremal holes (mass $M$ equaling charge $Q$) by an indirect argument.  At weak coupling an extremal state is just a bound state of the elementary objects in the theory --- strings and branes -- and its degeneracy can be computed, An index argument then sets a lower bound on the degeneracy of states at strong coupling where one expects a black hole \cite{Sen:1995in,Dabholkar:2004yr,Strominger:1996sh}. This bound is found to match Bekenstein's area expression \cite{Bekenstein:1973ur} for the classical hole with the same quantum numbers 
\begin{equation}
S_{micro}=S_{bek}=\frac{A} {4G}
\end{equation}
This agreement also extends to near-extremal states, which can emit radiation. One finds that the radiation rate from the weak coupling brane bound state agrees with the Hawking radiation rate from the corresponding hole \cite{Das:1996wn,Maldacena:1996ix}. Clearly string theory is on the right track to understand black holes, but we still have the information puzzle: the radiation from the weakly coupled brane bound state preserves unitarity, while the Hawing radiation from the corresponding hole does not.

To resolve the puzzle we must understand these brane bound states at strong coupling.  For the  simplest black hole -- the 2-charge extremal hole -- all the microstates were constructed at strong coupling, and found to be {\it fuzzballs}: horizon sized quantum objects with no horizon \cite{Lunin:2001jy,Mathur:2005zp,Bena:2022rna}. Some families of states have been constructed for more complicated holes; in every case the microstate has turned out to be. a fuzzball; i.e., there is no horizon. Fuzzballs radiate from their surface like normal bodies, so there is no information paradox. 

This resolution of the paradox raises a natural question. Semiclassical black hole physics manifests a  beautiful thermodynamics, which was traditionally attributed to the existence of a horizon. If quantum gravity removes the horizon, then do we also lose this thermodynamics? It was found in \cite{Mathur:2023uoe,Mathur:2024mvo} that in $d+1$ dimensions, if any object is sufficiently compact -- i.e., its surface is a distance $s$ outside the horizon radius with
\begin{equation}
s\ll \left ( \frac{M}{m_p}\right )^{\frac{2}{(d-2)(d+1)}} l_p
\end{equation}
then the object must, to leading order, have the same temperature, entropy and radiation rates as the semiclassical black hole. The radiation of course will now preserve unitarity, since it emerges from the surface of the object instead of a vacuum horizon region. 

This leaves a final issue, which is very important. How does  the process of gravitational collapse end up in a fuzzball instead of the traditional vacuum geometry with horizon? The collapse process can be studied using `good slices' which have low curvature everywhere. If we can trust semiclassical physics whenever the curvatures are low (${\mathcal R}\ll l_p^{-2}$), then we will end up with a horizon. Thus we need to find a second mode of breakdown of the semiclassical approximation, a mode that has nothing to do with the curvatures becoming large.

Such a mode of breakdown was described in \cite{Mathur:2024mtf,Mathur:2024ify}.  To understand the idea, first consider electrons and positrons. These exist as real particles with energy $mc^2$. But the {\it vacuum} with $E=0$ contains {\it virtual}  fluctuations corresponding to these particles; these virtual fluctuations turn into the on-shell particles of Hawking radiation  when space is stretched non-adiabatically in the process of black hole formation. 

Now note that the electron and positron also form a {\it bound} state: the positronium, with a size $\sim 1$~Å.  Should we also have virtual fluctuations of positroniums? A moment's thought tells us that including positronium fluctuations in the vacuum wavefunctional would be overcounting. Instead, the existence of the positronium shows up as a correlation between the locations of the electron and positron fluctuations: the energy of a configuration is slightly lower when these fluctuations are $\sim 1$~Å apart, and the amplitude for such configurations in the vacuum wavefuntional is correspondingly a little higher.

There are other bound states in the standard model, and they each leave their imprint on the vacuum. But these effects are typically small, and since they are on distances of order the size of the bound state, they are not  important at the $\sim 3\, Km$ length scale involved in the evaporation of a solar mass black hole. But are there other bound states in the theory that we might have forgotten? Yes -- the $\exp[S_{bek}(M)]$ microstates of black holes with $0<M<\infty$ ! The fuzzball microstates are  bound states of the nonperturbative objects in string theory -- like Kaluza-Klein monopoles -- so the relevant correlations are those between the planck scale fluctuations of nonperturbative objects.  For a quantum field with  mass $\sim m_p$, the correlations would fall off exponentially for distances $L\gg l_p$. But the fuzzballs are extended objects with radius $R\sim 2GM$, and they are very numerous, so we conjecture that the effect of these fuzzballs on the vacuum wavefunctional if to make these correlations fall off like a power law rather than an exponential.  

Correlations in quantum theory are described by entanglements, and we describe this entanglement fall-off as follows. The vacuum with optimal entanglements has energy $E=0$. If we have these optimal entanglements for distances $r<R$ but not  for distances $r>R$, then the energy of the state will be higher by some amount $\Delta E$.  Using the natural scale suggested by black hole physics $R=2GM$, we set
\begin{equation}
\Delta E \sim \frac{R}{G}
\label{three}
\end{equation}
We can now see how semiclassical physics works for slow processes like star formation, but not when a black hole forms. When space stretches, new fluctuations are created in between the fluctuations which are already entangled with each other. To entangle the new fluctuations in the optimal manner to get the vacuum, we have to first break the entanglements that exist. Star formation is a slow process, and when space stretches, the entanglements between Planck scale fluctuations have time to readjust to the form required for the vacuum on the new spacetime. But when a black hole forms, points inside the horizon cannot exchange signals with points outside, so these entanglement cannot readjust to their vacuum form. Such non-optimal entanglements happen across distances  $R\sim GM$, so (\ref{three}) gives
\begin{equation}
\Delta E \sim M
\end{equation}
Thus evolution along the `good slices' is not possible: the region on these slices that we traditionally assume to be in a  vacuum 
state does not in fact have the correlations appropriate to the vacuum. There is an extra energy $\sim M$ for every interval of length $\sim GM$ along the ever-stretching good slices. Thus semiclassical evolution breaks down, and the wavefunctional evolves to a linear combinations of fuzzballs instead. 

Interestingly, a similar effect occurs at the cosmological horizon, and the extra energy (\ref{three}) has the scale to agree with the dark energy we see today \cite{Mathur:2021zzr}. It would be interesting to explore these cosmological ideas further.

\providecommand{\href}[2]{#2}\begingroup\raggedright\endgroup

\title{Black Shields}
\author{Vitor Cardoso}
\institute{\textit{Center of Gravity, Niels Bohr Institute, Blegdamsvej 17, 2100 Copenhagen, Denmark},
\and
\textit{CENTRA, Departamento de F\'{\i}sica, Instituto Superior T\'ecnico -- IST, Universidade de Lisboa -- UL, Avenida Rovisco Pais 1, 1049-001 Lisboa, Portugal}}

\maketitle 

\begin{abstract}
I overview the problem of quantifying the evidence for black holes, and list a few outstanding challenges in this foundational venture.
\end{abstract}

\section{Introduction}
\hspace{1.8cm}
\parbox{0.8\textwidth}{{\small 
\noindent {\it ``The following calculation yields the exact solution of the problem. It is
always pleasant to avail of exact solutions of simple form.''}
\begin{flushright}
Karl Schwarzschild, On the Gravitational Field of a Mass Point according to Einstein’s Theory (2015)~\cite{Schwarzschild:1916uq}.
\end{flushright}
}
}
\vskip 2mm
\hspace{1.8cm}
\parbox{0.8\textwidth}{{\small 
\noindent {\it Shield: To protect (a person or object) by the interposition of some means of defence; to afford shelter to; to protect (an accused person, etc.) by authority or influence. }
\begin{flushright}
From the Oxford English Dictionary
\end{flushright}
}
}

\vskip 4mm

Scientific knowledge evolves and improves mostly by questioning existing paradigms and testing them to exhaustion. Healthy doses of curiosity, skepticism, extensive data collecting and intellectual daring tend to make for a robust scientific building. Historically, under normal circumstances, the discovery of a new object or living being, planet or star would lead us to their study and characterization. Eventually, the object would be placed in a catalog alongside similar entities. In this context, black holes are very particular: they were predicted to exist solely based on analysis of known (or extrapolated) equations of state, as the outcome of gravitational collapse and as exact (and unique) solutions of Einstein equations. Observational characterization of their properties was not an ingredient in making black holes part of our lexicon.

The classical field equations make no sense in black hole interiors, and there are puzzles associated with quantum field theories on spacetimes with horizons.
Questioning and quantifying their existence is {\it imposed} on us: in an era where data concerning the strong field regime of gravity is available, any other option would be foolish~\cite{Cardoso:2019rvt,Buoninfante:2024oxl,Carballo-Rubio:2025fnc}. The discussion below assumes that if the objects are not extremely compact (see discussion and terminology in Ref.~\cite{Cardoso:2019rvt}) well-known tools are in place to discriminate them from black holes. These will not be discussed much further. I implicitly assume that the geometry admits light rings, high redshifts, and that it is ``close enough'' to having horizons. Most of the discussion parallels a review on the subject~\cite{Cardoso:2019rvt}, new results are sometimes mentioned. 

\section{Steps to darkness}
\begin{figure}
    \centering
    \includegraphics[width=\linewidth]{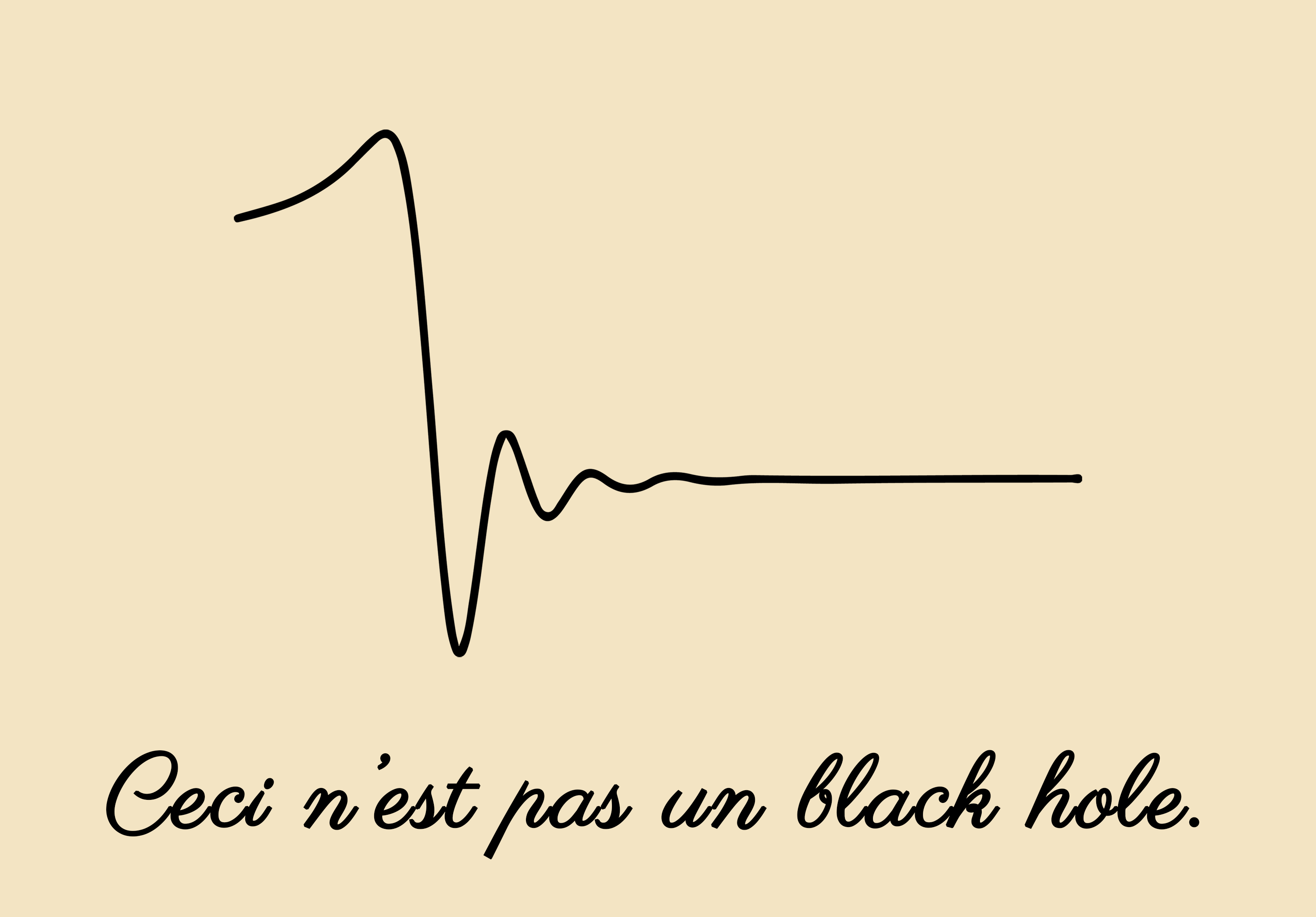}
    \caption{A muse on Ren\'e Magritte's art piece ``Ceci n'est pas une pipe.'' By Vitor Cardoso and Ana Carvalho (2016).}
    \label{fig:schematic_overview}
\end{figure}
When faced with the task of assessing the existence of horizons in our universe, there are two alternatives.

\begin{description}[font=$\bullet$~\normalfont\scshape\color{red!50!black}]

\item [consistency checks within paradigm] The first is to collect data of very high quality and see whether it describes black holes as predicted by General Relativity. This task requires both good detectors and a good knowledge of the predictions from General Relativity. The assumption that these objects are merging in vacuum also needs to be questioned.

\item [detailed understanding of alternatives] Although consistency checks within the theory are important, they should (because they can!) be complemented with searches for specific signatures or smoking guns of alternatives to black holes. This is also a fertile terrain for new theoretical developments and for understanding the challenges associated with compact objects without horizons. Roughly speaking this program would require:

$\bullet$ {\it Formulating extensions of General Relativity, or matter sector, admitting compact objects}. So far, satisfactory solutions include minimally coupled massive scalar or vector fields. These are satisfactory in the sense that they arise out of a simple, well-posed theory. Progress in simulating binaries of these objects has been superb~\cite{Cardoso:2016oxy,Palenzuela:2017kcg,CalderonBustillo:2022cja,Siemonsen:2024snb}.

The introduction of a new scale usually leads to a maximum mass, and a mass-radius relation that may differ substantially from that of a black hole in most of parameter space. Hence, explaining observations of compact objects whose mass differs by orders of magnitude is challenging.

I would like to make two observations here. The first is that even main-sequence stars satisfy $R\propto M^{0.8}$ and that it is not difficult to cook up an energy generation such that $R\propto M$\footnote{In particular, for a nuclear energy generation rate per unit mass $\epsilon_{\rm nucl}=\kappa \rho^{-3/2} T^\beta$ with $\kappa$ a constant, $\rho$ the density and $T$ the temperature of a radiative star, then the star radius scales with its mass~\cite{Kippenhahn:2012qhp}.}.

Secondly, all or almost all efforts focus on static or stationary solutions. It is not inconceivable that quasi time-periodic, or otherwise highly dynamical solutions bouncing off at some energy scale could also be relevant in the zoo and physics of ultracompact objects.

$\bullet$ {\it Formation mechanisms}, out of broad initial conditions, such that we would expect our universe to produce them in copious amounts. Given appropriate primordial abundance, light bosons are attractive candidates, and they can even be seeded by black hole superradiance~\cite{Brito:2015oca}. Formation mechanisms for any other credible alternatives are unknown or require exotic ingredients.

$\bullet$ {\it Stability on cosmological timescales}. Linear or nonlinear stability of horizonless objects is a fascinating topic, and shows how the quest for quantifying horizons can be fruitful. If the spacetime exterior to the object is very similar to the Kerr geometry (although there is no reason nor result that suggests so), then rotating compact objects should be plagued by ergoregion instabilities~\cite{Brito:2015oca,Friedman:1978ygc,Moschidis:2016zjy,Vicente:2018mxl}. For moderate rotation rates -- and assuming all compact objects are horizonless -- these would act as to produce a stochastic background of gravitational waves, which has not been measured~\cite{Barausse:2018vdb}. A proper assessment of the impact of dissipation on the ergoregion instability is still to be made. Ad hoc suppression of absorption show that one can suppress the instability~\cite{Maggio:2017ivp}, but it is unknown whether this corresponds to any possible physical mechanism. Indeed, thermodynamical arguments show that any spinning body should amplify incoming radiation~\cite{Bekenstein:1998nt} (also Section 3.5 in Ref.~\cite{Brito:2015oca}).

It has been conjectured that geometries with a light ring, even if non-spinning, can be unstable, albeit against a nonlinear mechanism~\cite{Keir:2014oka,Cardoso:2014sna,Zhong:2022jke}. Recent results on nonlinear partial differential equations question this result~\cite{Redondo-Yuste:2025hlv}, but it is clearly a subject where both exact results and nonlinear simulations are desperately needed\footnote{Particularly so to confirm or disprove results claiming an instability on short timescales, but with unknown dependency on initial data or resolution~\cite{Cunha:2022gde}.}. The challenge ahead is to estimate timescales for these and other putative instabilities of compact objects.

$\bullet$ {\it Observable signatures} that would allow them to pass current observations or to be discriminated in the future. It is possible to roughly divide imprints of new physics in two categories. One would be smoking guns, that stand out clearly in the signal. In the low-frequency band, inspiral stage, these could include the excitation of resonances of the inspiralling objects due to the absence of horizons: for ultracompact objects, the surface and the light ring could form a ``cavity'' where low-frequency gravitational waves can live~\cite{Cardoso:2019nis,Maggio:2021uge}. It is expected that the coalescence will excite the light ring and produce a series of bursts of {\it echoes} that are a universal feature of very compact horizonless objects~\cite{Cardoso:2019rvt,Cardoso:2016rao,Cardoso:2016oxy,Agullo:2020hxe,Siemonsen:2024snb}. Detection strategies and searches for these features are now developed and are an important tool to test the black hole paradigm~\cite{Abedi:2016hgu,Westerweck:2017hus,LIGOScientific:2020tif,LIGOScientific:2021sio,PhysRevD.108.104040}. On the other hand, persistent but small modifications to the General Relativistic vacuum black hole prediction are expected. These could show up as non-zero static tidal Love numbers of ``black holes''~\cite{Cardoso:2017cfl,Maselli:2017cmm,Chia:2023tle}, or anomalous tidal heating~\cite{Maselli:2017cmm,Datta:2019epe}. It is obvious that any change in the structure of compact objects will not manifest itself in only the way they vibrate or are tidally distorted: the full waveform will be affected. In particular, the most violent stage during the binary interaction -- the merger -- will carry a significant portion of the power. Unfortunately, unless an outstanding new theory of gravity to solve current issues in physics manifests itself, all we have are ad hoc, incomplete models to evolve through merger and ringdown. We need universal or robust signatures of generic theories to be able to move forward (or then, a better description than General Relativity).

It is important to keep in mind that electromagnetic observations of compact objects have also reached exquisite precision. In the assumption of thermal equilibrium between central object and surroundings, strong constraints can be imposed on a couple of supermassive compact objects~\cite{Broderick:2009ph,Cardoso:2019rvt,Ayzenberg:2023hfw}. 
\end{description}

\section{Discussion}
\hspace{1.8cm}
\parbox{0.8\textwidth}{{\small 
\noindent {\it ``There is a crack in everything. That's how the light gets in.''}
\begin{flushright}
Leonard Cohen, Anthem (1992).
\end{flushright}
}
}
\vskip 2mm
We are living privileged times: data of increasing quality, referring to an increasingly large number of compact objects, is available to us. All the data we have indicates that the weirdest predictions of General Relativity are materialized by nature.
In particular, innermost stable orbits, ergoregions, light rings and even horizons seem to be present in the cosmos. The quest to pinpoint and quantify the extent to which these features exist led us to a much more precise understanding of the theory, and of our ability to extract information from signals. Like Maxwell and others have pointed out, science lies in the next decimal place~\cite{Lawrence:1972}. It is our duty to go and dig it out.

\vskip 2mm
\noindent {\bf Acknowledgments.} 
I am thankful to Ana Carvalho and Jaime Redondo-Yuste for useful feedback and to Alex Nielsen for inadvertently providing me with a title to this contribution. The Center of Gravity is a Center of Excellence funded by the Danish National Research Foundation under grant No. 184.
We acknowledge support by VILLUM Foundation (grant no. VIL37766) and the DNRF Chair program (grant no. DNRF162) by the Danish National Research Foundation.
V.C.\ is a Villum Investigator and a DNRF Chair.  
V.C. acknowledges financial support provided under the European Union’s H2020 ERC Advanced Grant “Black holes: gravitational engines of discovery” grant agreement no. Gravitas–101052587. 
Views and opinions expressed are however those of the author only and do not necessarily reflect those of the European Union or the European Research Council. Neither the European Union nor the granting authority can be held responsible for them.
This project has received funding from the European Union's Horizon 2020 research and innovation programme under the Marie Sklodowska-Curie grant agreement No 101007855 and No 101131233.
\bibliographystyle{utphys}

\providecommand{\href}[2]{#2}
\begingroup\raggedright

\endgroup

\title{Towards precision observational constraints on black holes and black hole mimickers}
\titlerunning{Towards precision observational constraints}
\author{Alex B. Nielsen}
\institute{\textit{Institute of Matematics and Physics, \\ University of Stavanger, NO-4036, Stavanger, Norway}}

\maketitle 

\begin{abstract}
Gravitational wave observations and dynamical spacetimes entail profound challenges for black hole mimickers. Horizonless objects require large modifications in the physics near where black hole horizons would be. We recall how in dynamical spacetimes, different notions of horizon have different locations and so novel mimicker physics can arise in different places. We review unexpected observational progress in constraining black holes from their gravitational ringdown signal. The event GW190521 shows clear evidence of a multimodal ringdown structure and gravitational wave spectroscopy is likely to be an important observational tool in the coming years. 
\end{abstract}

Black holes are one of the most well-known objects in all of physics. In the past ten years, new observational tools have become available, that allow theoretical models of black holes to be confronted with a wealth of observational data. A primary question in interpreting this data, is whether a given astronomical source is indeed a black hole or not. A working definition of a black hole for many astronomers is a dark, compact object that is too massive to be a neutron star \cite{Visser:2008rtf}. Fundamental physics motivates us to go beyond this and study the difference between black holes and black hole mimickers -- objects that are astronomically dark and compact, but aren't actually black holes.

For fundamental physics, the key difference between black holes and black hole mimickers is the existence of horizons. Black hole mimickers require large differences in the physics near where black hole horizons would be. Standard textbook definitions of black holes define black holes in terms of event horizons \cite{Wald:1984rg}. Event horizons themselves are defined in terms of future null infinity and are by definition teleological. They cannot be located unless information is provided about the physics all the way to infinity. This poses minor challenges for numerical solutions of black hole spacetimes \cite{Cohen:2008wa}, but poses a major conceptual challenge for identifying astrophysical black holes. {\it Event horizons cannot be identified using finite measurements.} Since all actual observations are finite and only extrapolations can get us to infinity, we cannot actually identify black holes if we look for observational evidence of event horizons.

Several proposals have been made to circumvent this conceptual issue. One approach is to define black hole horizons based on Penrose's notion of a trapped surface, where all normal light rays are instantaneously converging \cite{Penrose:1964wq}. This approach includes trapping horizons \cite{Hayward:1993wb} and dynamical horizons \cite{Ashtekar:2004cn}. Common to these surfaces is that they are spacelike when growing in size. Trapping horizons have been shown to obey thermodynamic relations \cite{Hayward:1993wb,Nielsen:2008cr} and can readily be identified using only bounded measurements of the gravitational field. These surfaces coincide with event horizons in eternally static spacetimes like the Schwarzschild solution, but differ, often macroscopically \cite{Nielsen:2010gm}, from event horizons in dynamical situations such as accretion or during mergers.

Another approach, combining both the local property of trapped surfaces and the null property of event horizons, is conformal Killing horizons \cite{Nielsen:2017hxt}. These horizons inherit many of the useful properties of regular Killing horizons, but can be extended to dynamical spacetimes and play a role in the emission of Hawking radiation \cite{Nielsen:2012xu}.

It is in strongly dynamical spacetimes where these different definitions of a black hole are most apparent and where the modelling of black hole mimickers is most challenging. Dynamical models of black hole mimickers need to be clear about where the physics deviates from standard black holes. Thanks to gravitational wave astronomy, we now have observational data from dynamical compact binaries \cite{LIGOScientific:2016aoc,LIGOScientific:2020tif,LIGOScientific:2021sio}. Current gravitational wave data provide a number of avenues for testing black hole mimicker models; in the inspiral phase due to tidal deformations \cite{Johnson-Mcdaniel:2018cdu}, from the merger-ringdown phase and from late-time echo signals due to less than perfect absorption \cite{Cardoso:2016rao}.

In the ringdown phase after a single object has formed, the standard expectation for black holes is that they will rapidly settle down to a Kerr state through the emission of quasi-normal modes \cite{Teukolsky:1972my}. That a stationary, vacuum black hole with a regular horizon should be described by the Kerr metric is supported by the black hole uniqueness theorems \cite{Chrusciel:2012jk} summarised as {\it black holes have no hair}. This in turn implies that the frequencies and damping times of all emitted quasi-normal modes should only be functions of two parameters, the final mass and the final spin. If more than one such mode is measured, these two parameters will be overdetermined and one can test the assumptions of the uniqueness theorems and by extension, alternatives to Kerr black holes. The number of modes that can actually be measured depends on the parameters of the progenitors and the details of the non-linear merger in a way that is not yet fully mapped out \cite{Abedi:2023kot}. This all is the programme of black hole spectroscopy \cite{Dreyer:2003bv}.

\begin{figure*}
\includegraphics[width=\textwidth]{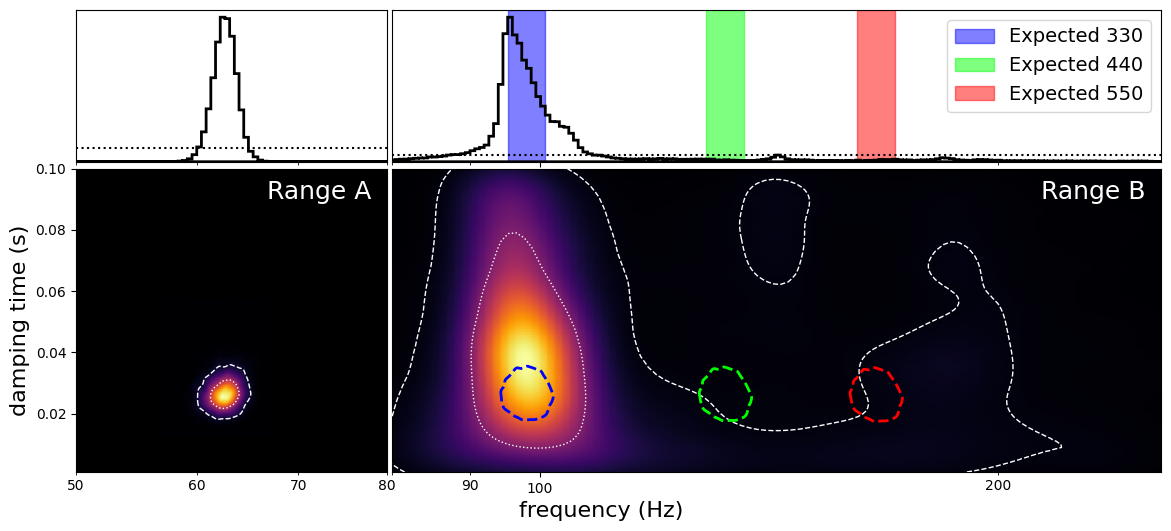}
\caption{Results of a search for damped sinusoid ringdown modes in the data of gravitational wave event GW10521, showing evidence of two distinct modes at frequencies $\sim63$Hz and $\sim98$Hz. These can be identified as the dominant 220 and sub-dominant 330 ringdown modes of a perturbed Kerr black hole. No evidence is found for higher 440 or 550 modes. Figure reproduced with permission from \cite{Capano:2021etf}.} 
\label{fig:agnostic6ms}
\end{figure*}

Initial estimates were that observational black hole spectroscopy would take some time to be realised and was unlikely to occur until the arrival of space-based detectors or next-generation detectors, sometime in the 2030's  \cite{Berti:2016lat,Cabero:2019zyt}. But nature has been kind to the gravitational wave astronomer and already now, observations are being made that identify multiple modes in the ringdown phase of merger events \cite{Isi:2019aib,Capano:2021etf,Siegel:2023lxl}. In the case of GW190521, a high mass system \cite{LIGOScientific:2020iuh}, the ringdown phase shows evidence of two distinct modes, see Fig.(\ref{fig:agnostic6ms}), with a Bayes factor of $56\pm1$ preferring two fundamental modes over one \cite{Capano:2021etf}. Such results have opened the door to black hole spectroscopy and testing black hole mimicker models by their gravitational wave response to merger-induced perturbations.

A validation of the multi-modal result for GW190521 was performed using simulated signals in \cite{Capano:2022zqm} and evidence for multiple modes was confirmed independently \cite{Siegel:2023lxl}. Although different analysis techniques for performing black hole spectroscopy are still being developed, they offer the possibility of testing a great variety of different waveform features in both the frequency domain and time domain \cite{Kastha:2021chr}. Even the conservative expectations of \cite{Berti:2016lat,Cabero:2019zyt} suggest that multi-modal spectra will be observable for many events per year when space-based detectors and next-generation ground-based detectors come online. These observations will provide precision constraints on alternatives to Kerr black holes that will be competitive with, and likely exceed, those possible using other techniques \cite{Ahmed:2024ykc}.

Spectroscopy has been an important tool for astronomy. The possibility of using gravitational waves to perform spectroscopy on perturbed black holes is very exciting. The challenge for black-hole-mimicker modellers is to show that their models are compatible with such observational results. These results will only improve in precision as detector sensitivity increases and new detectors are built.

\providecommand{\href}[2]{#2}\begingroup\raggedright\endgroup

\title{Black Hole Images with the Event Horizon Telescope}
\author{Andrew Chael}
\institute{\textit{Princeton Gravity Initiative, Princeton University, Princeton NJ, 08544}}

\maketitle 

\begin{abstract}
The Event Horizon Telescope (EHT) has made images of millimeter-wavelength radio emission on event-horizon scale around the largest apparent supermassive black holes M87* and Sgr A*. Black hole images are in principle unique probes of spacetime and the nature of astrophysical black holes. However, using these images to distinguish between black holes and black hole mimickers requires understanding both what features  are well constrained by EHT measurements and how the details of the astrophysical emission mechanism contribute to the observed image structure. 
\end{abstract}

\section{Summary of EHT Results}

The Event Horizon Telescope (EHT) is a global network of radio telescopes observing at a frequency of 230 GHz. Using Very-Long-Baseline Interferometry (VLBI) at the highest radio frequencies ever achieved \cite{PaperII,Raymond2024}, the EHT has produced resolved images of the hot plasma a few Schwarzschild radii from the supermassive black holes in the Galactic Center \cite{SgrAPaperI,SgrAPaperII,SgrAPaperIII,SgrAPaperIV,SgrAPaperV,SgrAPaperVI,SgrAPaperVII,SgrAPaperVIII} and the elliptical galaxy M87 \cite{PaperI,PaperII,PaperIII,PaperIV,PaperV,PaperVI,PaperVII,PaperVIII,PaperIX}. 
Despite a difference of three orders of magnitude in the black hole mass,
both rings have diameters $d\approx10\,GM/Dc^2$, where $M$ is the black hole mass and $D$ is the angular diameter distance. The diameters of both observed black hole images are close to the expected ``black hole shadow'' diameter, 
consistent with predictions from models of hot, near-horizon accretion \cite{PaperV}. Updates to both M87* and Sgr A* images in linear and circular polarization have highly constrained the magnetic field strength and geometry around both black holes \cite{PaperVIII,SgrAPaperVIII}. 

The EHT images of Sgr A* and M87* directly probe supermassive accretion and jet launching \cite{PaperV,PaperVIII,SgrAPaperV,SgrAPaperVIII} high-energy plasma physics \cite{Ho2024,Chael25}, and general relativistic light bending, redshift, and parallel transport \cite{PaperVI,SgrAPaperVII,PaperVIII,SgrAPaperVIII}. Because all of these effects are significant in determining the structure of EHT images, using EHT images to test theories of gravity or of black hole mimickers requires understanding the key features that are extractable from EHT images and their dependence on the standard astrophysical model of low-luminosity plasma accretion onto black holes. 

\section{Image Parameters}

\begin{figure*}[t!]
\centering
\includegraphics[width=0.9\linewidth,trim={0cm 0cm 0cm 0cm},clip]{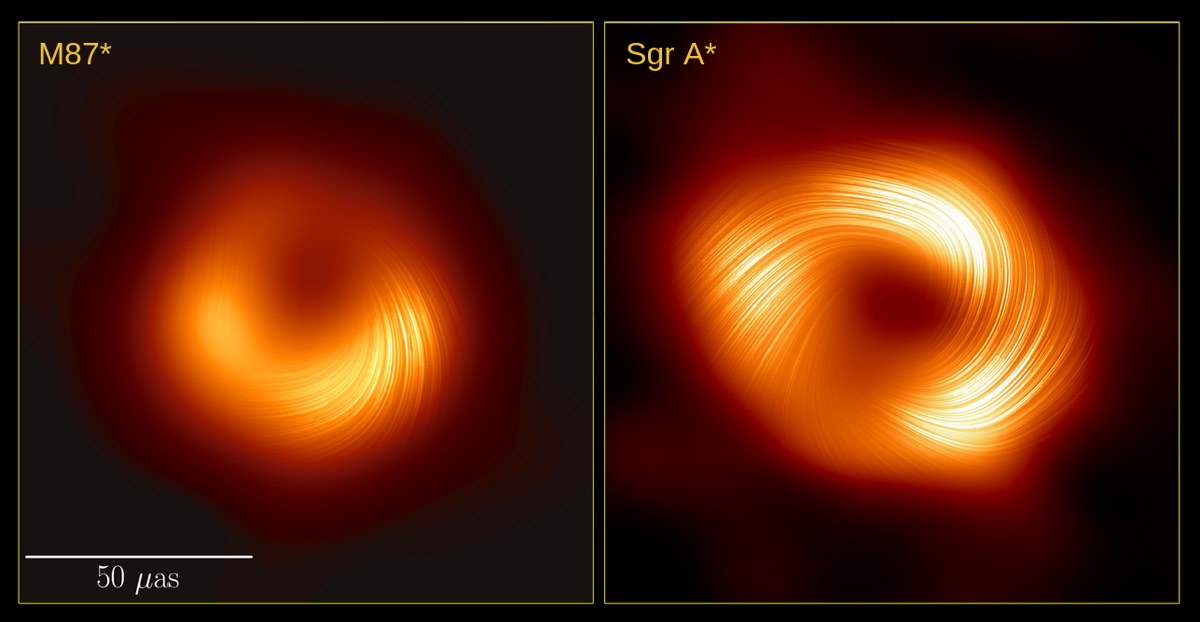}
\vspace{-0.3cm}
\caption{
Polarized images of the supermassive black holes M87* (left; \cite{PaperI,PaperVII}) and Sgr A* (right; \cite{SgrAPaperI,SgrAPaperVII}) observed by the Event Horizon Telescope (EHT) at 230 GHz. Despite three orders of magnitude difference in black hole mass $M$, both sources show a prominent ring of emission with a dim interior and a diameter $d\approx10 GM/Dc^2$, where $D$ the angular diameter distance to the source. The spiral linear polarization pattern of both images, here indicated by the thin lines, strongly constrains the geometry of near-horizon magnetic fields in both sources \cite{PaperVIII,SgrAPaperVIII}.  (Image Credit: EHT Collaboration)
}
\label{fig::ehtims}
\end{figure*}

EHT images are the end result of a long data pipeline where data are recorded at widely-separated telescopes \cite{PaperII}, correlated, calibrated \cite{PaperIII}, and imaged \cite{PaperIV}. While the EHT pipeline leverages over 50 years of experience in VLBI, the EHT faces unique challenges in the high data rates and rapid atmospheric phase fluctuation inherent to 230 GHz observations \cite{PaperII}.   
The EHT collaboration has relied on building and testing multiple analysis pipelines at each stage to ensure reliability. For instance, the final polarized images of M87* from the EHT's 2017 observations were produced by five independent image reconstruction and polarimetric calibration methods \cite{PaperVII}. 
 
The EHT images of both Sgr A* and M87* show ring-like structure with a relatively dim interior. The diameter of both images is close to that predicted for the ideal black hole shadow ($d_{\rm shadow}=2\sqrt{27}GM/Dc^2$). In M87*, the ring width is not well constrained, with an upper limit $w/d<0.5$ dependent on the image reconstruction method \cite{PaperIV}. In Sgr A*, the ring width is resolved, with $w/d \in (0.3,0.5)$ \cite{SgrAPaperIII}. The ring position angle $\eta$ is not constrained in Sgr A* \cite{SgrAPaperIII}, while in M87* the position angle is well constrained and is observed to shift from $\eta\approx170\deg$ in the EHT's 2017 observations \cite{PaperIII} to $\eta\approx208\deg$ in 2018 \cite{EHT2024M87}. 

Polarimetric images of both sources are more constraining of the accretion flow parameters than the total intensity structure. Both images show a characteristic spiral pattern of linear polarization vectors around the emission ring \cite{PaperVII,SgrAPaperVII}.
In both sources, circular polarization is detected but not confidently reconstructed in the image plane \cite{PaperIX,SgrAPaperVIII}. 
Sgr A* is more linearly polarized than M87*, with an average linear polarization fraction at the EHT's resolution scale of $\langle|m|\rangle\approx 26\%$ \cite{SgrAPaperVII} vs M87*'s $\langle|m|\rangle\approx8\%$. For both sources, the EHT characterizes the pattern of linear polarization in terms of its second azimuthal Fourier mode, $\beta_2$  \cite{Palumbo20}. After accounting for Faraday rotation, both sources have $\beta_2$ phase values with the same sign, indicating that a consistent helicity of the average linear polarization pattern; 
$\angle\beta_2\approx -150\deg$ for M87* and $\angle\beta_2\approx -170\deg$ for Sgr A*, after Faraday derotation. This sign of $\angle\beta_2$ is consistent with the expectation from models with electromagnetic energy outflow from the black hole where the black hole spin points into the plane of the sky \cite{Chael23}.

\section{Astrophysical Interpretation}

230 GHz emission from Sgr A* and M87* is produced by synchrotron radiation from relativistic electrons in the hot ($T>10^{10}$K), magnetized accreting plasma \cite{Ichimaru77,NarayanYi94,Yuan14}. The primary tools used by the EHT collaboration for investigating the structure and dynamics of these hot accretion flows are general relativistic magnetohydrodynamic (GRMHD) simulations \cite{Komissarov99,Gammie03}.
To constrain the properties of \m87 and Sgr A*, features from EHT images and multi-wavelength constraints were compared to large libraries of GRMHD simulation images spanning a range of different parameters, including the black hole spin, accumulated magnetic flux on the black hole, and ion-to-electron temperature ratio.

GRMHD scoring procedures by the EHT collaboration have found that while total intensity images of M87* are consistent with a wide variety of models \cite{PaperV}, polarimetric images strongly prefer models of magnetically arrested (MAD) accretion flows \cite{PaperVIII,PaperIX}, where the horizon-scale magnetic field is strong, ordered, and dynamically important \cite{Bisnovatyi1974,Narayan2003}. Total intensity and multi-wavelength data more strongly constrain Sgr A*'s properties, which also prefer MAD models \cite{SgrAPaperV}; this preference for strong magnetic fields in Sgr A* is confirmed by the polarimetric image analysis \cite{SgrAPaperVIII}. Sgr A* GRMHD models are generically more variable than historical sub-mm observed lightcurves of the source \cite{SgrAPaperV}; this so-called ``variability crisis'' poses a challenge to the GRMHD interpretation. 

EHT results from both M87* and Sgr A* prefer black hole spins pointed away from the line of sight \cite{PaperV,SgrAPaperVIII}; in Sgr A*, this orientation is also preferred by constraints on the angular velocity of near-infrared \cite{Gravity2018} and sub-millimeter \cite{Wielgus2022} flaring regions.

\section{Next Steps}

The next decade should see a sizable advance in the capabilities of black hole images for testing astrophysical and gravitational theories. 
Increasing the bandwidth and number of ground sites in the current EHT array \cite{Doeleman19,MidRange,Johnson23} will improve the array's dynamic range, enabling detection of the jet base in M87* and more stringent tests for the existence of a jet outflow in Sgr A*. Multi-frequency imaging with the expanded EHT at 86,230, and 345 GHz \cite{Chael23a} will resolve key degeneracies in the properties of the accretion flow \cite{Ricarte22}. High-dynamic-range, multi-frequency EHT images may also reveal the black hole's achromatic ``inner shadow,'' feature, which in MAD simulation models approaches the lensed direct image of the equatorial event horizon \cite{Chael21}, providing a new gravitationally lensed feature that can help resolve degeneracies in constraining the black hole mass, spin, and inclination from images \cite{Chael21}.  

Beyond ground-based VLBI, the proposed Black Hole Explorer (BHEX) NASA Small Explorers mission would triple the current EHT's resolution by providing interferometric baselines from Earth to mid-Earth orbit \cite{BHEX}. BHEX would enable the detection of the highly lensed photon ring \cite{Johnson20} and increase the number of resolvable black holes from two to ${\sim}$ten \cite{Zhang24}. BHEX measurements of Sgr A*, M87*, and other black holes will enable direct spin measurements from strong gravitational lensing as well as direct mass and spin measurements for an ensemble of local low-luminosity black holes \cite{BHEX}. 

Strong horizon-scale magnetic fields near black holes naturally produce powerful electromagnetic outflows. The Blandford-Znajek \cite{BZ} mechanism, where electromagnetic energy is extracted from a black hole's spin via magnetic fields that thread its event horizon, is widely believed to provide the launching power from extragalactic jets \cite{Begelman1984}, including the famous jet from M87, which has been observed in over a century of multi-frequency observations to deliver energy from horizon to galactic scales \cite{EHTMWL,Lu2023}. Future polarimetric observations with the EHT and BHEX may be able to probe the wind-up of black hole magnetic fields at multiple scales from close to the event horizon down the extended jet \cite{Gelles2025}, enabling a direct test of the Blandford-Znajek mechanism \cite{Chael23}. 

Finally, EHT images of black holes are sensitive to strong gravitational lensing and redshift and can be used to test the black hole metric \cite{Psaltis20,SgrAPaperVI}. Current EHT results show that the image size in M87* and Sgr A* is within ${\sim}10\%$ of that predicted under the Kerr metric under a wide range of astrophysical assumptions about the accretion flow \cite{SgrAPaperVI}. Furthermore, the observed size, central brightness depression, and broadband spectra of Sgr A* places stringent limits on the radius of any thermalizing or reflective surfaces above the event horizon \cite{Broderick2009,SgrAPaperVI}. Increasingly sensitive observations with the EHT and narrowing the allowed space of astrophysical accretion flow models will further enhance the power of these tests with near horizon images, and the extension of horizon-scale imaging to ${\sim}10$ sources with BHEX could enable imaging tests of the black hole metric and constraints on the existence of a horizon over a range of black hole mass scales. 

\providecommand{\noopsort}[1]{}
\providecommand{\href}[2]{#2}\begingroup\raggedright\endgroup

\title{Gravitational wave observations of black hole mimickers: 
where do we look, and what do we look for}
\titlerunning{Gravitational wave observations of black hole mimickers}
\author{Andrea Maselli}
\institute{\textit{Gran Sasso Science Institute (GSSI), I-67100 L’Aquila, Italy},
\and
\textit{INFN, Laboratori Nazionali del Gran Sasso, I-67100 Assergi, Italy}}

\maketitle 

\begin{abstract}
Gravitational wave (GW) observations of finite-size effects in the signals from 
coalescing binaries have opened a window into the internal structure of 
relativistic sources. Crucially, these observations provide a new avenue for investigating 
the existence of exotic objects—compact as neutron stars (NSs) and black 
holes (BHs)—that may coexist with them in the Universe. In this proceeding, we examine 
how finite-size effects influence the inspiral dynamics of binary mergers, their 
corresponding GW signatures, and their potential for testing the Kerr nature of BHs 
across a range of mass scales.
\end{abstract}

Observations of NS mergers by LIGO/Virgo/KAGRA have provided crucial 
insights into the structure of compact objects, offering a new perspective on the behavior 
of matter 
in extreme gravity regimes \cite{LIGOScientific:2020aai,LIGOScientific:2018hze,LIGOScientific:2017vwq}. 
The high sensitivity and event rates expected from next-generation ground- and 
space-based detectors \cite{Branchesi:2023mws,LISA:2024hlh,Abac:2025saz,LISA:2022kgy,Ajith:2024mie} 
promise to turn GWs into powerful probes of stellar interiors, linking the macroscopic 
properties of compact objects to their internal composition. This advancement opens 
up exciting possibilities for new tests of fundamental physics, aiming to probe the 
existence of novel families of compact objects—BH mimickers—whose 
compactness\footnote{The compactness ${\cal C}=GM/Rc^2$ is defined as the ratio 
of an object’s mass to its radius, where $G$ and $c$ are the gravitational constant and 
speed of light. Schwarzschild BHs have ${\cal C}=1/2$, while for NSs 
${\cal C}\sim 0.1$-$0.2$, depending on the equation of state.} lies between that of 
NSs and BHs \cite{Cardoso:2019rvt}.

Without focusing on specific BH mimicker models, we discuss here some 
general \textit{finite-size signatures}—i.e., effects related to the body's internal 
structure—that help distinguish different classes of compact objects. In particular, 
we examine how such signatures can used to test the uniqueness of the Kerr geometry 
\cite{Kerr:1963ud,Carter:1971zc,Robinson:1975bv} using the inspiral signal of 
binary coalescences.

Finite-size effects imprint distinctive signatures on GW signals and influence 
different stages of the inspiral. At low frequencies, sources behave as point particles, 
with internal structure effects providing subdominant contributions to the waveform. 
As the system evolves toward merger, finite-size effects become increasingly significant, 
affecting orbital dynamics and GW emission. These contributions can often be 
modelled using agnostic approaches that introduce waveform corrections independent 
of specific physics cases.

\section*{Agnostic signatures}

No-hair theorems and the uniqueness of Kerr geometry dictate that the entire set of 
multipole moments for stationary, isolated BHs in General Relativity (GR) is determined 
solely by their mass and spin angular momentum:
\begin{equation}
M^{\rm BH}_{\ell m} + i S^{\rm BH}_{\ell m} = M^{\ell +1}(i\chi)^\ell \,,
\end{equation}
where $M^{\rm BH}_{\ell m}$ and $S^{\rm BH}_{\ell m}$ are the mass and current multipole 
moments, $M=M_{00}$ and $J=S_{10}$ denote the BH mass and angular momentum, and 
$\chi=J/M^2$ is the dimensionless spin parameter \cite{Hansen:1974zz}. 
Given their equatorial symmetry, Kerr BHs have 
nonzero moments only for $m=0$, with mass (current) moments vanishing for 
odd (even) $\ell$. Multipole moments of BH mimickers are expected to be different from their 
Kerr counterpart and possibly break axisymmetry \cite{Bena:2020see,Bianchi:2020bxa,Loutrel:2022ant}, 
with corrections that can be 
parametrised as $M_{\ell m}^{\rm mim}=M^{\rm BH}_{\ell m}+\delta M_{\ell m}$ 
and $S_{\ell m}^{\rm mim}=S^{\rm BH}_{\ell m}+\delta S_{\ell m}$ \cite{Glampedakis:2017cgd,Raposo:2018xkf,Raposo:2020yjy,Herdeiro:2020kvf,Bena:2020uup,Vaglio:2022flq}.
The dominant contribution to the GW signal is encoded by the 
quadrupole moment $M^{\rm BH}_{2 0}$, which enters the post-Newtonian\footnote{Within the pN 
approach, the waveform is expanded in powers of the ratio $(v/c)^n$, where $v$ is the 
system’s velocity, and the expansion order is determined by $n/2$ \cite{Blanchet:2013haa}.} (pN) waveform phase at 2pN order.

Next-generation detectors, such as the Einstein Telescope (ET), will measure deviations from Kerr in the reduced quadrupole moment, 
$\kappa = -M^{\rm mim}_{20}/(M^3 \chi^3)$, where $\kappa^{\rm BH} =1$,
with precision down to $10^{-1}$ for stellar-mass binaries \cite{Abac:2025saz}. A similar 
level of accuracy is expected from the future space-based detector LISA, covering a 
broad mass range. For both detector families, tests based on measurements of the 
spin quadrupole and high-order moments, modeled using generic parametrizations 
of the mimickers’ multipolar structure, forecast constraints at the percent level or better \cite{Krishnendu:2017shb, Krishnendu:2018nqa, Kastha:2018bcr, Krishnendu:2019ebd, Kastha:2019brk}.

The most promising constraints on multipole moments are expected from LISA observations 
of systems with significant mass asymmetry, such as extreme and intermediate mass ratio 
inspirals (E/IMRIs), with mass ratios $q \sim 10^{-3}-10^{-6}$. These systems could allow 
measurements of the spin-induced quadrupole with an accuracy of up to one part in $10^4$ \cite{Barack:2006pq, Babak:2017tow}, and provide constraints on high-order multipole moments \cite{Krishnendu:2019ebd, Kastha:2019brk}, offering unique tests of the Kerr nature of the 
massive binary component.\\

Tidal interactions in the late inspiral stage, before merger, also imprint a distinctive 
signature on the emitted GW signal, carrying valuable information 
about the source’s internal structure. This effect is encoded by the Love numbers, 
a set of parameters that describe the body’s deformability properties \cite{Hinderer:2007mb,Binnington:2009bb,Damour:2009vw}. 
The dominant tidal correction to the waveform appears at the 5pN 
order \cite{Flanagan:2007ix} and is determined by the 
quadrupolar, electric-type Love number $k_2$, which acts as a coupling 
constant between the external tidal field and the object’s deformation.

A striking result in GR is that Love numbers vanish for BHs 
in vacuum \cite{Binnington:2009bb,Damour:2009vw,Poisson:2014gka,Pani:2015hfa,Landry:2015zfa,LeTiec:2020spy,Chia:2020yla,Riva:2023rcm,DeLuca:2023mio,Iteanu:2024dvx,Kehagias:2024rtz}. This property establishes Love 
numbers as unique probes of new physics. The detection of a nonzero value would serve 
as a smoking-gun signature of deviations from the Kerr solution, as horizonless BH 
mimickers, BHs in non-vacuum environments, or those with additional fields in modified 
gravity theories exhibit small but finite Love numbers \cite{Mendes:2016vdr,Maselli:2017vfi,Sennett:2017etc,Cardoso:2017cfl,Baumann:2018vus,Cardoso:2018ptl,Raposo:2018rjn,Brustein:2020tpg,Nair:2022xfm,Katagiri:2024fpn,Cannizzaro:2024fpz,Berti:2024moe,DeLuca:2025bph}.

Next-generation ground-based detectors could measure $k_2$ in the range $\sim 10^{-2}-10^{-1}$ 
for systems with total masses $M\lesssim 100M_\odot$ \cite{Cardoso:2017cfl,Branchesi:2023mws,Abac:2025saz}.  LISA could constrain Love numbers as small 
as $10^{-2}$ using observations of comparable-mass binaries with $M\sim 10^6M_\odot$ \cite{Maselli:2017cmm}. 
These measurements could allow to discriminate families of BH mimickers \cite{Maselli:2018fay}, 
although model selection would require events with high signal-to-noise ratios \cite{Addazi:2018uhd}.
Exquisite constraints could be achieved with LISA observations of E/IMRIs. In such systems, 
the Love number of the massive component affects the waveform at leading order in 
$q$, amplifying its contribution \cite{Pani:2019cyc}. Observations of these inspirals could 
constrain values of $k_2$ as small as $10^{-3}$, potentially probing Planck-scale deviations 
near the horizon \cite{Piovano:2022ojl}.\\

Finally, a unique signature characterizing BH dynamics during the inspiral phase is tidal heating \cite{Hughes:2001jr,Alvi:2001mx,Goldberger:2005cd,Porto:2007qi,Poisson:2009di,Cardoso:2012zn}. As natural absorbers, BHs allow a small fraction of the emitted radiation during coalescence to be lost through the event horizon. This energy dissipation is particularly effective for rapidly spinning objects\footnote{Within a 
pN inspiral waveform, the contribution of tidal heating for spinning (non-spinning) objects enters 
the phase at 2.5pN (4pN) order.}, leading to slight changes in both mass and spin. These changes can significantly contribute to the accumulated GW phase \cite{Bernuzzi:2012ku,Taracchini:2013wfa}.

In contrast, BH mimickers interact weakly with GWs, potentially resulting in 
negligible absorption during the inspiral. The (partial) absence of tidal heating thus 
serves as a key observational signature to distinguish Kerr BHs from horizonless mimickers \cite{Maselli:2017cmm}. 
Stringent constraints on absorption will come from sources detected by 
LISA, such comparable-mass binaries with high 
signal-to-noise ratios \cite{Maselli:2017cmm}. EMRIs and IMRI evolving in the 
LISA band will provide the tightest bounds, 
constraining mimicker reflectivity down to levels as low as $10^{-5}$ \cite{Datta:2019euh,Datta:2020rvo,Maggio:2021uge,Zi:2023geb,Datta:2024vll}.

\section*{A coherent waveform model}

While agnostic parametrizations provide 
the most flexible tools for searching for hints of new physics, 
approaches built on specific models of compact objects offer deeper 
insights into possible deviations from the Kerr solution. Top-down 
approaches allow to exploit powerful analogies with binary NS 
mergers and develop coherent waveforms that account for the 
simultaneous contributions of multiple finite-size effects 
\cite{Pacilio:2020jza}. For a given 
model, such effects depend on the same parameters that determine 
the underlying stellar structure---such as the equation of state of dense 
matter for a NS. This, in general, reduces the number of 
additional quantities required beyond the BH baseline, 
mitigating degeneracies and improving the measurement accuracy 
of the entire parameter set.

A concrete application of this strategy was explored in Refs.~\cite{Pacilio:2020jza,Vaglio:2023lrd}, 
which investigated the detectability of scalar boson star (BS) binaries with quartic self-interactions. 
These are condensates of a complex field governed by the Lagrangian
${\cal L}_{\varphi}=-\frac{1}{2}\partial_\mu\varphi^\star\partial ^\mu\varphi
-\frac{1}{2}\mu^2\vert\varphi\vert^2-\frac{1}{4}\sigma \vert\varphi\vert^4$, 
whose structure, in the strong coupling limit  $\sigma\gg \mu$, is determined 
solely by the ratio $M_{\rm B}=\sqrt{\sigma}/\mu^2$.

\begin{figure}[hbpt!]
\centering \includegraphics[width=0.58\columnwidth]{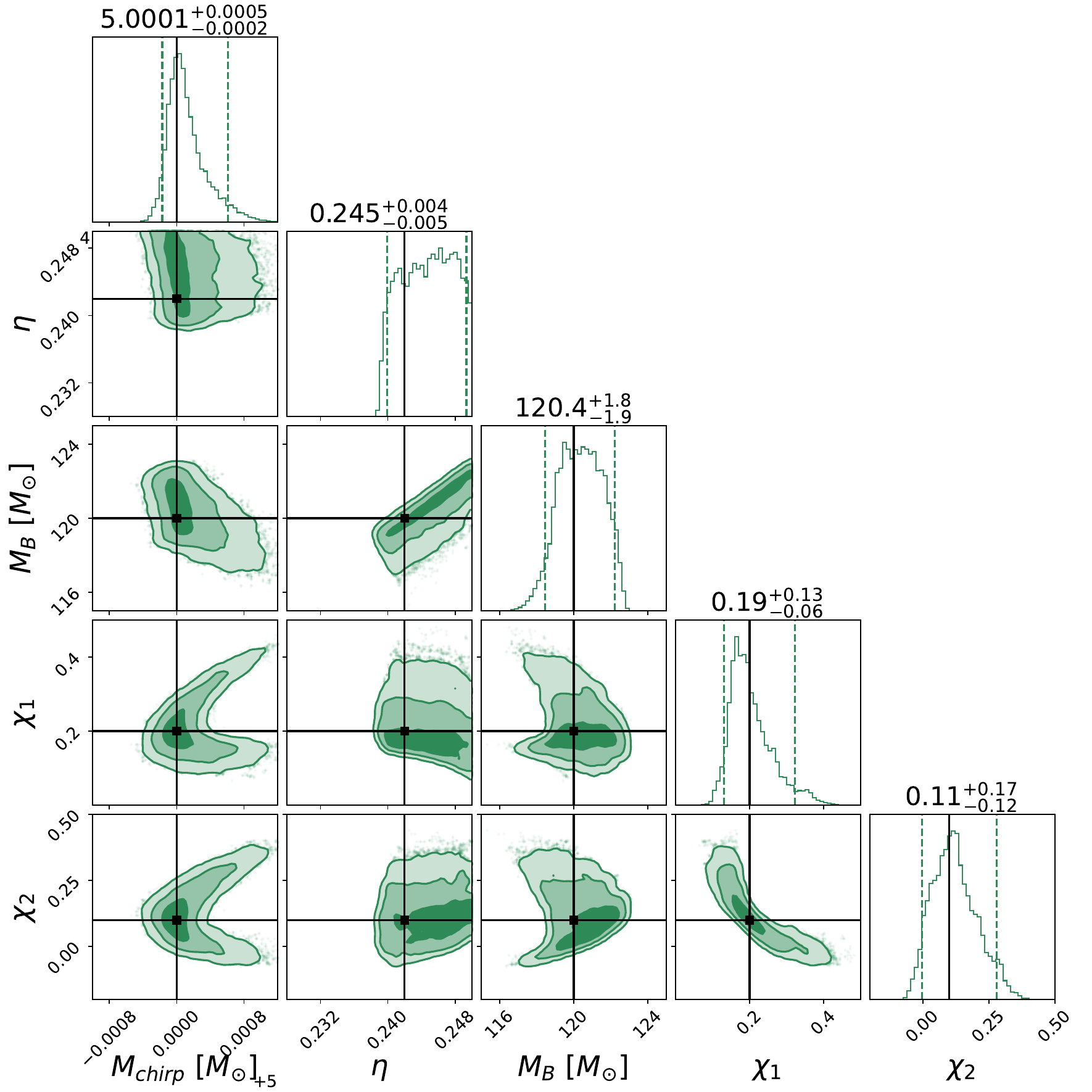}
\caption{Contour plots of the posterior distribution for a GW signal emitted by a BS 
binary with chirp mass ${\cal M} =\eta^{3/5}(m_1+m_2)= 5M_\odot$, symmetric 
mass ratio $\eta=m_1m_2/(m_1+m_2)^2 = 0.2423$, spin parameters $\chi_1 = 0.2$ and $\chi_2 = 0.1$, 
and coupling ratio $M_B = 115M_\odot$. The signal is detected by ET 
with a signal-to-noise ratio of $\sim 130$. Diagonal (off-diagonal) panels 
display marginal (2D joint) posteriors. Median values and 90\% credible intervals for 
each parameter are reported at the top of each panel. Solid lines indicate the injected 
values of the source parameters. Taken from \cite{Vaglio:2023lrd}.}\label{fig:posteriorsBS}
\end{figure}

In this setup, Ref.~\cite{Vaglio:2023lrd} developed a waveform model 
incorporating the BS spin-induced quadrupole moment and tidal 
deformability parameters within a pN template designed 
for the inspiral evolution of BH binaries. This waveform was used 
to assess\footnote{Ref.~\cite{Johnson-Mcdaniel:2018cdu} performed a similar analysis focusing on a simpler model of non-rotating BSs, described in terms of a perfect fluid with a polytropic equation of state.} 
the ability of next-generation detectors to recover the properties of binary BSs, 
and in particular, the coupling 
ratio $M_{\rm B}$. This analysis 
demonstrated that $M_{\rm B}$ can be constrained at the percent 
level, with the most precise measurements obtained from low-mass binaries, 
for which the inspiral part of the signal is more dominant.
Figure~\ref{fig:posteriorsBS} shows the recovered posterior of 
the source parameters for a case analyzed. 
The analysis of Ref.~\cite{Vaglio:2023lrd} also revealed that 
neglecting one of the finite-size contributions within the recovery template 
leads to biases in parameters if the injected signal represents the 
full waveform.

While valid only in the inspiral stage, this study highlights 
the relevance and necessity of developing top-down studies based on specific 
models of BH mimickers, with the goal to 
construct inspiral-merger-ringdown models, extending the 
signal evolution into the merger domain through numerical relativity 
simulations \cite{Liebling:2012fv,Palenzuela:2007dm,Palenzuela:2017kcg,Bezares:2017mzk,Sanchis-Gual:2018oui,Bezares:2022obu,Siemonsen:2023hko}.

\section*{Acknowledgments}
This work is partially supported by the MUR PRIN Grants No. 2022-Z9X4XS and No. 2020KB33TP, and by the INFN TEONGRAV initiative.

\bibliographystyle{utphys}
\providecommand{\href}[2]{#2}\begingroup\raggedright\endgroup

\title{EHT Tests of Gravity: What we’ve learned so far and what's to come}
\author{Lia Medeiros}
\institute{\textit{Center for Gravitation, Cosmology and Astrophysics, Department of Physics,\\ University of Wisconsin–Milwaukee, P.O. Box 413, Milwaukee, WI 53201, USA}}

\maketitle 

\begin{abstract}
The event-horizon-scale images of the black hole in M87 and the Galactic Center black hole, Sagittarius A* (Sgr A*), published by the Event Horizon Telescope (EHT) have enabled strong-field tests of the Kerr metric in a previously unexplored regime. I will briefly review how EHT observations can be used for gravitational tests and discuss past EHT results. I will also describe how improvements in data analysis algorithms may further improve these constraints.
\end{abstract}

The theory of general relativity has been tested countless times since passing its first test in 1919 \cite{Dyson1920}. When testing fundamental theories, it is crucial to test the theory in different regimes and to test different aspects of it. The Event Horizon Telescope (EHT) probes the strong field regime near black holes (at $\sim 3GM/c^2$ \cite{Psaltis2021}). However, since the EHT observes supermassive black holes, it probes lower curvatures than those studied, for example, with current LIGO/Virgo/KAGRA detections. Another distinction between EHT tests and those with gravitational waves is that the spacetime is effectively stationary for EHT tests since the mass in the accretion disk is small compared to the black hole. Therefore, the EHT can only perform metric tests and is not sensitive to the dynamical aspects of gravitational theories. In contrast, tests with gravitational waves are sensitive to both the dynamical aspects and the metric.
EHT tests can thus be considered tests of the no-hair theorem, since, according to the theorem, Kerr is the only stationary, axisymmetric, asymptotically flat spacetime that is free of pathologies and consistent with general relativity \cite{Kerr1963,Israel1967,Israel1968,Carter1968,Carter1971,Hawking1972,Price1972a,Price1972b,Robinson1975}.

\begin{figure}[t]
    \includegraphics[height=.395\columnwidth]{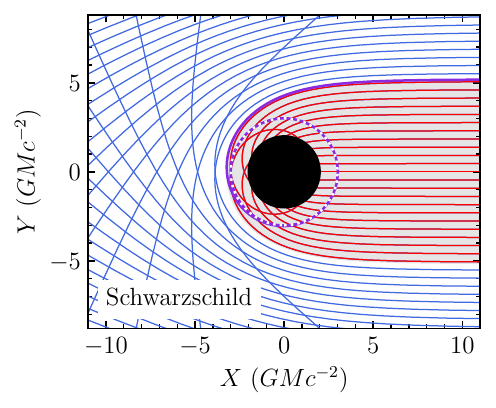}
    \includegraphics[height=.395\columnwidth]{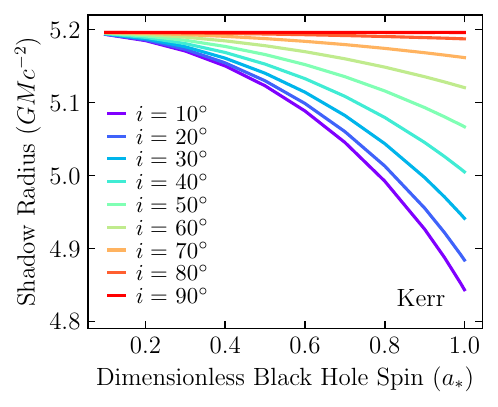}
    \caption{\textit{(left)} Photon trajectories through a Schwarzchild spacetime. Photons shown in red (blue) fall into (escape) the black hole. The shadow boundary is defined by the critical impact parameter between the blue and red photons. 
    A photon with critical impact parameter (purple) is tangent to the photon orbit (dotted white circle).    The black disk shows the event horizon. Figure generated with \texttt{Mahakala} \cite{Sharma2023}. \textit{(right)} Radii for all Kerr shadows as a function of spin and inclination, Figure adapted from \cite{Medeiros2020}.
    } 
    \label{fig:shadow_def}
\end{figure}

The gravitationally relevant feature that I focus on here is the black hole shadow, whose boundary is defined as the critical impact parameter between photons that fall into the black hole and those that escape, as viewed by an observer at infinity (see \cite{Bardeen1973,Luminet1979,Falcke2000,SAP6} and the left panel of Figure~\ref{fig:shadow_def}). This definition is equivalent to defining the shadow boundary as the image of the lensed photon orbit; a photon at the critical impact parameter will be tangent to the photon orbit. The shadow radius for a Schwarzchild black hole is $\sqrt{27}GM/c^2$ and the radius ranges from $\approx 4.8GM/c^2$ to $5.2 GM/c^2$ for all Kerr shadows (see the right panel of Figure~\ref{fig:shadow_def} and e.g. \cite{Medeiros2020,Johannsen2010}). This range is only about $\pm4\%$, which, as we will see below, is small compared to current EHT error bars on shadow size. 

Although the shadow depends only on the spacetime metric, the EHT image depends on astrophysics as well. In \cite{M87P6,SAP6} the authors identify three requirements and argue that if these are met, then the observational ring of emission will be close to (within an error bar) the black hole shadow boundary (see e.g., \cite{Jaroszynski1997,Johannsen2010,Narayan2019,Ozel2021,Bronzwaer2021,Kocherlakota2022}). First, we need enough photons to illuminate the black hole, second we need these photons to be near the photon orbit so they will be lensed by the black hole, and finally, we need the surrounding plasma to be optically thin at the observed wavelength. Theoretical work showed that these requirements should be met for both Sgr~A$^*$ and M87 at mm-wavelengths \cite{Ozel2000,SAP5}.

In \cite{SAP6}, the authors perform a calibration between the ring size measured in EHT images and the shadow boundary to determine an error bound for EHT shadow size measurements. The authors separate the error budget into \textit{(i)} the theoretical uncertainty ($\alpha_1$), the uncertainty between the shadow size and the size of the emission ring in high-resolution simulated images, and \textit{(ii)} the observational uncertainty ($\alpha_2$), the uncertainty between the size of the emission ring in simulated images and the ring size in EHT-like observations. We constrain the fractional deviation parameter ($\delta$), defined by \cite{SAP6}
\begin{equation}
d_m = \alpha_c (1+\delta)d_\mathrm{sh, Sch}, 
\end{equation}
where $\alpha_c = \alpha_1\times\alpha_2$, $d_m$ is the measured ring diameter in EHT images, and $d_\mathrm{sh, Sch}= \sqrt{27}GMc^{-2}$ is the shadow diameter for a Schwarszchild black hole.

\begin{figure}[t]
    \includegraphics[height=.345\columnwidth]{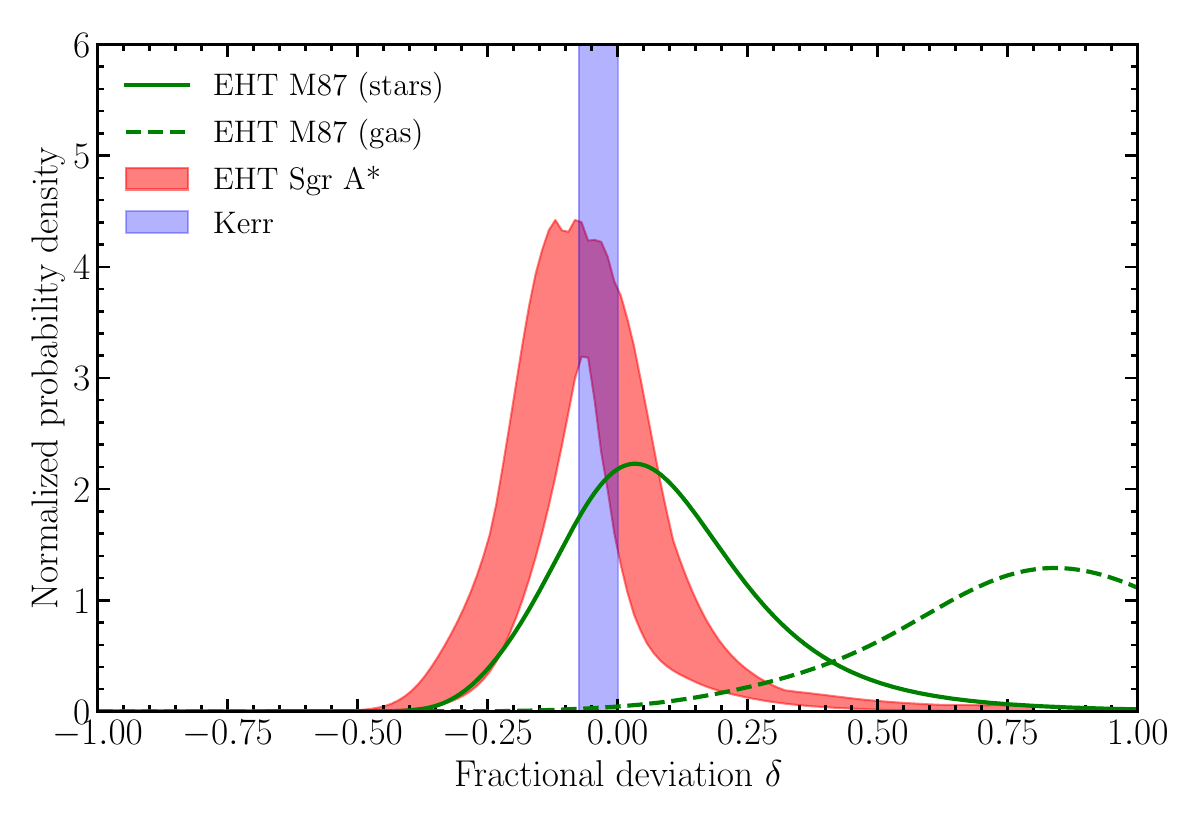}
    \includegraphics[height=.345\columnwidth]{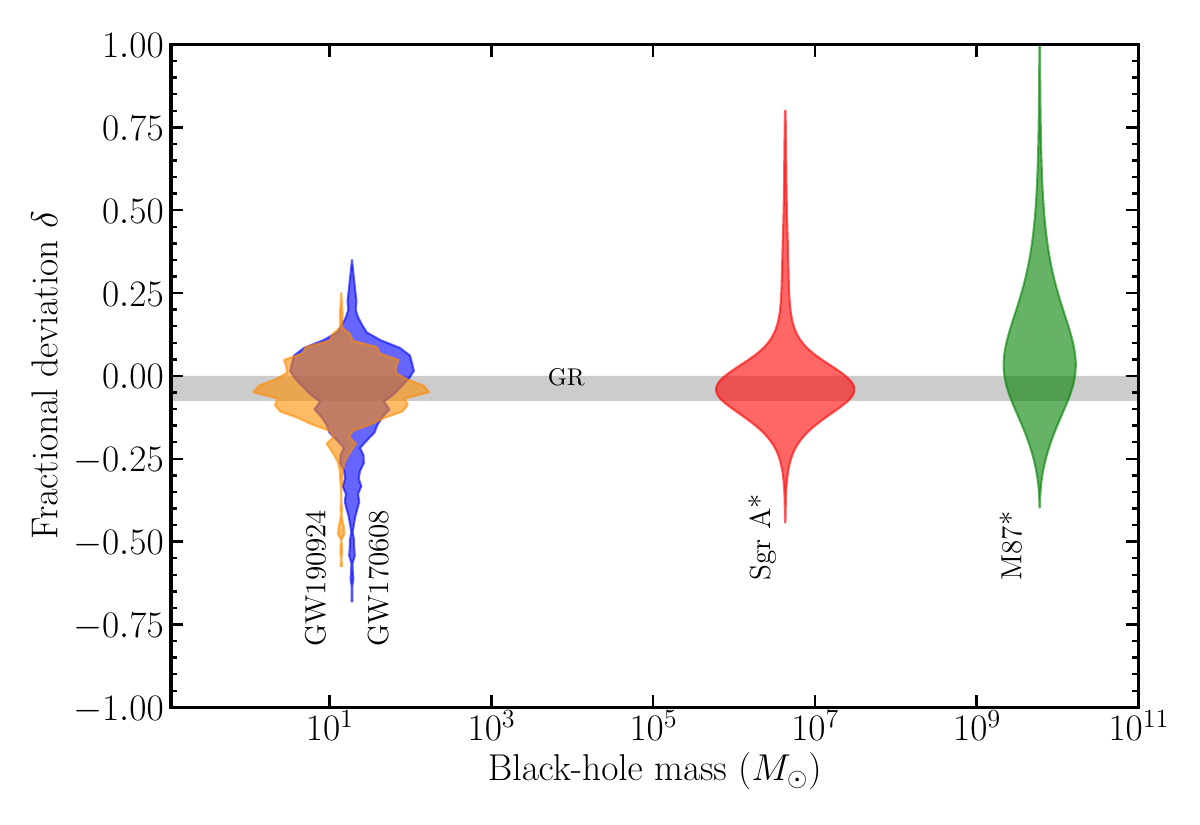}
    \caption{\textit{(left)} Summary of EHT constraints on the fractional deviation ($\delta$) for Sgr~A$^*$ (pink shaded region) and M87 (solid and dashed green curves) compared to the Kerr prediction (blue shaded region, includes all spins and inclinations). \textit{(right)} Violin plot comparing EHT constraints to constraints from gravitational waves (gravitational wave constraints are projected onto $\delta$ constraints, see \cite{Psaltis2021,M87P6} for details on this comparison). Figures reproduced from \cite{SAP6}.
    } 
    \label{fig:bounds}
\end{figure}

The left panel of Figure~\ref{fig:bounds} summarizes the constraints on the fractional deviation parameter ($\delta$) for the Sgr~A$^*$ results presented in \cite{SAP6}. The pink shaded region contains the results of 18 analyses, each using different \textit{(i)} mass over distance measurements \cite{gravity2022,Do2019}, \textit{(ii)} theoretical error estimates (based on general-relativistic magnetohydrodynamic (GRMHD) simulations, analytic accretion models in Kerr spacetimes, and analytic accretion models in non-Kerr spacetimes), and \textit{(iii)} imaging algorithms, i.e., ways to reconstruct an image from the sparse interferometric data (each making their own assumptions and resulting in different observational errors). All analyses are consistent with the Kerr prediction. The figure also compares Sgr~A$^*$ results to EHT constraints from M87. Since EHT constraints rely on a prior mass over distance measurement, constraints from Sgr~A$^*$ are more stringent than those for M87 primarily due to the better constraints on mass and distance for Sgr~A$^*$ (see also \cite{Psaltis2020}).

The bounds on $\delta$ shown in Figure~\ref{fig:bounds} cannot rule out any part of the Kerr parameter space. However, these bounds can rule out some alternatives to Kerr (see, e.g., Figures 6-8 in \cite{Medeiros2020} for the shadow radii of several non-Kerr parametrized metrics). As shown in \cite{Psaltis2020}, in the limit of spherical symmetry and in areal coordinates, the radius of the unstable photon orbit is:
\begin{equation}
r_{\mathrm{ph}} = \sqrt{-g_{tt}}\left( \frac{d\sqrt{-g_{tt}}}{dr} \Big|_{r_{\mathrm{ph}}}\right)^{-1},
\end{equation}
i.e., it depends only on $g_{tt}$. Since the boundary of the shadow is equivalent to the lensed photon orbit, the shadow radius is 
\begin{equation}
r_{\mathrm{sh}} =  \frac{r_{\mathrm{ph}}}{\sqrt{-g_{tt}(r_{\mathrm{ph}})}}.
\end{equation}   
By constraining the size of the black hole shadow with EHT observations, we are constraining the time-time component of the black hole metric.

The EHT constraints are based on parametrized metrics that were developed for the strong field regime. However, to compare to other tests, we can also expand the $g_{tt}$ component using the Parametric Post Newtonian (PPN) expansion:
\begin{equation}
g_{tt} = -1 + \frac{2}{r} - 2\left(\frac{\kappa_1}{r^2}\right) + 2\left( \frac{\kappa_2}{r^3} \right) - 2\left( \frac{\kappa_3}{r^4} \right) + \mathcal{O}(r^{-5}),
\end{equation} 
where $\kappa_1$, $\kappa_2$, and $\kappa_3$ are first, second, and third order deviation parameters respectively, and $r$ is the radius in areal coordinates. For all parametrized metrics considered in \cite{SAP6} the bounds on $\kappa_1$ ($\kappa_2$) are between $-0.3$ and $0.9$ ($-4.0$ and $1.5$). These second-order constraints are hundreds of times more stringent than current solar system tests, and comparable to LIGO/Virgo constraints when \cite{SAP6} was published (see also \cite{Psaltis2021}). The right panel of Figure~\ref{fig:bounds} compares the EHT constraints from Sgr~A$^*$ and M87 to gravitational wave constraints by LIGO/Virgo.

In addition to the constraints on the time-time component of the metric based on the shadow size measurement, EHT results can also constrain alternatives to an event horizon. In \cite{SAP6} the authors consider a few specific examples of naked singularities, a thermal surface in equilibrium, and a perfectly reflective surface. Here I will focus on the thermal surface model, which has been well developed in the literature \cite{Narayan1998,Narayan_2002,Broderick_Narayan_2006,Broderick_Narayan_2007,Narayan_McClintock_2008,Broderick+2009}. This model assumes: \textit{(i)} matter in the compact object at the center of Sgr~A$^*$ satisfies energy conservation, \textit{(ii)} it obeys the laws of thermodynamics, in particular, it approaches statistical equilibrium in steady state, and \textit{(iii)} it couples to and radiates in all electromagnetic modes. 

Since Sgr~A$^*$ is a hot accretion flow, the accretion disk cannot efficiently radiate away heat, resulting in brightness temperatures above $\sim 10^9$K \cite{SAP3,SAP6}. Due to Sgr~A$^*$'s low accretion rate, its luminosity is much lower than $\dot{M}c^2$, i.e., most of the energy/heat gets advected inward and falls onto the central object \cite{Narayan1995,Yuan2014}. Given the EHT constraint on the size of the emission surface, the model predicts a considerable amount of infrared radiation, orders of magnitude higher than observations (see Figure~14 in \cite{SAP6}). A black hole mimicker would need to convert all incident matter and energy into material that is electromagnetically inactive to escape electromagnetic detection. Exceptions to this constraint may be possible if an initial assumption is violated, such as the surface not having reached thermodynamic equilibrium due to time-dilation (this may be possible for surfaces that lie between $10^{-23}$ and $10^{-14}$ cm above the event horizon).

\begin{figure}[t]
    \includegraphics[height=.3\columnwidth]{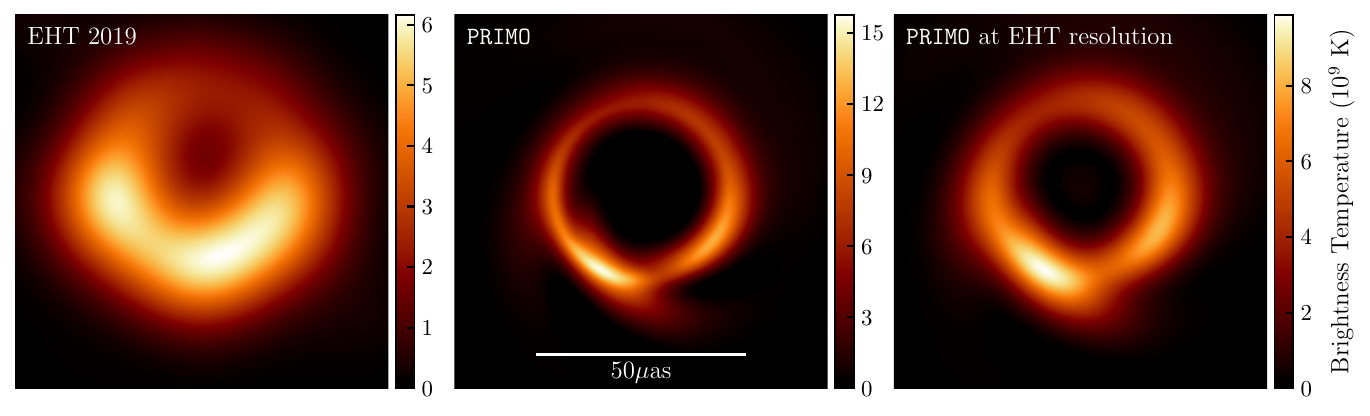}
    \caption{\textit{(left)} The image of M87 published by the EHT in 2019. \textit{(middle)} The image of M87 reconstructed by \texttt{PRIMO}. \textit{(right)} The \texttt{PRIMO} image of M87 blurred to approximate the EHT resolution. Figure adapted from \cite{Medeiros2023b}.
    } 
    \label{fig:primo}
\end{figure}

The results above are based on the 2017 EHT observations, which included 8 telescopes at 6 locations. 
Since then, the Greenland Telescope, the Kitt Peak Telescope, and NOEMA have joined the EHT array. 
These additions improved coverage in the Fourier domain and should improve constraints on image morphology and model parameters. In addition to hardware advances, improvements to data-analysis software can also result in considerable improvements. \texttt{PRIMO}\footnote{\texttt{PRIMO} is a dictionary learning algorithm based on principal component analysis (see \cite{Hogg2024} for a discussion on how to define machine learning).}, a machine-learning algorithm trained on GRMHD simulations, reconstructed a new version of the M87 image and achieved the maximum allowed resolution of the array (see Figure~\ref{fig:primo} and \cite{Medeiros2023a,Medeiros2023b,Psaltis2024}).
The ring diameter in the \texttt{PRIMO} image is consistent with previous results, and the improved resolution lowered the ring-width constraint by a factor of two. The brightness depression (dark spot) at the center of the \texttt{PRIMO} image is considerably darker than previous images. Although a calibration of \texttt{PRIMO} is still underway, it may lead to improved constraints on the observational uncertainty, accretion parameters (due, for example, to the improved constraint on ring width, or to using the \texttt{PRIMO} framework to compare simulations to EHT data), and on the albedo of a possible reflecting surface, since a reflecting surface should result in brightness within the shadow (see e.g. Figure 15 in \cite{SAP6}).

\section*{Ackhowledgements}
L. M. thanks the EHT collaboration, all those who contributed to the results highlighted here, and to the workshop organizers.  L. M. gratefully acknowledges support from NSF AST-2407810.

\providecommand{\href}[2]{#2}\begingroup\raggedright\endgroup

\title{Frozen Stars as Black Hole Mimickers}
\author{Ram Brustein}
\institute{\textit{Department of Physics, Ben-Gurion University,
		Beer-Sheva 84105, Israel}}

\maketitle
\begin{abstract}
	Frozen stars are ultra-compact objects which are free of singularities and lack a formal horizon, but yet they share many of the same observable properties with the Schwarzschild or Kerr black holes predicted by general relativity.  General relativity predicts singularities within black holes,  signaling a breakdown of the theory. The frozen star model proposes an alternative, suggesting that astrophysical black holes may instead be horizon-scale compact objects with an internal quantum structure that prevents their further collapse and the formation of singularities. The quantum effects can be modeled by a geometry which includes modifications to the Schwarzschild or Kerr geometries on  horizon scales rather than at the Planck scale. This paradigm requires the existence of exotic matter beyond the Standard Model.
\end{abstract}

\section{ Introduction}
The frozen star (FS) model \cite{Brustein:2018web,Brustein:2021lnr,Brustein:2023cvf,Brustein:2024yyc,Brustein:2023wyr,Brustein:2024gia} is based on two fundamental assumptions: first, that black holes (BHs) are non-singular objects which do not collapse under their own gravity, and second, that strong quantum effects smear the would-be singularity over horizon-size length scales. This leads to two key consequences. First,  excitations appear at the horizon scale rather than at Planck length. Second, these modifications lead to a significant departure from the predictions of semi-classical general relativity (GR), requiring exotic matter sources \cite{Brustein:2019bou}, such as open strings in a spoke-wheel configuration or highly excited closed strings at the Hagedorn temperature.\\

The FS exhibits a unique internal structure. Each radial layer of the object behaves like an ``almost-horizon,'' but avoiding classical geometric singularities and infinite redshifts. This is a geometric realization of a property of the polymer model \cite{Brustein:2016msz,Brustein:2016xzw}: the maximal entropy of highly excited closed strings inside each radial layer. The entropy generates an effective universal quantum pressure capable of preventing collapse of matter into a singularity. This is reminiscent of the quantum Fermi pressure which counteracts gravitational collapse of neutron stars. In the case of FSs, this quantum pressure is replaced by a geometric modeling of the maximal entropy which amounts to having maximally negative radial pressure.\\

A key prediction of the FS model, which is mentioned only briefly here, but discussed extensively elsewhere \cite{Brustein:2017koc,Brustein:2017kcj,Brustein:2017nis,Brustein:2023gea,Brustein:2024sah,Brustein:2020tpg,Brustein:2021bnw,Avitan:2023txy}, concerns the gravitational waves emitted when two such objects collide. In addition to the ringdown phase of classical BHs, which can be attributed to the decay to equilibrium of the surrounding spacetime,  FSs  exhibit an additional branch of ringdown modes,  with a longer decay time and lower amplitude than the primary signal. This effect results from the excitation of the internal fluid-like structure of the FSs during their merger. Additionally, the internal spectrum induces a large, negative Love number. These observational signatures provide a potential method for distinguishing FSs from traditional BHs by their gravitational wave signatures and thus potentially confirming their existence.

\section{Frozen star geometry and matter}
The FS geometry is remarkably simple,
\begin{equation*}
	ds^2 = -\varepsilon^2 dt^2 + \frac{1}{\varepsilon^2} dr^2 + r^2 d\Omega_2^2,
\end{equation*}
where the constant $\varepsilon^2 \ll 1$ should be thought of as the ratio of some fundamental scale, such as the Planck scale, to the radius of the star, which is slightly larger than the would be horizon,
$
R \approx 2GM(1 + \varepsilon^2).
$
For $ r>R$, the metric is exactly the Schwarzschild metric.
This geometry needs to be regularized at the core and smoothed at the outer boundary \cite{Brustein:2021lnr,Brustein:2023cvf}.
The FS geometry is sourced by a highly anisotropic energy-momentum tensor with the negative radial pressure $p_r$ related  to the energy density $\rho$, by $p_r = -\rho = - (1 - \varepsilon^2/(8\pi G r^2)$,
while the transverse pressure vanishes.
The mass inside every radial shell is almost equal to its Schwarzschild mass,
$2G m(r) = r(1 - \varepsilon^2)$.\\

The matter which sources the FS geometry can be identified with the string fluid resulting from the decay of an unstable $D$-brane or a brane-antibrane system at the end of open-string tachyon condensation \cite{Sen:2004nf,Gibbons:2000hf,Yee:2004ec} (See also \cite{osti_4477262,Letelier:1979ej}.). The string fluid corresponds to flux tubes emanating from the center and ending at the Schwarzschild radius of the star as shown in the Figure,
\begin{figure}[h]
	\vspace{-.3in}
	\begin{minipage}[c]{0.5\textwidth}
		\center{\includegraphics[height=4cm]{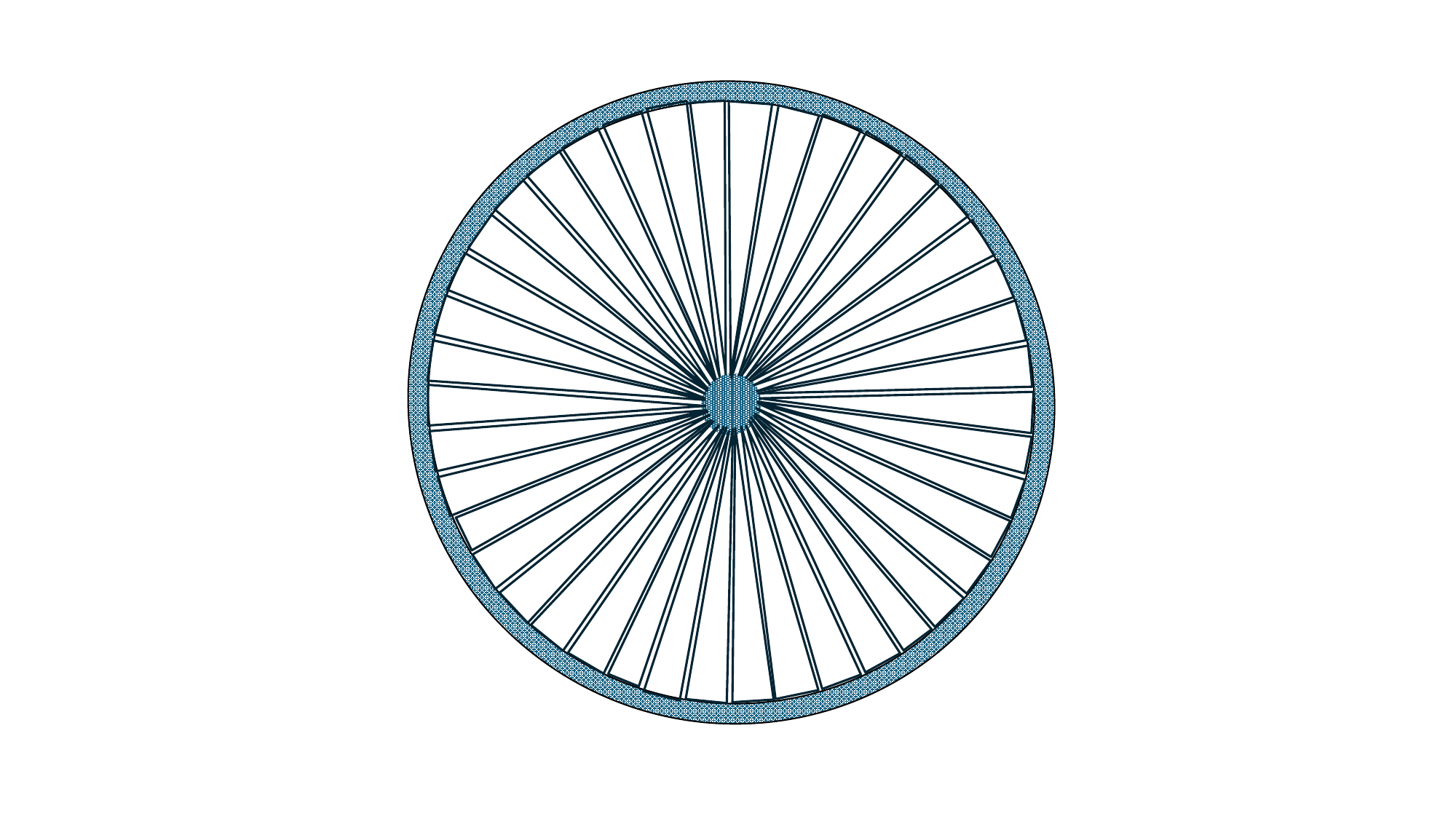}}
	\end{minipage}\hfill
	\begin{minipage}[c]{0.5\textwidth}
		\vspace{-.3in}
		\caption{A frozen star sourced by a fluid of electric flux lines that are emanating radially from a  point-like charge in its core and ending on its outer layer which is oppositely charged. The electric force between the charges is offset because the mass of the star is fixed.}\label{fig1}
	\end{minipage}\hfill
\end{figure}
The effective Lagrangian for this fluid can be recast into a Born-Infeld form. When the string fluid is coupled gravity, the static, spherically symmetric solutions are shown to be the same as those describing FSs . Frozen stars can therefore be modeled as gravitationally confined flux tubes.

\section{Thermodynamics and dynamical formation}
Several arguments show a Hawking flux is generated if an object has a large enough redshift at the outer surface so that it is externally indistinguishable from a Schwarzschild BH for all practical purposes, as is the case for the FS. In \cite{Barcelo:2010xk}, these were formalized. The FS was shown both requirements \cite{Brustein:2023hic}. It follows that the FS emits Hawking radiation at the Hawking temperature.\\

An additional argument relies on the Schwinger pair production effect, discussed in \cite{Brustein:2015eba} and more recently in \cite{Wondrak:2023zdi}.  Yet another argument due to Mathur and Mehta in \cite{Mathur:2023uoe}, relies on  the properties of the object's thermal atmosphere which can be used to fix the temperature of radiated modes at precisely the Hawking value. Having identified the temperature and energy of the FS, we have used the first law of thermodynamics to show that it possesses an entropy equal to the Bekenstein-Hawking entropy of GR BHs. This can also be shown directly by using Euclidean methods.  \\

An important challenge to any model of ultra-compact alternatives to GR BHs, such as the FS model, is explaining  how they are dynamically form during the collapse of ``normal'' matter. Dynamical formation has to occur before the collapsing matter enters and an apparent horizon is formed. Here, again, the large entropy of  FSs is an essential ingredient. We were able to realize Mathur’s idea \cite{Mathur:2008kg} in the FS model. Inspite of having an exponentially small transition probability amplitude to any of the microstates of the BHs $ \Gamma \sim e^{-S_{\text{BH}}}$, having an exponentially large phase space $ e^{+S_{\text{BH}}}$ leads to a total transition probability $\sim 1 $. Since the Einstein’s equations are blind to entropy an interim solution is to use Euclidean methods, which we have done \cite{Brustein:2023hic}.

\section{Conclusion}
The FS model proposes a modification to classical BH descriptions by replacing the singularity with a horizon-scale, a spoke-wheel configuration of open strings possessing maximally negative radial pressure. The FS is an ultra-compact, regular solution of the Einstein equations and it is black and ultra-stable.
The matter that sources the FS obeys the null energy condition and can be associated with fundamental strings. Its thermodynamics matches the thermodynamics of GR BHs and consequently its large entropy facilitates its dynamical formation.

\section*{Acknowledgemnts}
I thank my collaborators Shani Avitan, Hagar Meir, Yotam Sherf,  Tom Shindelman, Tamar Simhon, Kent Yagi and Yoav Zigdon. I would like to thank Suvendu Giri and Frans Pretorius for discussions and suggestions. Last but not least, I would like to express my gratitude  and thanks to my longtime collaborator Joey Medved. This presentation is based on our joint work. The research is supported by the German Research Foundation through a German-Israeli Project Cooperation (DIP) grant ``Holography and the Swampland'' and by VATAT (Israel planning and budgeting committee) grant for supporting theoretical high energy physics. \\

\bibliographystyle{utphys}

\providecommand{\href}[2]{#2}\begingroup\raggedright\endgroup

\title{Testing the nature of compact objects through gravitational-wave observations}
\author{Anuradha Gupta}
\institute{\textit{Department of Physics and Astronomy, The University of Mississippi,\\ University, Mississippi 38677, USA}}

\maketitle 

\begin{abstract}
Gravitational-wave observations of coalescing binaries provide an excellent platform to test the nature of compact objects and to detect exotic compact objects, if present in the Universe. The existence of these exotic compact objects has been theorized in the literature, and they can have masses and compactness comparable to those of black holes—hence, they are often referred to as black hole mimickers. There are several ways to distinguish between compact objects using gravitational waves, and one such method involves spin-induced quadrupole moments, which affect the dynamics of binaries. This method has been applied to high significant binaries observed up to the third observing run of the LIGO-Virgo-KAGRA collaboration; however, no significant evidence for a binary containing black hole mimickers has been found so far.
\end{abstract}

Various theories of gravity predict the existence of exotic compact objects that are massive and compact enough for the gravitational-waves (GWs) from their binaries to closely resemble those of binary black holes in general theory of relativity~\cite{Cardoso:2019rvt,Maggio:2021ans}. Hence, these exotic objects are often referred to as {\it black hole mimickers}. Examples of such black hole mimickers include boson stars~\cite{PhysRev.172.1331,PhysRev.187.1767}, fuzzballs~\cite{Mathur:2009hf,Bena:2022rna}, gravastars~\cite{Mazur:2004fk}, ultracompact anisotropic stars~\cite{1974ApJ...188..657B,Raposo:2018rjn}, elastic stars~\cite{Alho:2022bki,Alho:2023mfc,Alho:2023ris}, wormholes~\cite{1988AmJPh..56..395M,Visser:1995cc,Lemos:2003jb}, strange quark stars~\cite{1986A&A...160..121H,1986ApJ...310..261A}, asymmetric dark matter stars~\cite{Kouvaris:2015rea}, frozen stars~\cite{Brustein:2016msz}, and others.

Several methods exist to identify the true nature of compact objects using GWs, allowing us to distinguish between black holes and their mimickers. The first of these methods utilizes the compact object's response to the tidal field of its companion in a binary system, characterized by the tidal deformability parameter. For a black hole, the tidal deformability is zero, whereas for a black hole mimicker, it is non-zero~\cite{Uchikata:2016qku,Cardoso:2017cfl,Sennett:2017etc,Chakraborty:2023zed}. This is a promising method, as it does not require the object to be spinning, since the tidal interaction between compact objects is present even in Newtonian gravity.

However, when compact objects are spinning, a series of spin-induced multipole moments are generated that affect the binary's dynamics. This provides the second method.
The dominant moment, the spin-induced quadrupole moment of an object of 
mass $M$ is given as $Q = - \kappa M^3 \chi^2$, where $\chi = S/M^2$ is the dimensionless spin parameter, 
with $S$ being the spin angular momentum of the object and
$\kappa$ is the dimensionless quadrupole parameter that quantifies the amount of distortion in the gravitational field outside the star due to its spin. 
The $\kappa$ value depends sensitively on the detailed properties of the object. For instance, for Kerr black holes $\kappa = 1$ \cite{doi:10.1063/1.1666501,RevModPhys.52.299}, for slowly spinning neutron stars it varies between $\sim$2 and $\sim$14 \cite{Pappas:2012qg,Pappas:2012ns,Harry:2018hke}, for spinning boson stars it is between $\sim$10 and $\sim$150 \cite{PhysRevD.55.6081}, and for gravastars it could be negative \cite{Uchikata:2016qku}.
In the post-Newtonian (PN) approximation, the effect of the leading order spin-induced multipole moment (i.e., mass-type quadrupole) appears as a 2PN phase correction in the gravitational waveform. The PN corrections to this first appear at 3PN order~\cite{Bohe:2015ana}, and the sub-leading spin-induced multipole moment (current-type octupole) starts to contribute at the 3.5PN order~\cite{Marsat:2014xea}.

Other methods include using the excitation of resonant modes in compact objects~\cite{Asali:2020wup}, GW echoes~\cite{Barausse:2014tra,Cardoso:2016oxy}, tidal heating~\cite{Mukherjee:2022wws}, and testing the black hole no-hair theorem~\cite{Hansen:1974zz,Carter:1971zc,Gurlebeck:2015xpa}. Here, we will focus on the spin-induced quadrupole moment as a probe to search for black hole mimickers in the observed binary black hole (BBH) population. 

Krishnendu et al.~\cite{Krishnendu:2017shb,Krishnendu:2019tjp} developed a Bayesian inference framework to infer deviations from the BBH nature in binary signals observed through GWs. In this framework, deviation parameters were introduced in the inspiral phase of the IMRPhenomPv2 waveform model~\cite{Hannam:2013oca,Husa:2015iqa,Khan:2015jqa}, such that the spin-induced quadrupole moment parameters $\kappa_1$ and $\kappa_2$ for the primary and secondary components of the binary can be expressed as $1 + \delta \kappa_1$ and $1 + \delta \kappa_2$, respectively. For a BBH, both $\delta \kappa_1$ and $\delta \kappa_2$ vanish, whereas for a binary involving black hole mimickers, one or both of these parameters can be non-zero. Krishnendu et al.~noted that the correlation between mass and spin parameters makes the simultaneous measurement of $\delta \kappa_1$ and $\delta \kappa_2$ challenging. Therefore, they proposed to measure the symmetric and antisymmetric combinations of $\delta \kappa_1$ and $\delta \kappa_2$ instead:
$$
\delta \kappa_s = \frac{(\delta \kappa_1 + \delta \kappa_2)}{2}; \quad
\delta \kappa_a = \frac{(\delta \kappa_1 - \delta \kappa_2)}{2}
$$
Again, for a BBH, both $\delta \kappa_s$ and $\delta \kappa_a$ vanish, whereas for a non-BBH system, one or both of these parameters can be non-zero. However, to reduce the dimensionality of the problem, Krishnendu et al.~measure only $\delta \kappa_s$ along with the other binary parameters, while assuming $\delta \kappa_a = 0$ (as is true for BBHs). This is a strong assumption, implying that this method will only be applicable to binaries consisting of compact objects with identical spin-induced deformations. Nevertheless, a follow-up investigation will be needed if the data suggest a $\delta \kappa_s$ posterior significantly different from zero.

This method has been applied to the potential GWTC-3~\cite{LIGOScientific:2020tif,LIGOScientific:2021sio} BBHs observed by LIGO~\cite{LIGOScientific:2014pky} and Virgo~\cite{VIRGO:2014yos} detectors that satisfy the following criteria: the signal must have a false alarm rate less than 1/1000 per year, must have been observed in two or more detectors, the network signal-to-noise ratio in the inspiral phase must be equal to or greater than 6, and the posterior of the effective inspiral spin parameter $\chi_{\rm eff}$ inferred from standard parameter estimation must exclude zero at the 68\% credible interval. The first and second criteria were applied to all events analyzed in~\cite{LIGOScientific:2020tif,LIGOScientific:2021sio}, but the third and fourth are specific to this analysis. This is because in the current implementation of the method, the effects due to spin-induced quadrupole moments are only present in the inspiral part of the waveform model. Therefore, to reduce systematic biases from the merger-ringdown phase of the signal, this method needs to be applied only to inspiral-dominated events with a sufficiently high inspiral signal-to-noise ratio. Furthermore, since the method relies on the presence of spin in at least one compact object in the binary (otherwise there would not be any spin-induced moments), the fourth criterion ensures that the method is applied to systems with significant spin to obtain meaningful constraints on $\delta \kappa_s$. Applying these criteria yields a total of $13$ events in GWTC-3. 

\begin{figure}[h]
\centering
\includegraphics[width=\textwidth]{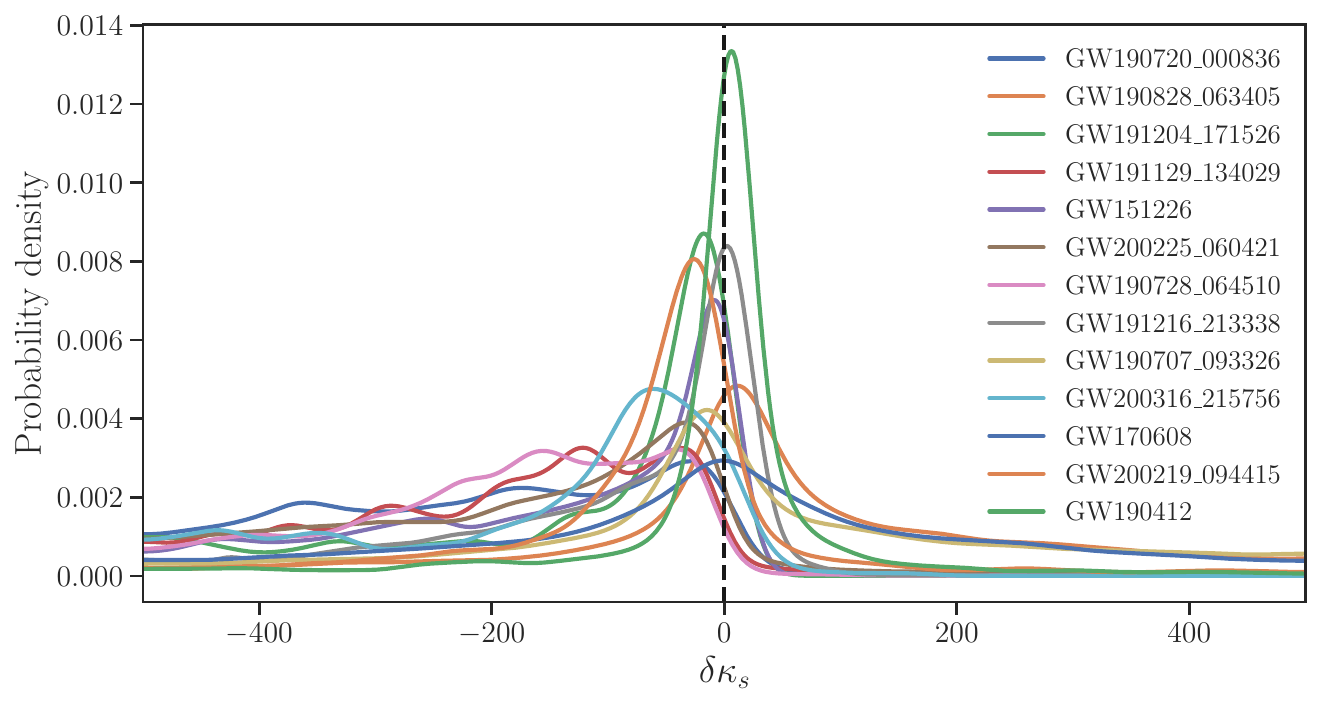}
\caption{Posterior probability distribution of $\delta \kappa_s$ for GWTC-3 binaries satisfying the selection criteria for this analysis. This figure is prepared using public data and code from~\cite{ligo_scientific_collaboration_virgo_coll_2022_7007370,dcclink}.}
\label{ks_single_events}
\end{figure}

Figure~\ref{ks_single_events} shows the posterior distribution of $\delta \kappa_s$ for the 13 events that satisfy the selection criteria, assuming a uniform prior on $\delta \kappa_s \in [-500, 500]$. All $\delta \kappa_s$ posteriors are consistent with zero within the 90\% credible interval, suggesting no strong evidence for any black hole mimicker.

\begin{figure}[h]
\centering
\includegraphics[width=\textwidth]{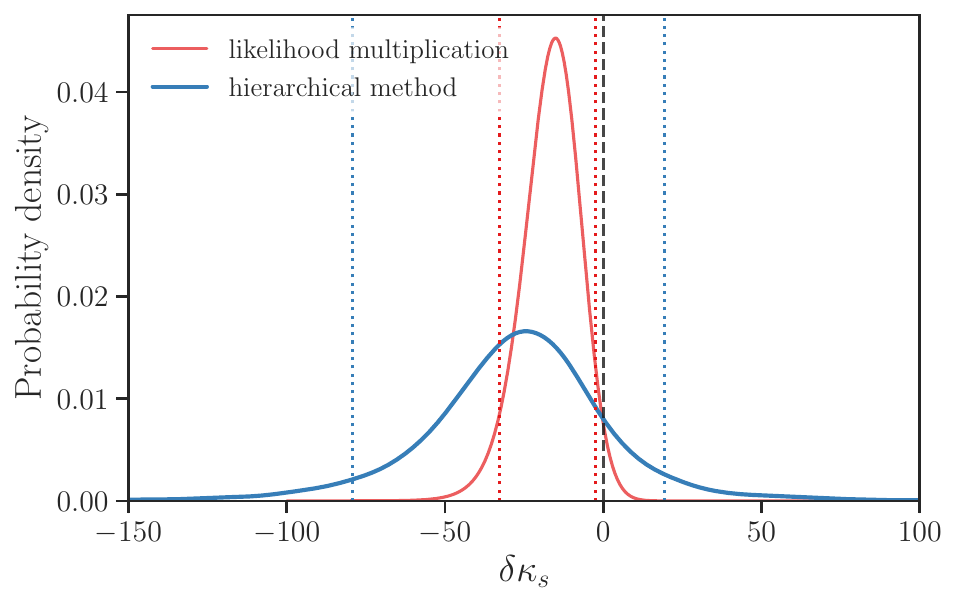}
\caption{Combined posterior probability distribution of $\delta \kappa_s$ from 13 GWTC-3 events that satisfy the selection criteria. Black dashed line represents the BBH value $\delta \kappa_s=0$ while the dotted lines represent the 90\% symmetric credible intervals. This figure is prepared using public data and code from~\cite{ligo_scientific_collaboration_virgo_coll_2022_7007370,dcclink}.}
\label{ks_combined}
\end{figure}

The posteriors from individual events were also combined to obtain a population-level bound on $\delta \kappa_s$ using two methods. The first method computes the combined posterior by simply multiplying the likelihoods of each individual event, assuming that all events in the population share the same $\delta \kappa_s$ value. The second method, called hierarchical analysis, assumes that the $\delta \kappa_s$ values for events in the population follow a Gaussian distribution. The combined posteriors on $\delta \kappa_s$ using the two methods are shown in Fig.~\ref{ks_combined}. The $\delta \kappa_s$ posterior obtained using the hierarchical method is consistent with zero within the 90\% credible interval, implying no evidence for black hole mimickers at the population-level either. 
However, the $\delta \kappa_s$ posterior from the simple likelihood multiplication method excludes zero. This shift can be attributed to weak individual-event constraints on $\delta \kappa_s$ in the negative prior region, due to waveform degeneracies. Very recently, Divyajyoti et al.~\cite{Divyajyoti:2023izl} extended the method of Krishnendu et al.~to include the effects of double spin-precession and higher harmonics on spin-induced quadrupole moment measurements. This updated method is expected to break waveform degeneracies in the negative $\delta \kappa_s$ region and improve constraints on $\delta \kappa_s$ in the future.

\section*{Acknowledgment}
AG is supported by NSF grants PHY-2308887 and AST-2205920. LIGO Laboratory and Advanced LIGO are funded by the United States National Science Foundation (NSF) as well as the Science and Technology Facilities Council (STFC) of the United Kingdom, the Max-Planck-Society (MPS), and the State of Niedersachsen/Germany for support of the construction of Advanced LIGO and construction and operation of the GEO600 detector. Additional support for Advanced LIGO was provided by the Australian Research Council. Virgo is funded, through the European Gravitational Observatory (EGO), by the French Centre National de Recherche Scientifique (CNRS), the Italian Istituto Nazionale di Fisica Nucleare (INFN) and the Dutch Nikhef, with contributions by institutions from Belgium, Germany, Greece, Hungary, Ireland, Japan, Monaco, Poland, Portugal, Spain. The construction and operation of KAGRA are funded by Ministry of Education, Culture, Sports, Science and Technology (MEXT), and Japan Society for the Promotion of Science (JSPS), National Research Foundation (NRF) and Ministry of Science and ICT (MSIT) in Korea, Academia Sinica (AS) and the Ministry of Science and Technology (MoST) in Taiwan. Unless otherwise specified, the contents of this release are licensed under the Creative Commons Attribution 4.0 International License. To view a copy of this license, visit http://creativecommons.org/licenses/by/4.0/ or send a letter to Creative Commons, PO Box 1866, Mountain View, CA 94042, USA.

\providecommand{\href}[2]{#2}\begingroup\raggedright\endgroup

\title{Exotic compact objects in their astrophysical environment}
\author{Héctor R. Olivares-Sánchez}

\institute{
	\textit{
		Departamento de Matem\'{a}tica da Universidade de Aveiro and Centre for Research and Development in Mathematics and Applications (CIDMA), Campus de Santiago, 3810-193 Aveiro, Portugal}
		}

\maketitle

\begin{abstract}
Several extensions to general relativity and the standard
model of particle physics predict the existence of exotic
compact objects that can reach a compactness high enough
to be considered as black hole mimickers.
Understanding their astrophysics is essential to
determine their feasibility as mimickers and to identify signatures
to distinguish them.
Bosonic stars are some of the most well understood
\acp{ECO}, which makes them interesting models
to study the observational appearance of more general \acp{ECO}.
Here we review a selection of the literature
on bosonic stars interacting with their environment,
with a special focus on magnetized accretion and general relativistic magnetohydrodynamic
(GRMHD) simulations.

		\noindent\textbf{Keywords:} black hole physics -- accretion discs -- gravity -- modified gravity

	\end{abstract}

	\section{Introduction}
	\label{sec:intro}
	
	In ``Lunar reflection'' (original: 
	{\it Reflejo Lunar}\footnote{\url{https://remedios-varo.com/old/reflejo-lunar-1957/}})
	by the Spanish-Mexican painter
	Remedios Varo, we see a mysterious object floating over a water pond in the middle of a forest at night.
	Two thoughts capture my attention as a viewer, which I can relate to ideas in the search
	for exotic compact objects.
	The first one is that despite the mystical nature of the object,
	its reflection on the water surface is that of a familiar object: the moon.
	A personal interpretation is that when we observe an object's image
	it reveals some of its properties, but it may at the same time hide
	another important part of its true nature.
	The second one is that the object represented is immersed in the natural
	world. We see that it casts light on the trees around it,
	and the trees interact with it by trapping in their branches
	some of the smaller light orbs that travel with it.
	The realization that each of the fascinating objects predicted by
	theoretical physics, if part of the real world, has to interact
	with an environment governed by familiar physical laws,
	played a crucial role in shifting our view on \acs{BH}
	from mathematical curiosities to part of the astrophysical reality.
	
	Several extensions to \ac{GR} and the \ac{SM} of particle physics
	predict the existence of \acp{ECO} known as \ac{BH} mimickers,
	that can reach a compactness high enough to be confused
	with \acp{BH}, but possess no event horizon.
	It is worth mentioning that many of these objects
	were not conceived originally as \ac{BH} mimickers,
	but have an independent theoretical support.
	This is, for example, the case for {\it bosonic stars},
	which are simply predictions of any particle physics model
	that introduces	new bosons with very weak interactions
	with the \ac{SM} particles, %
	but which turn out to also behave as \ac{BH} mimickers.
	
	Bosonic stars are actually some of the most well understood
	\acp{ECO}. They have a known formation mechanisms and
	well defined dynamics, which makes them interesting models
	to study the observational appearance of more general \acp{ECO}.
	After a very brief introduction on bosonic stars (Section \ref{sec:bosonic-stars})
	In this work, we will revise the existent literature
	on modeling the interactions between them and
	their astrophysical environment, in particular,
	in accretion scenarios.
	Section \ref{sec:stationary}
	describes scenarios based on semianalytic models
	where boson stars act as \ac{BH} mimickers.
	Section \ref{sec:hydro-simulations} reviews works
	that performed \ac{GRHD} simulations in bosonic
	star spacetimes.
	Section \ref{sec:mri} discusses the main process
	that produces angular momentum transport in highly
	ionized accretion discs and a mechanism that can suppress
	it in \ac{ECO} spacetimes.
	\Ac{GRMHD} simulations of accretion onto bosonic stars
	and the realization of the aforementioned mechanism
	are discussed in Section \ref{sec:grmhd-accretion}.
	Finally, in Section \ref{sec:conclusion}
	we conclude and mention some possible research directions
	on the subject.
	
    \section{Bosonic stars}
    \label{sec:bosonic-stars}
    Bosonic stars are self-gravitating stationary configurations made of coherent bosonic fields.
    These fields are classical, and depending on whether these are
    scalar or vector, stars are known as {\it boson} or {\it Proca} stars.
    The dynamics of these fields follow from the actions
    
    \begin{align}
    	\label{eq:scalar-action}
    	S_{\rm EKG}%
    	&= \int dx^4 \sqrt{-g}
    	\left(
    	                  \frac{R}{16\pi}
    	                 -\partial^\mu \bar{\phi} \partial_\nu\phi
    	                 - V(|\phi|^2) \right) \\
    	\label{eq:Proca-action}
    	S_{\rm EP}%
    	&= \int dx^4 \sqrt{-g}
    	\left(
    	                  \frac{R}{16\pi}
    	                 -\frac{1}{4} F_{\mu\nu} \bar{F}^{\mu\nu}
    	                 - V(|\mathbf{A}|^2) \right)
    \end{align}
	where $g$ is the metric determinant, $R$ is the Ricci scalar,
	$\phi$ is the scalar field, $\mathbf{A} = A_\mu dx^\mu$ is the vector field,
	$F_{\mu\nu}=\partial_\mu A_\nu - \partial_\nu A_\mu$ is the field strenght,
	and the bar denotes complex conjugation.
	
	Depending of the form of the potentials $V(|\phi|^2)$ and $V(\mathbf{A})$, it is possible to find
	different families of bosonic stars.
	For instance, boson stars with a simple quadratic potential $V=m^2|\phi|^2$ are known as {\it mini boson stars}
	while potentials that include quartic and sextic terms can form {\it solitonic stars} or {\it Q-stars}.
	
	Bosonic stars are very interesting for several reasons. As mentioned above, they have a well known
	formation mechanism, known as {\it gravitational cooling} and a well defined dynamics described
	by the Einstein-Klein-Gordon or the Einstein-Proca equations, which follow from actions
	\eqref{eq:scalar-action} and \eqref{eq:Proca-action}, respectively.
	The stability properties of some families have been well studied, revealing that in general
	they have stable and unstable branches separated by a critical mass that depends on the
	mass of the bosonic particle.
	In general, small masses produce large bosonic stars
	(for ultralight scalars, these can reach sizes comparable to galactic halos).
	This allows boson particles of different masses to produce structures that can mimic \acp{BH} in different mass ranges,
	from stellar to supermassive.
	For modeling their interaction with the astrophysical environment,
	there are three important qualitative properties that have to be considered:
	(i) they have no event-horizon, (ii) they are regular at the origin (possess no singularity), and (iii) if the bosonic particle
	has a very weak interaction with the \ac{SM} sector, they can be considered to have no surface.
	These properties mean that baryonic matter can in principle pass through the star feeling only its gravitational
	potential. It also means that matter does not disappear behind a horizon as it is the case for \acp{BH},
	so it will continue to influence its surroundings and to emit radiation that can in principle reach a distant observer.
	
	More information on the different possibilities studied in the literature can be found in the excellent
	review by \citet{liebling_dynamical_2012}. The original description of Proca stars
	is presented in \citet{brito_proca_2016}.
	
    \section{Bosonic stars as black hole mimickers}
    \label{sec:stationary}
    Since the discovery of Proca star solutions is more recent,
    most of the literature on bosonic stars as \ac{BH} mimickers
    refers to boson stars.
    It was soon realized that boson stars could act as \ac{BH} mimickers,
    which motivated the study of their
    observational properties in different astrophysical scenarios.
    \citet{guzman_accretion_2006} and \citet{Guzman2009} studied the 
    electromagnetic spectrum of thin relativistic
    accretion disks \cite{Novikov1973}
    around boson stars. They found that for self-interacting boson stars
    (with a quartic potential) it was possible to find a combination of parameters
    almost perfectly mimicking the spectrum of a \ac{BH} of a given mass.
    However, in general thin disks around boson stars are expected to behave
    differently than for \acp{BH}.
    While for \acp{BH} the accretion disk is expected
    to terminate at the \ac{ISCO},
    the boson stars considered in Refs. \citep{guzman_accretion_2006,Guzman2009}
    are in general not
    compact enough to produce this feature, and stable orbits extend down to the
    origin. In fact, many \acp{ECO} admit stable low angular momentum orbits that
    are impossible for \acp{BH}, as they would become capture orbits.
    Interestingly, this includes periodic orbits with zero angular momentum.
    This makes necessary to resort to arguments beyond thin disk theory
    in order to determine the inner edge of the accretion disk.
    Although Refs. \citep{guzman_accretion_2006,Guzman2009}
    took the conservative assumption of
    extending the accretion disk down to the origin, it may be expected
    from conservation of mass that the disk is truncated at some
    radius, where material either escapes as an outflow or accumulates
    into a central structure sustained by thermal pressure.
    
    Another way in which boson stars could mimic black holes is
    by strongly lensed images as those obtained through
    \ac{VLBI} experiments as the \ac{EHT}.
    This possibility was first studied
    by \citet{vincent_imaging_2016}, who found that ray-traced images
    of synchrotron emission from toroidal plasma configurations
    around boson stars show \acp{CBD} that can mimic in shape and size
    the shadows of Kerr \acp{BH}.
    Although these flow configurations can be considered as
    an approximation to the thick disks produced in hot accretion flows
    \citep{Abramowicz1978},
    such as those expected in the vicinity of the main targets
    of the \ac{EHT}, \sgra and \mess,
    the model does not prescribe a value for the inner radius of
    the tours.
    This is particularly important because it is the
    the lensed image of the hollow region inside the torus
    what determines the geometry of the \ac{CBD}.
    A natural question is whether the issue of the termination
    radius of the accretion disk can be answered by
    going beyond stationary configurations trough direct numerical
    simulations. The next sections will therefore be dedicated
    to revise the efforts in simulating fluids in motion
    around bosonic stars.

    \section{Simulating fluids around boson stars}
    \label{sec:hydro-simulations}
    A pioneer work on \ac{GRHD} simulations of accretion onto
    boson stars was that by \citet{Meliani2016}.
    They performed simulations of spherical
    (Bondi-Michel \citep{bondi_spherically_1952,Michel1972})
    accretion onto boson stars in 1D and 2D.
    For the 2D simulations, matter was accreting following the
    profile of the Bondi solution only on a wedge-shaped region
    around the equatorial plane.
    As it could be expected, they observed the fluid getting shocked
    and escaping along the polar directions.
    Other interesting works in \ac{GRHD} are those of
    \citet{meliani_tidal_2017} and \citet{teodoro_tidal_2021}.
    Motivated by the passage of the G2 cloud near \sgra
    (see e.g., \citep{shcherbakov_properties_2014}),
    they simulated a close encounter between a gas cloud and a boson star.
    They observed important differences in its trajectory and lightcurve
    with respect to the expectations from the \ac{BH} scenario.
    Most notably, they showed the possibility of observing
    oscillations and residual emission
    from matter trapped in the potential well of the boson star
    (that would have disappeared behind the horizon for the case of a \ac{BH}),
    as well as unusual orbits
    (including very-low-angular-momentum ones as mentioned in Sec. \ref{sec:stationary}).
    
    However, in order to study accretion scenarios such as those
    expected near the \ac{EHT} targets and X-ray binaries, it is
    crucial to take into consideration de role of magnetic fields.
    In the next Section we explore existent simulations of magnetized accretion
    around boson stars.

    \section{Suppression of the magneto-rotational instability}
    \label{sec:mri}
    One of the most important mechanisms responsible of angular momentum
    transport in highly ionized accretion disks is the \ac{MRI}
    \citep{balbus_powerful_1991}.
    This instability occurs in very general magnetized accretion
    scenarios\footnote{Although is relative importance with respect to
    other magnetic instabilities is disputed, for instance, in the magnetically
	arrested disk (MAD) scenario \citep{begelman_what_2022}.},
	as its operation depends mainly on the accretion disk having
	a rotation profile that decreases with distance,
	that is,  $d\Omega/dR < 0$, where $\Omega$ is the
	fluid angular velocity and $R=r\sin{\theta}$ is the cylindrical radius.
	This is the case for Keplerian accretion disks
	and for circular equatorial geodesics in the the Kerr spacetime.
	However, once we consider nonvacuum spacetimes,
	it becomes relatively easy to violate this condition.
	This can be seen straightforwardly in Newtonian dynamics by considering
	a particle orbiting a spherically symmetric object.
	The rotation velocity needed to maintain the particle
	in a circular orbit at radius $r$ is given by
	
    \begin{equation}
    	\Omega^2(r) = \frac{Gm(r)}{r^3} \,,
    \end{equation}
    where $G$ is Newton's constant and $m(r)$ is the mass enclosed by
    a sphere of radius $r$.
    From this, it is possible to notice that for any
    $m(r)$ that grows faster than $m \propto r^3$ over some range in $r$
    (that is, faster than constant density),
    we will have $d\Omega/dR \ge 0$ and the \ac{MRI} will be suppressed.
    Given that after the angular momentum is redistributed
    (for instance, by the \ac{MRI}),
    the rotation profile of an accretion disk approaches that of circular
    geodesics, the above criterion becomes useful to predict
    whether we can expect the \ac{MRI} to be suppressed in a given spacetime.
    This means that, if the rotation profile of circular geodesics $\Omega_{\rm K}(r)$
    possess a local maximum at $r_{\rm max}$, we may expect material
    to accrete for $r>r_{\rm max}$ and stop at $r_{\rm max}$,
    where a portion of it will be repelled by the centrifugal barrier and
    escape at higher latitudes, and another portion will
    stall at that location, providing a termination radius for the accretion disk.
    For cases where $\Omega_{\rm K}(r)$ is monotonically decreasing with radius,
    matter will simply accrete and stall at the origin, where it will
    form a pressure-supported structure.

    \section{Magnetized accretion onto bosonic stars}
    \label{sec:grmhd-accretion}
    In \citet{olivares_how_2020}, my collaborators and I simulated
    accretion onto two members of the family of mini-boson stars.
    One of the models (model A) possess a maximum of $\Omega_{\rm K}(r)$,
    while for the other (model B) it is monotonically decreasing.
    The simulation of boson star model A shows the formation of a mini-torus
    at the expected location, while that of model B
    shows a nearly spherical accumulation of mass around the origin.
    The analysis of the forces acting on fluid elements on the equatorial plane
    and of the rotation velocity profile is consistent with the scenarios
    described in Section \ref{sec:mri} \citep{olivares_how_2020}.
    
    \begin{figure}
    	\centering
    	\includegraphics[width=0.3\linewidth]{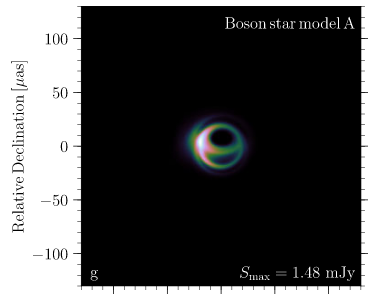}
    	\caption{Ray-traced image of accretion onto boson star model A, showing the lensed image of the hollow region inside the mini-torus.}
    	\label{fig:bosonstar}
    \end{figure}

    Interestingly, ray-traced images of the simulation of model A display a
    \ac{CBD} that is formed in the same way as those in
    Ref.~\citep{vincent_imaging_2016} (Fig. \ref{fig:bosonstar}).
    For spherically symmetric spacetimes, it is possible to obtain a simple
    expression that relates the inner radius of the mini-torus $r_{\rm in}$ in
    Boyer-Lindquist-type coordinates to the impact parameter
    of photons grazing it, $b(r_{\rm in})$. This is
    \begin{equation}
    	\label{eq:impact_parameter}
    	b(r_{\rm in}) = \frac{r_{\rm in}}{\alpha(r_{\rm in})} \,,
    \end{equation}
    where $\alpha(r_{\rm in})$ is the lapse function of the metric.
    
    Since $b(r_{\rm in})$ gives the radius of the \ac{CBD},
    it is possible to use equation \eqref{eq:impact_parameter}
    to estimate which members of a family of \acp{ECO} may
    produce a \ac{CBD} of size comparable to a \ac{BH} shadow,
    namely $3\sqrt{3}\ M$ for a Schwarzschild \ac{BH}.
    In this way, it is possible to find that the only members
    of the mini-boson star family that produce a \acp{CBD} are on
    the unstable branch, and none of them can produce one
    of a sufficiently large size to be mistaken by a \ac{BH}
    of the same mass. This limits the possibilities
    of mini-boson stars as black hole mimickers.
    
    The possibility of predicting whether an \ac{ECO}
    can mimic a \ac{BH} by the mechanism described above
    solely by analyzing its metric motivated
    an ``imitation game challenge'' to find
    possible mimickers.
    \citet{herdeiro_imitation_2021}
    analyzed several families of bosonic stars, finding
    a mini-Proca star ($V(|\mathbf{A}|^2) = \mu^2 |\mathbf{A}|^2 /2$)
    solution for which $r_{\rm max} = 6\ M$,
    which could mimick the appearance of a Schwarzschild
    \ac{BH} surrounded by a thin accretion disk
    terminating at the \ac{ISCO}.
    This model is in the non-relativistic (Newtonian)
    branch of the solutions, which was considered to be stable.
    Later, however, it was shown that spherically symmetric
    Proca stars are excited states and decay
    to the true ground state of the family, which is prolate,
    making a very interesting case of minimally coupled
    relativistic stars where staticity and stability
    implies non-sphericity \cite{herdeiro_non-spherical_2024}. 
    
    In a work in preparation, \citet{Meneses2025},
    sampled numerous families of $Q$-stars, 
    which are produced by the potential
    
    \begin{equation}
    	\label{eq:qstar_V}
    	V(\Phi) = \mu^2|\Phi|^2 + \frac{1}{2}\lambda|\Phi|^4 + \frac{1}{3} \nu|\Phi|^6\,,
    \end{equation}
    where $\lambda$ and $\nu$ are parameters of the theory.
    It was found that a region of the parameter space has
    models that are stable and produce a $b(r_{\rm in})$
    that approaches the value $3\sqrt{3}$ of Schwarzschild \ac{BH}
    shadows. \Ac{GRMHD} simulations confirm the scenario described
    above with the size of the hollow only slightly smaller than
    the prediction given by the maximum of $\Omega_{\rm K}(r)$.
    The latter is to be expected, as the rotation profile is not
    perfectly geodesic and the pressure gradient pushes
    to mini-torus material towards the origin.

    \begin{figure}
    	\centering
    	\includegraphics[width=0.3\linewidth]{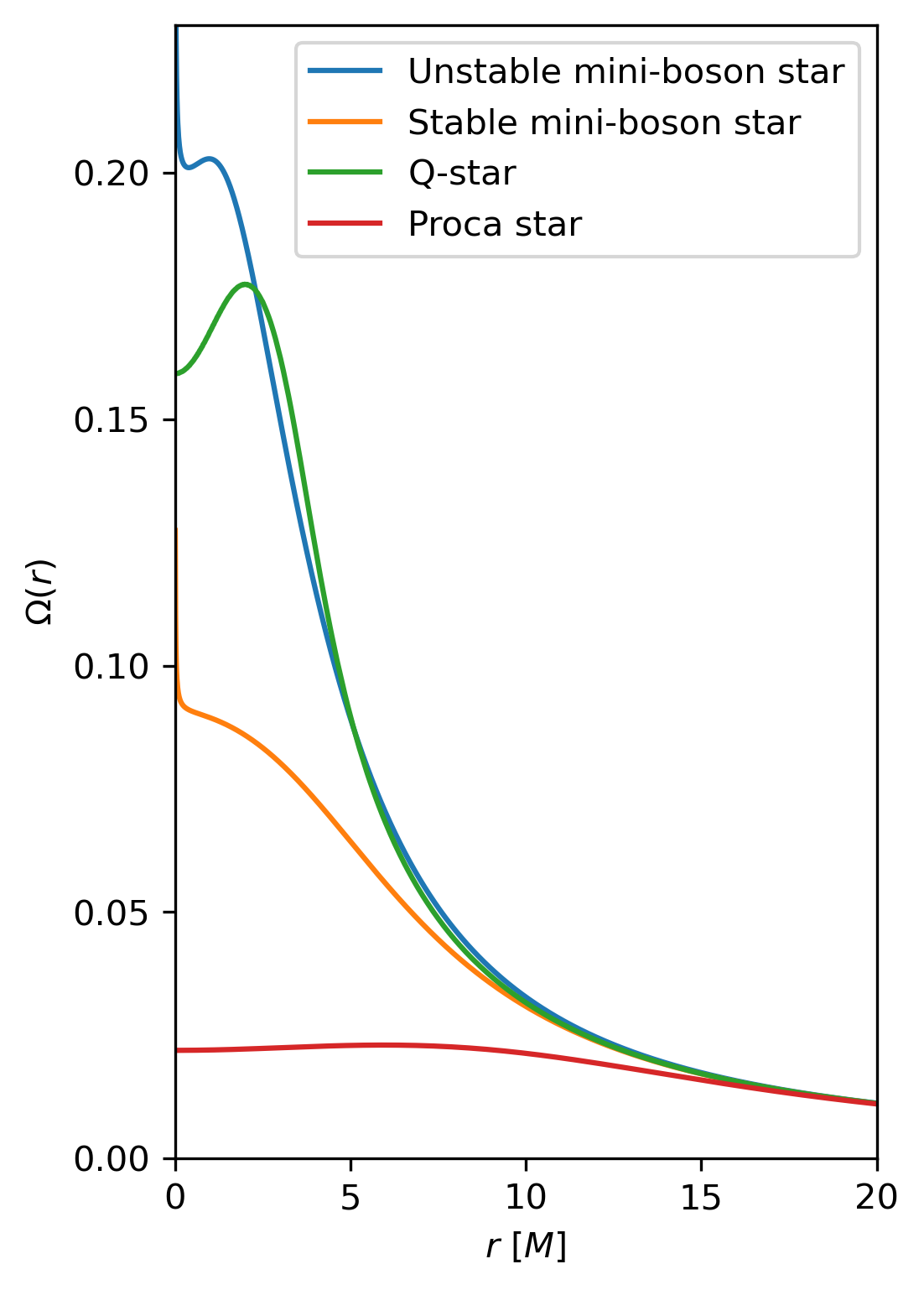}
    	\includegraphics[width=0.3125\linewidth]{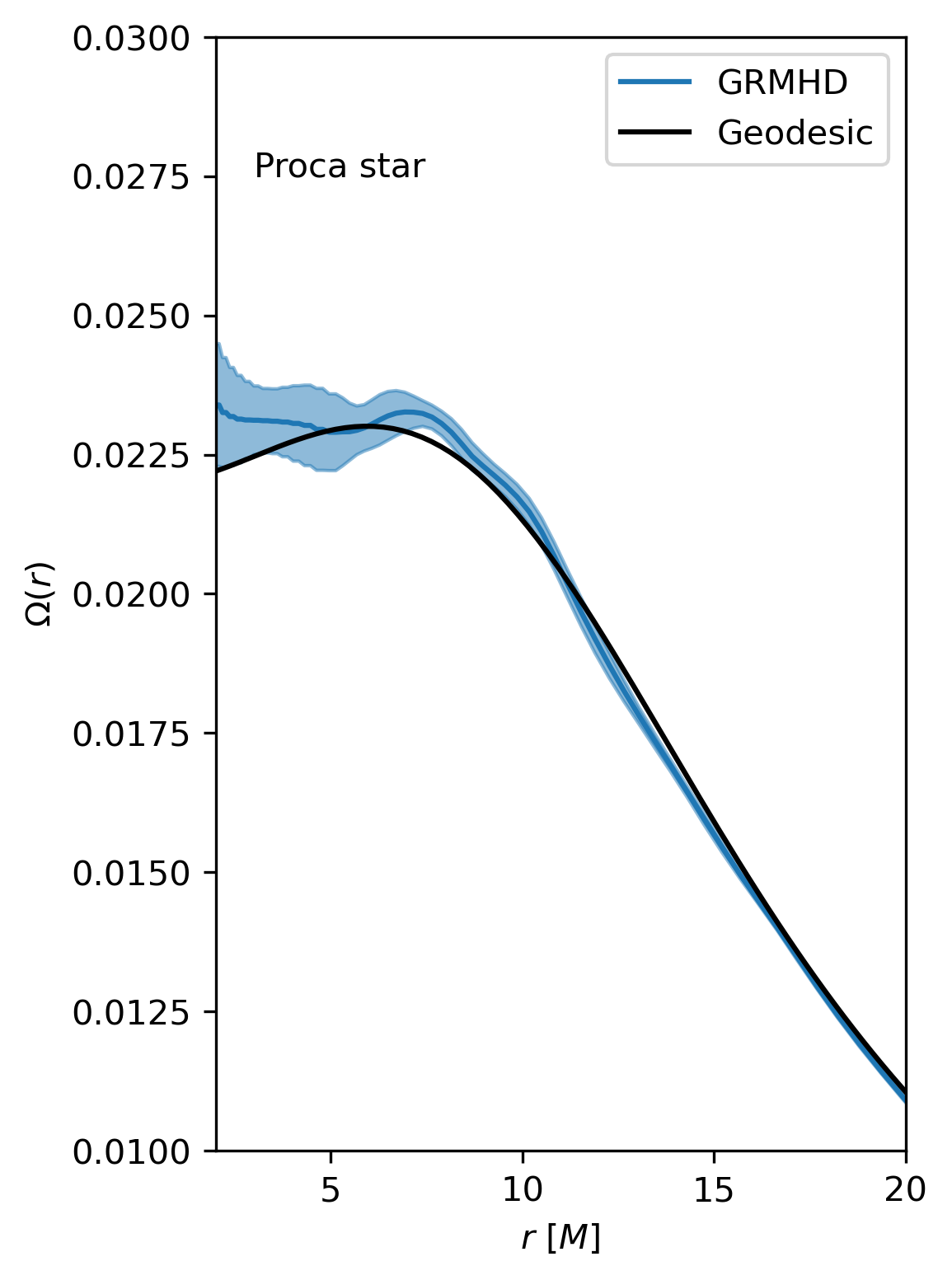}
    	\caption{{\it Left:} Rotation profiles of circular geodesics for all the bosonic stars
    	considered in Section \ref{sec:grmhd-accretion}. {\it Right:} Comparison of the
    	rotation profile of circular geodesics in the Proca star spacetime with the actual
	    accretion profile of the simulated accretion flow. The solid line represent
    	a time average over 5000 $M$, and the filled region represents one standard deviation.}
    	\label{fig:rotation_profile}
    \end{figure}

    Even though the Proca star model from Ref. \citep{herdeiro_imitation_2021}
    has been shown to be unstable, it is interesting to
    verify whether \ac{GRMHD} simulations confirm
    the formation of the hollow at $r_{\rm max}=6\ M$.
    These simulations were carried out by \citep{Olvera2025}.
    It turns out, however, that although initially the material
    stalls at $r_{\rm max}$, it slowly makes its way towards the center
    until the hollow is filled.
    The reason behind this can be understood by comparing the
    profiles of $\Omega_{\rm K}(r)$ for the different objects
    mentioned in this work (Fig. \ref{fig:rotation_profile}, left panel).
    It can be noticed that the maximum of this profile for the
    Proca star is almost flat in comparison with those
    of the $Q$-star and of boson star model A.
    By comparing $\Omega_{\rm K}(r)$ with the
    time average and standard deviation of the actual
    rotation profile obtained from the simulation
    (Fig. \ref{fig:rotation_profile}, right panel), it can be observed that
    the amplitude of the turbulent fluctuations is larger than the
    height of the maximum of $\Omega_{\rm K}$ with respect to its
    neighborhood, which makes it negligible for the accretion flow.
    This result gives important information on the robustness
    of \acp{CBD} produced by the suppression of the \ac{MRI}.
    
    \section{Conclusion and outlook}
	\label{sec:conclusion}
	We have reviewed a selection of works (semianalitic and simulation-based),
	that have investigated the observational properties of bosonic
	stars, and in particular their possible role as \ac{BH} mimickers.
	The most important lesson from this walk through the literature
	is that understanding the astrophysics of \acp{ECO} is essential to
	determine their feasibility as \ac{BH} mimickers and to identify signatures
	to distinguish them. We have also learned that interactions
	between \acp{ECO} and their environments can lead to unexpected effects,
	which highlights the importance of performing numerical simulations.
	
	An interesting direction to continue this research
	is the study of rotating objects, as all of the objects
	in this work for which \ac{GRMHD} accretion has
	been simulated are stationary and spherically symmetric.
	Rotation is expected to produce a much richer phenomenology.
	For instance, it has been suggested that the non-existence
	of circular geodesics in some rotating Proca star spacetimes
	could be another way of producing a termination radius of
	the accretion disk, which could be accompanied by a
	\ac{CBD} in ray-traced images \citep{sengo_imitation_2024}.
	A work in preparation by \citet{Sengo2025}
	aims to test this prediction by means of \ac{GRMHD} simulations.
	Another direction is a more systematic exploration of
	synthetic observables produced by these simulations.
	For instance, differences between a true \ac{BH} shadow
	and a \ac{CBD} produced by the mechanism described here could be
	revealed by multi-frequency or polarized observations.

    \vspace*{0.5cm}
    \noindent
    {\large \bf Acknowledgments}

	\noindent
    The author is supported by the Individual CEEC program - 5th edition funded by the Portuguese Foundation for Science and Technology (FCT).
    This work is supported by the Center for Research and Development in Mathematics and Applications (CIDMA) under the FCT Multi-Annual Financing Program for R\&D Units.
    The author acknowledges support from the projects PTDC/FIS-AST/3041/2020, CERN/FIS-PAR/0024/2021 and 2022.04560.PTDC. This work has further been supported by the European Horizon Europe staff exchange (SE) programme HORIZON-MSCA-2021-SE-01 Grant No. NewFunFiCO-10108625.

\title{Black Holes are the Best Mimickers}
\author{Steven L. Liebling}
\institute{\textit{Long Island University, Brookville, NY 11548}}

\maketitle 

\begin{abstract}
Instead of studying black holes by considering some compact object
in the highly compact limit, I discuss a couple recent projects
in which the black holes behave exotically. In particular, I look
at the gravitational waveform for a black hole emitting a smaller black
hole, and a different project examining the observability of
a magnetized primordial black hole that collides with a neutron star.
Finally, I discuss an on-going project to map boson stars to neutron stars.
\end{abstract}


Instead of picking some compact object and considering its highly compact
limit,
one can start with a black hole in general relativity~(GR) and consider
what happens if it does not act like a classical black hole--i.e. one consistent with GR. If observable consequences of
such behavior can be predicted, then appropriate searches of observational data may discover
signs of new physics or, more likely, produce null tests of GR.

Such reasoning is not meant to give license 
to 
study every non-standard theory out there. But if, without undue effort,
 some esoteric idea is found to produce  observable effects, then surely it makes sense to consider it, even if one has no faith that an effect will be found.
As an example, many years ago we modeled near-horizon fluctuations in a black hole merger
arising from a nonviolent non-locality proposal from the high energy community,
and were able to show that such fluctuations produce very significant deviations
from the GR prediction and were therefore excluded~\cite{Liebling:2017pqs}.

As it turns out,\footnote{In other words, I did not set out with a plan to investigate \textit{black holes behaving badly} (the alternative title for this piece).} I have considered black holes behaving outside the bounds of classical GR a couple times in the past few years. Below I briefly describe these, focusing on the exciting possibility of an observable while providing references for the details. I close with a brief description of work in progress towards a mapping of boson stars to neutron stars,
themselves both black hole mimickers.

\section{Black hole emission}
\textbf{What would the gravitational waveform for a black hole emitting a smaller black hole look like} and would the waveform be simple to find?  

Black holes in GR do not emit other black holes, and so this is a very exotic possibility that nevertheless does have a number of motivations.
One such motivation comes from high energy papers which suggest that large extremal
black holes may be unstable in keeping with the \textit{weak gravity conjecture.}~\cite{Cheung:2018cwt,McInnes:2021zlt} or may be subject to fragmentation~\cite{Xia:2024nmp}.

Arguably more conventional motivations exist. The emission here is just some small, compact object, and so the emission could be a particularly massive particle as part of the black hole's Hawking radiation. 

Ultra-spinning black holes in higher dimensions are known to be unstable and so perhaps we could observe such an instability~\cite{Emparan:2003sy,Figueras:2017zwa}. This would be a classical GR solution, albeit with additional dimensions. 

Finally, because GR is
time symmetric in the sense that the time reversal of a given solution is also a solution, the time reversal of a black hole merger is also a solution.\footnote{Thanks to Luis Lehner and Arvind Borde for clarifying that the time reversal of a black hole merger consists of a white hole emitting a smaller white hole.
As discussed by Hawking and Ellis~\cite{Hawking:1973uf} and by Wald~\cite{Wald:1984rg}, the \textit{bifurcation} of a black hole results in a naked singularity and is therefore excluded.}

It turns out that black hole emission, if one ignores the earliest moments of emission, is essentially an outspiral event which is predicted for binaries
that lose mass. Some traditional examples include stars in an extreme mass ratio binary that lose material~\cite{Dai_2013,Linial:2017hep}.
A more exotic possibility consists of
a rotating black hole bound in a binary with some other compact object. Superradiance may drive the formation of a cloud around the black hole. The cloud and the black hole retain the original black hole mass, and the binary remains bound. However, if the cloud is subsequently radiated, the black hole loses mass, and could become unbound and outspiral~\cite{Kavic:2019cgk} (see also~\cite{Cardoso:2011xi,Cao:2023fyv}).

At first glance, one might consider simply
time reversing the waveform resulting from an extreme mass ratio inspiral, 
but such a reversal  would imply that one is irradiating a black hole merger with gravitational waves. Instead, working with Tousif Islam and Gaurav Khanna,
we worked with linearized gravity~\cite{Islam:2024wqf}. The first step was calculating a plunging geodesic for the secondary which we then reversed to get the geodesic for an outgoing particle. Solving the Teukolsky equation for this
\textit{reverse plunge} yielded our waveform. 

The natural question asks how different this reverse plunge waveform is from simply reversing the waveform obtained from the plunge. We do find some oscillations, which become more striking in the higher modes. However, there is significant overlap, and, in particular, it is not surprising that we see a \textit{reverse chirp}.
In this light, \textbf{we argue in support of searches of such reverse chirps, either by considering the time reversal of merger wave models or otherwise with generic, unmodeled searches for both LIGO-Virgo-Kagra and LISA observatories.}

\section{Magnetic primordial black holes}
\textbf{What if small black holes do not evaporate out of existence quickly?}
Some looking to solve the black hole information problem posit long-lived
remnants, but a more conventional mechanism exists via magnetic monopoles.
Such magnetic charges are widely expected to have formed in the early universe
with cosmic inflation diluting them into unobservability. But if instead
these monopoles fell into primordial black holes, they would be very hard to
neutralize, unlike an electrical charge, because of the expected large mass
of monopoles.

Magnetized primordial black holes~(mPBHs) would Hawking radiate towards extremality
at which point they would achieve zero temperature and cease
radiating. My colleagues and I studied the interaction of such mPBHs with
neutron stars~\cite{Estes:2022buj} and argue for two effects.

The first involves a collision of the mPBH with a pulsar. Previous studies
have found that an unmagnetized PBH encountering a neutron star would
traverse the star entirely, possibly returning if bound. However,
the electromagnetic interactions with the magnetic monopole would stop
the transiting mPBH almost immediately, transferring all of its angular momentum
and producing an anti-glitch. 
Anti-glitches have been observed, although the
first anti-glitch in a rotation powered pulsar was only recently
discovered~\cite{Tuo:2024pvf}.

The other effect is the possibility that a mPBH inside a neutron star
may form a long-lived star. Generally, a PBH inside a star will accrete
the entire star~\cite{East:2019dxt,Chen:2024qke}. However, the mPBH, once it accretes,
will then radiate, forming a stabilizing feedback mechanism. With
some significant simplifying assumptions, we solve TOV-like equations
for such equilibrium solutions. The effect of the central mPBH would be to
heat the star and to distort the star's magnetic field, both of which
potentially observable.

\section{Mapping between boson stars and neutron stars}
Boson stars have a long history as black hole mimickers, in part because they have a reasonable formation mechanism and require nothing more exotic than a scalar field~\cite{Liebling:2012fv}. Recently, a couple groups have made some remarkable success
in evolving the mergers of highly compact boson stars numerically~\cite{Ge:2024itl,Evstafyeva:2024qvp,Evstafyeva:2023kfg,Evstafyeva:2022bpr,Croft:2022bxq}
and
\cite{Siemonsen:2024snb,Siemonsen:2023age,Siemonsen:2023hko}.

However, I instead want to describe a project aimed at \textbf{studying boson stars mimicking neutron stars} that my student began three years ago. Initially, the idea was to use a neural network to ``learn'' the mapping from a set of mass and radius observations of neutron stars to the equation of state for dense stellar matter.
Lindblom had studied this inverse-mapping problem using other methods~\cite{Lindblom:2025kjz,Lindblom:2018ntw,Lindblom:2014sha,Lindblom:2013kra,Lindblom:2012zi}, but a neural network seemed well suited for the task. The idea of
doing similarly for boson stars came next. In particular, the structure of boson stars
is dictated by its scalar potential in much the same way that the equation of state dictates the structure of neutron stars. 
In this way, one then has a map between boson stars and neutron stars. 
My student made good progress with both these mappings, but issues remained requiring further work.

I have had a few students pick up this problem, but none have made progress. In the meantime, many have used neural networks to invert the neutron star part of the map~\cite{Soma:2022qnv,Krastev:2023fnh,Farrell:2023ojk,Ventagli:2024xsh,Morawski:2020izm,Sun:2024nye,Ofengeim:2024toq}. I briefly describe my approach.

In one python script, I generate a large number of piecewise polytropic equations of state with three segments and fixed transition densities. For each of these, I use an open-source TOV solver~\cite{tov} to compute the respective mass-vs-radius curve for each family. For those satisfying certain conditions, I create a dataset containing the parameters of the equation of state and the mass-vs-radius curves. I had initially found Chebyshev interpolants to the mass and radius each as functions of central density, and asked the neural network to find those coefficients. However, that did not work well, and instead adopted an approach consistent with other approaches; namely, I use four points from the mass-vs-radius curve at fractions (1.0,0.9,0.8,0.7) of the maximum radius. Not knowing good parameters for the neural net, I appreciated that some of the references included such details; I will report the parameters I use when I find an optimal solution.

In another python script, I use PyTorch to create a neural network and train it using the generated dataset. The neural network takes as input the eight values corresponding to the four points on the mass-vs-radius curve, and outputs the three polytropic indices of the equation of state. 
The trained network is then saved to file.

For the boson star, I parameterize the scalar potential as 
$V(|\phi|^2)=a |\phi|^2 
            +b |\phi|^4
            +c |\phi|^6$ so that the set of coefficients $\left\{ a,b,c\right\}$
are the output of the neural network. The training script is a minimally modified version of that used for the neutron stars. 

The final script loads each of the
two neural networks, and feeds the same mass-vs-radius points to each. One then has a scalar potential which maps to a piecewise polytropic equation of state.

All this is in place, but the mapping is not good. With improvements, hopefully this work will make clear where any boundaries are; in other words, for which mass-vs-radius curves, is there no such mapping? 
A wider class of potentials and boson stars could also be included; as described in my review with Carlos Palenzuela, \textbf{there is a huge variety of boson stars}~\cite{Liebling:2012fv}. Future work will: \textbf{include tidal deformability}, which may significantly shrink the scope of where these two compact objects are ``degenerate,'' and  \textbf{extend this mapping to other compact objects} either through the use of a solver or simply a training dataset.


%
\bibliographystyle{utphys}

\providecommand{\href}[2]{#2}\begingroup\raggedright\endgroup

\title{Some Clues on Black Hole Mimickers}
\author{Paolo Pani}
\institute{\textit{Dipartimento di Fisica, Sapienza Università 
	di Roma, \& INFN, Sezione di Roma, Piazzale Aldo Moro 5, 00185, Roma, Italy}
    }

\maketitle 

\begin{abstract}
We provide a short and informal overview of some challenges associated with testing the classical black hole paradigm and explore the concept of exotic compact objects as black hole mimickers. We discuss some key clues in distinguishing exotic compact objects from black holes, including maximum compactness, reflectivity, their stability/dynamics, and scale invariance. Specific models such as boson stars, anisotropic stars, and fuzzballs are critically examined under these lenses, along with their theoretical and observational implications.
\end{abstract}

\section{Introduction}

Black holes (BHs) are now a cornerstone of modern astrophysics, and their existence is strongly supported by both electromagnetic and gravitational-wave~(GWs) observations. However, the classical description of BHs in General Relativity~(GR) faces several unresolved theoretical challenges, including the entropy problem, the Hawking information paradox, and the presence of singularities~\cite{Bena:2022ldq,Carballo-Rubio:2025fnc}. These issues suggest the possibility of new physics at the horizon scale and motivate the exploration of alternative models for dark compact objects, also known as exotic compact objects~(ECOs)~\cite{Cardoso:2019rvt}.

Some of the key features of a BH are: i)~the presence of (unstable) photon spheres in the spacetime; ii)~an infinite redshift surface (the event horizon); iii)~their scale invariance: the BH mass is a free parameter not associated with any particular scale.
The latter property is also linked to the fact that the gravitational collapse and BH formation is essentially inevitable in any relativistic, classical theory of gravity, as nicely encoded by Thorne's hoop conjecture~\cite{Klauder:1972je}.

While the second property above is the defining feature of a BH, it is natural to ask at which level can ECOs  mimic various aspects of the BH phenomenology.
Among the various ECO models, we define \emph{ultracompact objects~(UCOs)} has those featuring a photon sphere, while we define \emph{BH mimickers} as those featuring a sufficiently high redshift as to actually mimic a BH horizon.

Mimicking the whole BH phenomenology is very challenging and it is important to stress that not all ECOs are UCOs and not all UCOs are viable BH mimickers.
In fact, given the inevitability of BH formation, one might question whether BH mimickers can really replace BHs \emph{in toto}, at least within a classical relativistic theory.

In this document we will briefly and critically discuss the clues that may help distinguish ECOs and BH mimickers from classical BHs.

\section{The Clues}

The search for distinguishing features between BHs and ECOs can be framed in terms of a few clues:

\subsection{Maximum Compactness/Redshift}  
The compactness of an object can be defined as $C \equiv \frac{GM}{Rc^2}$, where $M$ is the mass and $R$ is the (proper circumferential) radius. In spherical symmetry, BHs have $C=1/2$. 
Assuming the exterior of the object is (approximately) vacuum GR, the photon sphere is located at $r\approx 3M$, so by definition the compactness of a UCO is $C\gtrsim 1/3$.
 Having a photon sphere is a necessary but not sufficient condition to mimic the BH phenomenology, most of which is associated with the infinite redshift at the horizon. 
 Thus, a genuine BH mimicker should have an effectively large redshift, which would in particular prevent the interior to interfere with the physics of the photon sphere. This occurs if the time radiation takes to probe the interior of the object is sufficiently long, as it might happen for $C\to1/2$ or if the interior spacetime has very high redshift.
 
Buchdahl's theorem~\cite{Buchdahl:1959zz} sets a limit on the maximum compactness of perfect fluid stars: $C\leq4/9$, which is saturated by incompressible fluids featuring infinite sound speed.
Causality makes this bound more stringent: $C\lesssim0.364$, saturated by perfect fluids with a linear equation of state between the pressure and the density~\cite{Christodoulou:1995,Boskovic:2021nfs,Alho:2022bki}. This means that a causal perfect fluid in GR can marginally be a UCO, but not a BH mimicker.

To exceed the causal Buchdahl's bound and potentially allow for genuine BH mimickers, it is necessary to include \emph{anisotropies}, \emph{exotic matter}, or \emph{beyond-GR} effects.
For example, anisotropic stars within GR can reach the BH compactness, but only for unrealistic configurations, i.e. those violating causality or the dominant energy condition~\cite{Raposo:2018rjn,Alho:2022bki}. The highest compactness of physically admissible models  is reached for elastic (anisotropic) configurations~\cite{Alho:2023ris}, with $C_{\rm max}\approx 0.462$~\cite{Alho:2022bki}. However, further requiring such configurations to be radially stable reduces the maximum compactness to $C_{\rm max}\approx 0.389$, not much higher than the causal Buchdahl's bound on perfect fluids.

The most studied model of ECO are boson stars, self-gravitating solutions to GR coupled with a complex bosonic field, possibly subject to a self-interaction potential~\cite{Liebling:2012fv}. The maximum compactness of boson stars increases when the bosonic self interactions are large, but the field equation in this limit reduce to those of a perfect fluid~\cite{Boskovic:2021nfs,Vaglio:2022flq}, and hence the compactness is limited by causal Buchdahl's bound.  Thus, also boson stars in GR can marginally be UCOs, but not BH mimickers.

It is also important to stress that the compactness is a useful measure for the \emph{exterior} of the object, but does not tell anything about its interior. For example, while an incompressible fluid star has $C\leq 4/9$, its redshift at the center and the time radiation takes to probe the interior actually \emph{diverges} as $C\to 4/9$~\cite{Pani:2018flj}. 
Although this is clearly an unrealistic configuration, it illustrates the general principle that while the exterior may never be as compact as a Schwarzschild BH, the interior can compensate for the difference. Quantifying the compactness by the maximum redshift also avoids issues with a gauge-independent definition of the radius, or the fact that a sharp radius might not even exist, as in the case of boson stars.

All these considerations lead to the interesting open question:
\begin{tcolorbox}[width=\linewidth, sharp corners=all, colback=lightgray!30, boxrule=0.5pt]
\emph{What's the maximum compactness/redshift of a spherically symmetric, physically viable, and stable object in GR?}
\end{tcolorbox}

\subsection{Reflectivity}
Classical BHs are perfect absorbers, meaning that their horizon has zero reflectivity for any type of waves of any frequency.
Any deviation from this perfect absorption, i.e., any reflectivity, would indicate the absence of a classical horizon and, in turn, would be a smoking gun for ECOs.

For the sake of the argument, let's assume that wave propagation on the curved spacetime is described by an equation of the form
\begin{equation}
\frac{d^2\Psi}{dx^2}+(\omega^2-V(x))\Psi=0\,,
\end{equation}
for some master variable $\Psi$ and where $\omega$ is the frequency of the wave. If the ECO is sufficiently compact, the potential near its surface is negligible (as near the BH horizon), so the general solution is a superposition of ingoing and outgoing plane waves:
\begin{equation}
    \Psi\sim e^{-i\omega(x-x_0)}+{\cal R}(\omega)e^{i\omega(x-x_0)}
\end{equation}
where $x_0=x(R)$ is the location of the radius in the coordinate $x$. The reflectivity is the complex, frequency-dependent quantity ${\cal R}(\omega)$. Classical BHs have ${\cal R}=0$ for any frequency, while virtually any other object would have ${\cal R}\neq0$. In fact, since GWs interact very little with ordinary matter, one might even argue that $|{\cal R}|\approx 1$ is the rule rather than the exception, since $|{\cal R}|\approx 1$ simply means that radiation passes through the object without being absorbed. This is certainly true for low-frequency waves, such that 
\begin{equation}
    {\cal R}(\omega)=1+i\omega {\cal R}_1 +{\cal O}(\omega^2)\,. \label{reflectivity}
\end{equation}
Here, ${\cal R}_1$ is a model-dependent complex coefficient. Note that a large gravitational redshift inside the object would be accounted for by the \emph{phase} of the reflectivity: even a perfectly reflecting object ($|{\cal R}|=1$, as it is the case of any perfect-fluid star) could introduce a large phase difference of the reflected wave, effectively mimicking the BH condition, ${\cal R}=0$, for a certain amount of time.

Actually, one might even argue that \emph{BH boundary conditions are overrated}: more than the zero reflectivity, what matters is that radiation is infinitely redshifted and takes infinite time to reach the horizon. If dynamical perturbations take infinite time to reach the inner boundary, the actual boundary conditions should not matter.

Note that Eq.~\eqref{reflectivity} is effectively the boundary condition for linear perturbations of a given ECO model (parametrized in terms of ${\cal R}_1$ and higher-order coefficients). Thus, it dictates the quasinormal-mode spectrum, the time-domain ringdown response, including the presence of echoes~\cite{Maggio:2020jml}, and the tidal heating~\cite{Datta:2019epe,Maggio:2021uge}, but also quantities that can be defined in the static limit, such as the tidal deformability~\cite{Chakraborty:2023zed}.
All these quantities generically deviate from those of Kerr BHs and leave a direct imprint in various observables~\cite{Cardoso:2019rvt,Maggio:2021ans,DeLaurentis:2025nsy}.

Given its importance to describe deviations from the BH properties in a parametric, model-agnostic way, a natural question is:
\begin{tcolorbox}[width=\linewidth, sharp corners=all, colback=lightgray!30, boxrule=0.5pt]
\emph{What's the effective reflectivity of BH mimickers? How does this impact various observables?}
\end{tcolorbox}

\subsection{Dynamics \& Stability} 
This clue encompasses a range of features related to the dynamical behavior of ECOs and BH mimickers, including their stability, response to perturbations, and potential for mimicking BHs in certain observations. 

The dynamical behavior of ECOs, including their response to perturbations and their stability, is crucial for determining their viability as BH mimickers. 
Clearly, the first requirement is that viable ECOs are stable under linear perturbations (or at most unstable over sufficiently long time scales).

Due to the absence of a horizon, ECOs are at least partially reflective. For sufficiently compact objects (namely, UCOs), this yields to low-frequency, long-lived quasinormal modes, which are quasi-trapped by their effective potential.
This modes are prone to instabilities~\cite{Cardoso:2014sna}. Spinning UCOs might develop an ergosphere and hence be subject to the ergoregion instability at the linear level~\cite{Friedman:1978ygc} (see~\cite{Brito:2015oca,Cardoso:2019rvt,Maggio:2021ans} for some reviews). This instability rules out sufficiently fastly spinning UCOs which are perfectly reflective. However, a small amount of absorption (through
viscosity, dissipation, matter mode excitation, etc) can quench the instability completely~\cite{Maggio:2017ivp,Maggio:2018ivz}.

However, the long-lived modes (associated with the presence of a stable light ring~\cite{Cardoso:2014sna,Cunha:2017qtt}) can lead to a nonlinear instability even in the static case~\cite{Keir:2014oka,Cardoso:2014sna}.While there have been attempts to study this process~\cite{Cunha:2022gde,Redondo-Yuste:2025hlv} the time scale of this putative nonlinear instability is expected to be very long and more studies are needed to confirm it, characterize its time scale, and assess whether it is a generic phenomenon.

In this context, studies of the dynamics of viable ECO models that can be more compact than boson stars or perfect-fluid stars would be very useful. 
A promising example are topological stars: regular, horizonless solitons arising from dimensional compactification of Einstein-Maxwell theory in five dimensions, which could describe qualitative properties of microstate geometries for astrophysical BHs~\cite{Bah:2020ogh}. 
In the simplest model, they form a one-parameter family of solutions that smoothly interpolates between BHs, ECOs, UCOs, up to being arbitrarily close to BHs and hence being a model for BH mimickers.

Accordingly, their linear response to perturbation can exhibit both BH-like quasinormal modes, late time echoes and long-lived modes typical of UCOs, and star-like modes typical of less compact stars, depending on their parameters~\cite{Dima:2024cok,Dima:2025zot}.
Furthermore, topological stars were shown to be linearly stable across the most interesting region of the parameter space~\cite{Bena:2024hoh,Dima:2024cok,Dima:2025zot}.
Finally, since they arise from a well-defined, workable theory, they are prone to fully nonlinear studies. In particular, they provide a concrete testbed to assess whether the aforementioned nonlinear instability takes place in BH mimickers, and whether ECOs arbitrarily close to BHs can form dynamically in a well-motivated theory.  

Some outstanding questions concerning the ECO dynamics are:
\begin{tcolorbox}[width=\linewidth, sharp corners=all, colback=lightgray!30, boxrule=0.5pt]
\emph{Are BH mimickers stable? If not, what's their instability time scale? Can they form dynamically?}
\end{tcolorbox}

\subsection{Scale invariance} 
A key property of BHs that is typically overlooked is their scale invariance: being the Schwarzschild (or Kerr) metric a solution to \emph{vacuum} GR, their mass is a free parameter not associated with any fundamental scale.
It is extremely difficult to mimic this property with an ECO, if only because coupling to matter does introduce extra scales.

Therefore, to replace all BH candidates in the universe --~spanning from a few to billions of solar masses~-- with horizonless, regular alternatives, one must consider scale-invariant models.
To the best of our knowledge, in the context of GR this is possible only with specific anisotropic\footnote{The only scale-invariant perfect-fluid model in GR requires incompressible matter and it is therefore unphysical.} matter~\cite{Alho:2023mfc}. Beyond GR, one might attempt to find scale-free solitons, as it is the case of BH microstates in supergravity~\cite{Bena:2006kb,Bena:2016ypk,Bena:2017xbt,Bah:2021owp,Bah:2022yji} and topological stars~\cite{Bah:2020ogh}.

\subsection{Higher-dimensional camouflage: Fuzzballs}

Fuzzballs, a concept rooted in string theory, propose that BH microstates are horizonless, regular solutions with intricate topology that inherently exist in higher dimensions~\cite{Mathur:2009hf,Bena:2022rna,Bena:2022ldq}. The structure at the horizon scale is provided by \emph{microstate geometries} --~solitonic solutions with the same mass and charges as a BH, but where the horizon is replaced by a smooth, horizonless cap~\cite{Bena:2006kb,Bena:2016ypk,Bena:2017xbt,Bah:2021owp,Bah:2022yji}.
This framework is particularly compelling, as it offers a unified resolution to the BH entropy problem, the information-loss paradox, and the singularity issue.

These solutions crucially rely on two key aspects of string theory: the existence of non-perturbative objects (D-branes) whose mass decreases with increasing gravitational strength, and the introduction of numerous additional degrees of freedom. These degrees of freedom, which may account for BH entropy, enhance quantum tunneling effects and might prevent the formation of an event horizon~\cite{Kraus:2015zda,Bena:2015dpt}.

A fundamental feature of this proposal is its inherently higher-dimensional nature, where nontrivial topologies stabilize the horizon-scale structure, preventing its collapse. The microstates responsible for this structure manifest as smooth, topologically nontrivial geometries in ten dimensions. However, when Kaluza-Klein reduced to four dimensions, they appear to exhibit curvature singularities~\cite{Balasubramanian:2006gi,Bena:2022fzf}.

Additionally, dimensional reduction introduces Kaluza-Klein fields that are nonminimally coupled to gauge fields. Regardless of the specific nature of the horizon-scale structure, the vast Hilbert space associated with BH entropy includes coherent states, which resemble classical solutions in low-energy effective theories coupled to gravity. Consequently, from a four-dimensional perspective, one is left with GR coupled to various matter fields, including gauge fields and scalars, along with potential singularities that remain well-behaved in a higher-dimensional framework.

This paradigm provides a well-motivated, workable framework to study BH mimickers and, at the same time, it shows that potential pathologies of ECOs are harmless if the model has a consistent higher-dimensional embedding.
This suggests that:
\begin{tcolorbox}[width=\linewidth, sharp corners=all, colback=lightgray!30, boxrule=0.5pt]
\emph{Higher dimensions and nontrivial topologies might be key ingredients for viable models of genuine BH mimickers.}
\end{tcolorbox}

\section{Conclusion and Outlook}
The exploration of ECOs and BH mimickers provides a fascinating window into some fundamental questions surrounding BHs and gravity. Since BHs are arguably the most remarkable solutions to GR, the quest to test the BH paradigm continues to drive progress in both theory and observation.
Future GW observatories like LISA~\cite{LISA:2024hlh,Barausse:2020rsu,LISA:2022kgy} and Einstein Telescope~\cite{Abac:2025saz} will play a crucial role in probing the nature of dark compact objects and potentially revealing the observable signatures of quantum BHs~\cite{Carballo-Rubio:2025fnc,DeLaurentis:2025nsy}.

We conclude with a list of personal reflections worth highlighting:

\begin{itemize}
    \item Not all ECOs can be ultracompact, and not UCOs are genuine BH mimickers. In particular, boson stars are the ECO model \emph{par excellence}, but they can hardly be ultracompact and definitely cannot really mimic all aspects of BH phenomenology: their key observables (ringdown, multipolar structure, tidal deformability) are significantly different from those of a BH and can be easily distinguished. Most importantly, as many ECO models, boson stars have a mass scale dictated by the couplings of the underlying field theory. Thus, at variance with BHs, they cannot describe compact objects spanning diverse orders of magnitude in mass, from stellar-origin objects to supermassive ones.
    \item A genuine BH mimicker should ideally arise from a theoretically sound and fully workable theory, allowing for scale-invariant and stable ultracompact solutions. Furthermore, it should be possible to formulate the underlying theory as a well-posed initial value problem, in order to address questions such as the nonlinear stability of an isolated solution, or the merger of ECO binary systems.
    \item If the ECO spacetime has sufficiently high redshift, its different boundary conditions relative to a BH would not make a difference for observable quantities, at least within the time scales needed to probe the high-redshift object interior. In other words, if radiation is effectively trapped within the object, the latter would practically behave as a BH even in the total absence of absorption.
    \item Arguably, \emph{BH formation is inevitable within any classical relativistic theory of gravity}. All ECO models have extra matter fields coupled to gravity. Even for specific models where scale invariance can be achieved, a question remain whether the coalescence of two BH mimickers would produce a BH mimicker of the same class or a BH. This answer can be addressed in certain workable models (e.g. BH microstates) and it is of utmost importance.
    \item If the answer to the above turns out that a BH can always form in any classical relativistic theory, the only way out BH formation is by invoking quantum effects at the horizon scale. The latter have been studied in the context of BH microstates~\cite{Kraus:2015zda,Bena:2015dpt} but are, however, poorly understood in general.
    \item Various expectations one might have on the requirements for physically viable BH mimickers are challenged by models that are intrinsically higher dimensional, like fuzzballs. One key lesson from fuzzballs is that a solution might appear pathological in four dimensions while being completely regular in the higher dimensional uplift. Similarly, the presence of nontrivial topologies --~permitted for higher-dimensional objects~--offers a valuable and poorly explored mechanism for generating viable BH mimickers.
\end{itemize}

\section*{Acknowledgments}
This work is partially supported by the MUR PRIN Grant 2020KR4KN2 ``String Theory as a bridge between Gauge Theories and Quantum Gravity'', by the FARE programme (GW-NEXT, CUP:~B84I20000100001), and by the INFN TEONGRAV initiative.

\providecommand{\href}[2]{#2}\begingroup\raggedright\endgroup

\title{The black shell – a stringy black hole mimicker}
\author{Ulf Danielsson}
\institute{\textit{Institutionen för fysik och astronomi,\\
		Uppsala Universitet, Box 803, SE-751 08 Uppsala, Sweden}}

\maketitle 

\begin{abstract}
The black shell is black hole mimicker inspired by string theory with concrete, and potentially measurable, predictions. This is a brief review of the most important properties of black shells.
\end{abstract}

\section{Introduction}

The history of black holes goes all the way back to the English astronomer John Michel in 1783, and works by the French scholar Pierre-Simon Laplace from 1796 and 1799. Laplace even calculated the Schwarzschild radius, even though he did not call it like that. When Schwarzschild more than a century later wrote down his metric, \cite{Schwarzschild:1916uq}, nobody really understood what it meant. The connection to the earlier results was completely unclear. The Swedish optician Allvar Gullstrand \cite{Gullstrand:1921} and the French mathematician Paul Painlevé \cite{Painlevé:1921} even argued that general relativity was ambiguous and lacked in predictivity. 

The first to have the full conceptual picture of what kind of an object the Schwarzschild metric represented was Robert Oppenheimer in 1939 \cite{Oppenheimer:1939ue}, while Roger Penrose finalized its mathematical description a quarter of a century later \cite{Penrose:1964wq}.

When taking quantum mechanics into account, the problems started all over again, with Stephen Hawking discovering that black hole evaporate quantum mechanically, \cite{Hawking:1975vcx}. Classically, a black hole has no hair and retains no memory of how it once was formed. This suggests that the radiation will not carry any information either, which would mean that information is destroyed. Ironically, this is in contradiction with quantum mechanics.

String theory suggests that black holes are not featureless but carry just enough properties to account for the entropy and save quantum mechanics. While this is satisfactory from a scattering point of view, it remains a puzzle how to view local physics near the horizon. Proposals have been made, \cite{Susskind:1993if}, that also have been criticized, \cite{Almheiri:2012rt}.

The idea behind the black shell, and black hole mimickers in general, is to take these concerns seriously, \cite{Danielsson:2017riq}. Perhaps the correct solution to the paradox is to get rid of the horizon altogether?

\section{The black shell}

The key ingredient from string theory used in the black shell model, is the proliferation of AdS-vacua. We assume that our Minkowski vacuum is unstable against decay to a vacuum with a negative cosmological constant with a very long life time. When a black hole threatens to form, the probability increases so that an AdS bubble inevitably nucleates and catches the infalling matter. The reason for this is that matter on top of the shell has an enormous entropy compared to ordinary matter. The huge phase space compensates for the otherwise low probability. This is the same kind of argument familiar from other black hole mimickers such as fuzz balls, \cite{Mathur:2024ify}.

\begin{figure}
    \centering
    \includegraphics[width=7cm]{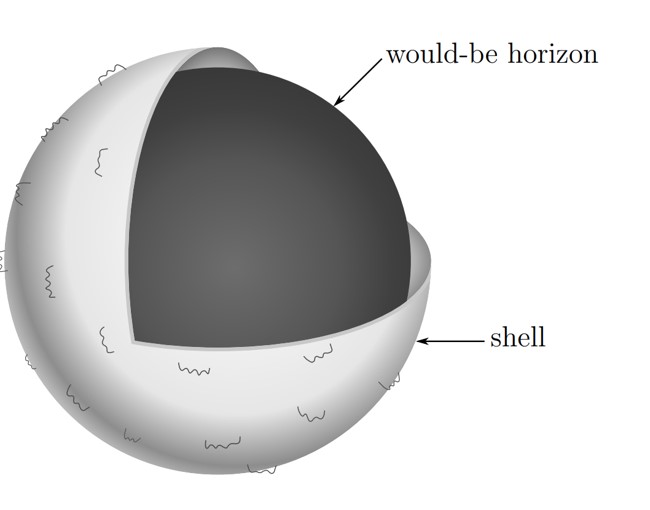}
    \caption{A black shell in cross-section.}
\end{figure}

As argued in \cite{Danielsson:2017riq}, the Israel junction conditions across the shell suggest that this process naturally occurs if the shell nucleates at the Buchdahl radius at $r=\frac{9GM}{4}$. This, then, is the radius of the horizonless black shell that replaces the black hole.

\section{Rotating black shells}

For a non-rotating black hole the metric is the unique and spherically symmetric Schwarzschild metric. The rotating metric is not unique, unless you assume the existence of a horizon and require that the metric is non-singular there. This gives the Kerr metric, which is fully characterized by its mass and angular momentum.

Since the black shell does not have a horizon, we a priori do not know which metric to use in the rotating case. In particular, there is no reason for it to be Kerr.

To find a unique metric and shape, we need to make a couple of assumptions. First, the matter on top of the brane is assumed to be traceless. This is what selects the Buchdahl radius in the non-rotating case. This constraint turns out to be very strong in the rotating case. In fact, we find that we must allow for a non-perfect fluid to find a solution. That is, radiation with viscosity and heat flow. Second, the interior is pure AdS that will rotate realative to the distant universe. Remarkably, this fixes the exterior metric, as well as the radius and shape of the black shell, \cite{Danielsson:2023onu}.

Interestingly, this problem is very similar to the one about rotating frames and Mach's principle, posed by Einstein in 1912, \cite{Einstein:1912}, and discussed by Josef Lense and Hans Thirring a few years later, \cite{Lense:1918zz}.  The idea was to realize Mach's principle such that the space time in the interior of a shell was perfectly rotating with respect to the surrounding universe. It was not solved until the 1980's, \cite{Pfister_1985,Pfister_1986}.

The construction outlined gives unique predictions that can be measured in the not too distant future. The quadrupole moment is predicted to be just over 1 percent larger than that of Kerr.

\section{Why are they black?}

The black shells are assumed to absorb everything that hits them. This is reasonable given that they carry so much entropy. The thermodynamical properties are very similar to those of a black hole. In \cite{Giri:2024cks} the electromagnetic properties were investigated, and it was explained how large permeability and permittivity can make the black shell a very efficient absorber without any reflection. The methods are similar to those of transformation optics that are important for meta-materials. The same construction also shows that the black shell can act like an electric generator just as well as a black hole.

When it comes to imaging, the inner shadow of a black shell will, in the non-rotating case, be 12.5 percent larger. This is currently not possible to measure, but there is hope in the not too distant future.

\section{Stability}

In order for the black shell not to collapse or explode, there needs to be a subtle transfer of energy between the fluid and the tension of the brane, \cite{Danielsson:2021ykm}. There is a span of parameters describing such a transfer where you obtain stability. In this regime, you can numerically throw mass at the black shell and see how it relaxes back to a shell of larger radius.

The black shell is special in the sense that it is sufficiently well defined so that this is possible.

\section{Outlook}

There are two important open questions. 

First, the black shell needs to nucleate through tunneling when matter is threatening to form a black hole. How can this be modeled? Is there an issue of causality? Tunneling will be needed also when sufficient amount of mass fall onto the black shell, or if two black shells collide. Hence, in order to describe the collision and merger of black shells, tunneling will be relevant.

Second, even though the model is inspired by string theory, it is not derived from first principles. In particular, it would be important to find a detailed explanation for the energy transfer that is so crucial for stability.

\bibliographystyle{utphys}
\providecommand{\href}[2]{#2}\begingroup\raggedright\endgroup

\title{Testing Black Holes with X-ray Observations}
\author{Cosimo~Bambi}
\institute{\textit{Center for Astronomy and Astrophysics,\\Center for Field Theory and Particle Physics, and Department of Physics,\\Fudan University, Shanghai 200438, China},
\and
\textit{School of Natural Sciences and Humanities, New Uzbekistan University,\\Tashkent 100007, Uzbekistan}}

\maketitle 

\begin{abstract}
This is a short review on how X-ray observations can test the nature of astrophysical black holes. 
\end{abstract}

\begin{figure}[t]
\centering
\includegraphics[width=0.45\linewidth]{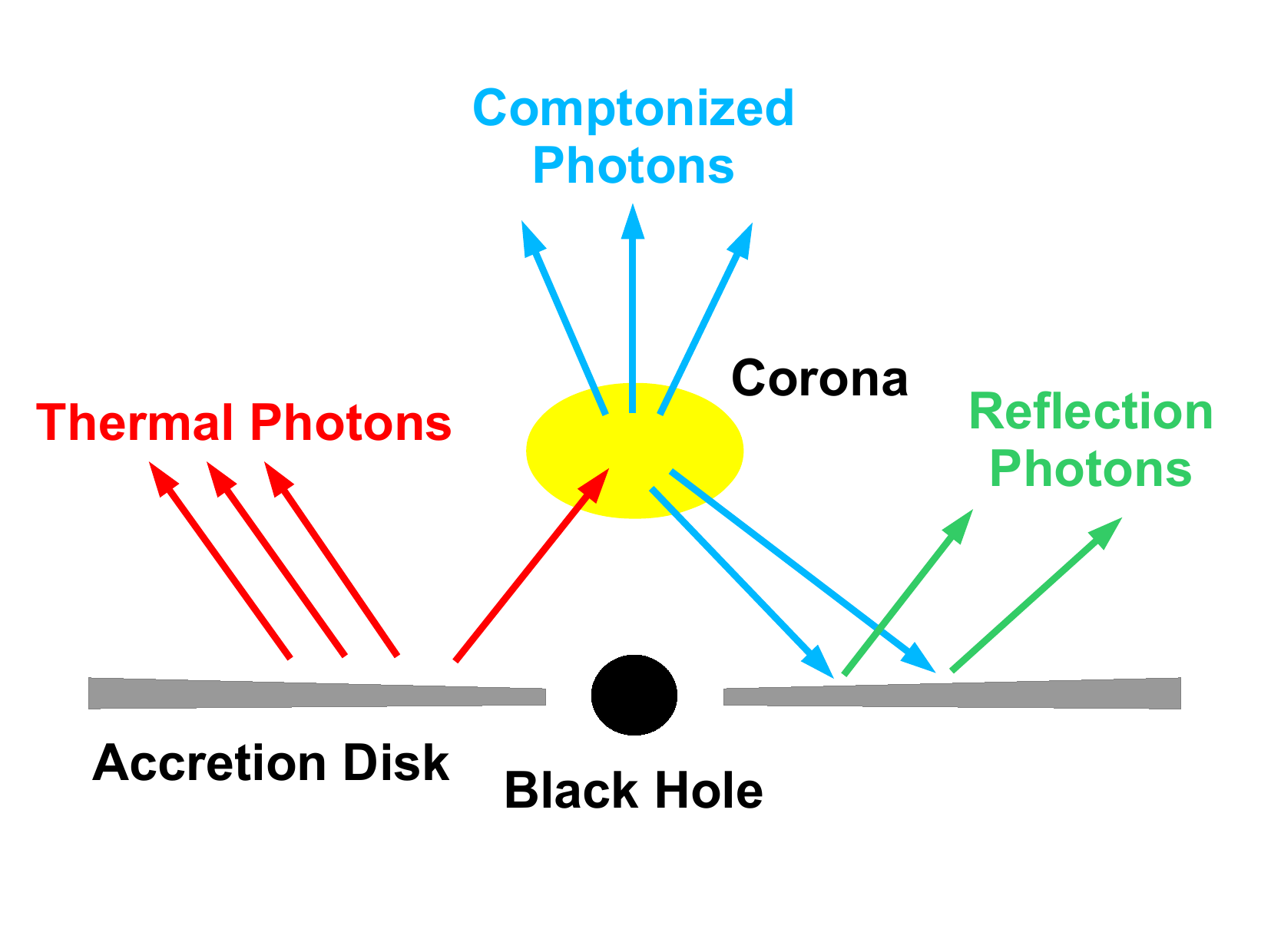}
\hspace{0.5cm}
\includegraphics[width=0.45\linewidth]{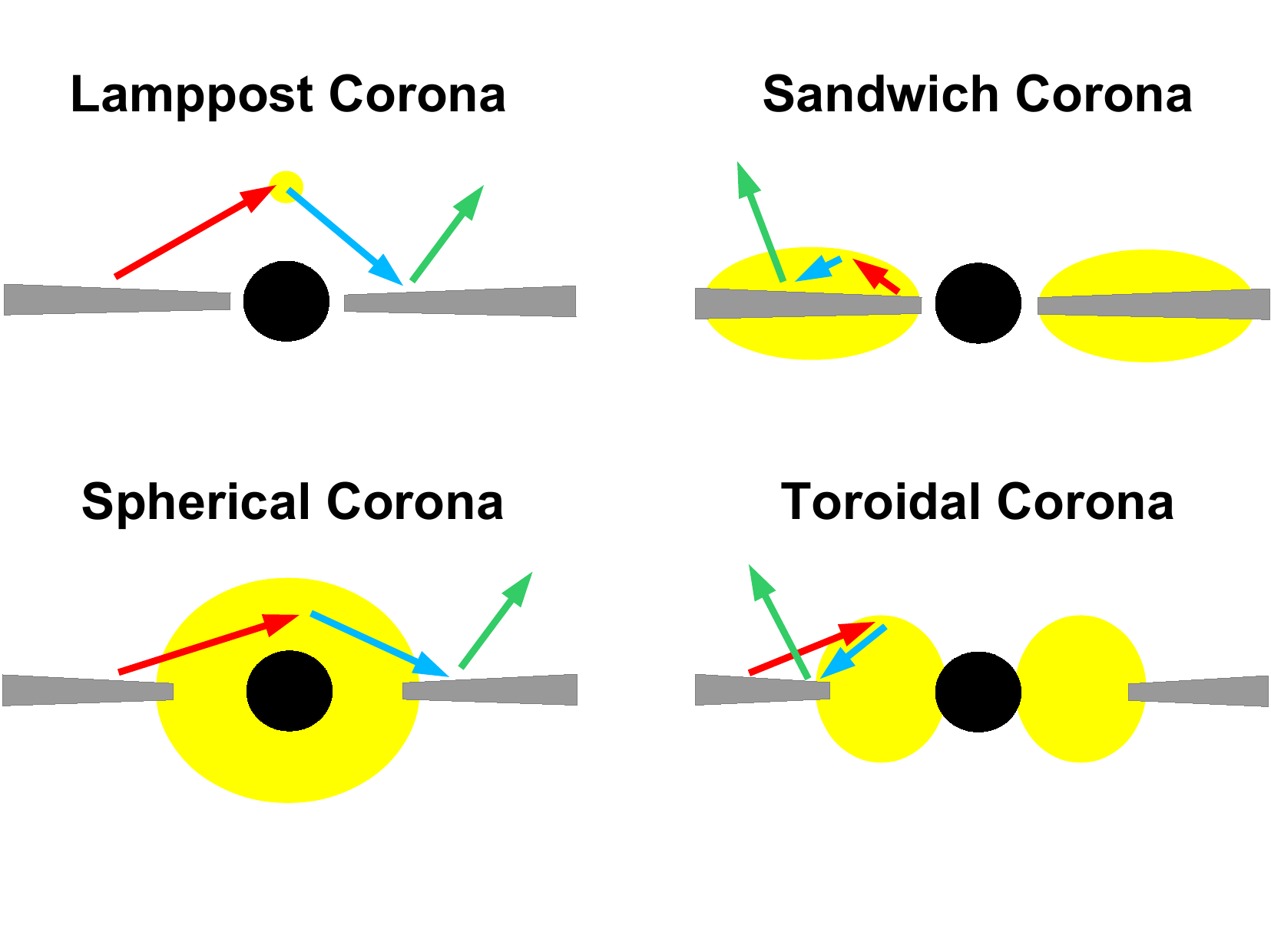}
\vspace{-0.5cm}
\caption{Left panel: Disk-corona model. Right panel: Examples of possible coronal geometries. Figure from Ref.~\cite{Bambi:2024hhi}.}
\label{f-corona}
\end{figure}

X-ray observations can be a powerful tool to test the nature of astrophysical black hole~\cite{Bambi:2015kza,Bambi:2022dtw}. The prototype of astrophysical system for these X-ray tests is shown in the left panel of Fig.~\ref{f-corona} and is usually referred to as the {\it disk-corona model}~\cite{1991A&A...247...25M}. The black hole can be either a stellar-mass black hole in an X-ray binary system or a supermassive black hole in an active galactic nucleus. The crucial ingredient is that the accretion disk is {\it cold}, geometrically thin, and optically thick. In these conditions, the thermal spectrum of the disk is peaked in the soft X-ray band (0.1-10~keV) in the case of stellar-mass black holes in X-ray binary systems and in the UV band (1-100~eV) in the case of supermassive black holes in active galactic nuclei~\cite{Shakura:1972te,Page:1974he}. The ``corona'' is some {\it hot} material (with an electron temperature $\sim 100$~keV) close to the black hole and the inner part of the accretion disk. The base of the jet, the hot atmosphere above the accretion disk, the material in the plunging region between the inner edge of the disk and the black hole, etc. may act as coronae~\cite{Haardt:1991tp,Titarchuk:1994rz,Dove:1997ei,Markoff:2005ht}. Possible coronal geometries are shown in the right panel in Fig.~\ref{f-corona}. Two or more coronae may coexist at the same time.

Since the disk is cold and the corona is hot, thermal photons from the disk can inverse Compton scatter off free electrons in the corona. The spectrum of the Comptonized photons can be normally approximated by a power law spectrum with a high-energy cutoff and a low-energy cutoff~\cite{Haardt:1991tp}. A fraction of the Comptonized photons can illuminate the disk: Compton scattering and absorption followed by fluorescent emission generate the reflection spectrum (green arrows in Fig.~\ref{f-corona}). The {\it non-relativistic reflection spectrum}, namely the reflection spectrum in the rest-frame of the material of the disk, is characterized by narrow fluorescent emission lines in the soft X-ray band and a Compton hump with a peak around 20-50~keV~\cite{Ross:2005dm,Garcia:2010iz}. The most prominent emission feature is usually the iron K$\alpha$ complex, which is at 6.4~keV in the case of neutral or weakly ionized iron atoms and can shift up to 6.97~keV in the case of H-like iron ions. The {\it relativistic reflection spectrum}, namely the reflection spectrum of the whole disk as detected far from the source, is blurred due to relativistic effects in the strong gravity region (Doppler boosting because of the motion of the material in the disk and gravitational redshift because of the gravitational well of the black hole)~\cite{Bambi:2017khi}. X-ray reflection spectroscopy refers to the analysis of these relativistically blurred reflection features and can be a powerful tool to test the nature of the compact object.

In the past years, my group at Fudan University has developed two models, {\tt relxill\_nk}~\cite{Bambi:2016sac,Abdikamalov:2019yrr,Abdikamalov:2020oci}\footnote{{\tt relxill\_nk} is an extension of the {\tt relxill} package~\cite{Dauser:2013xv,Garcia:2013lxa} developed by Thomas Dauser and Javier Garcia.} and {\tt nkbb}~\cite{Zhou:2019fcg}, which are specifically designed to test the Kerr hypothesis\footnote{The spacetime metric around astrophysical black holes is thought to be approximated well by the Kerr solution of General Relativity: this is the Kerr hypothesis~\cite{Bambi:2015kza}. Macroscopic deviations from the Kerr geometry are possible if General Relativity is not the correct theory of gravity, in the presence of macroscopic quantum gravity effects, as well as in the presence of exotic matter fields. An observation of a violation of the Kerr hypothesis would thus be a signature of new physics (even if not necessarily a breakdown of General Relativity).} from, respectively, the analysis of the reflection features and the thermal spectrum of a source. {\tt relxill\_nk} and {\tt nkbb} assume that the spacetime geometry around the compact object is more general than the Kerr solution. Deviations from the Kerr geometry are quantified by one or more {\it deformation parameters}. From the comparison of the theoretical predictions of the reflection and thermal spectra in this more general spacetime with the X-ray observations of a specific source, we can measure these deformation parameters and thus verify the Kerr hypothesis. The first test of the Kerr hypothesis using X-ray reflection spectroscopy was reported in Ref.~\cite{Cao:2017kdq} and was also the first test of the strong gravity region around a black hole with electromagnetic observations.

\begin{figure}[t]
\centering
\includegraphics[width=0.95\textwidth,trim=3.5cm 0.0cm 3.5cm 1.0cm,clip]{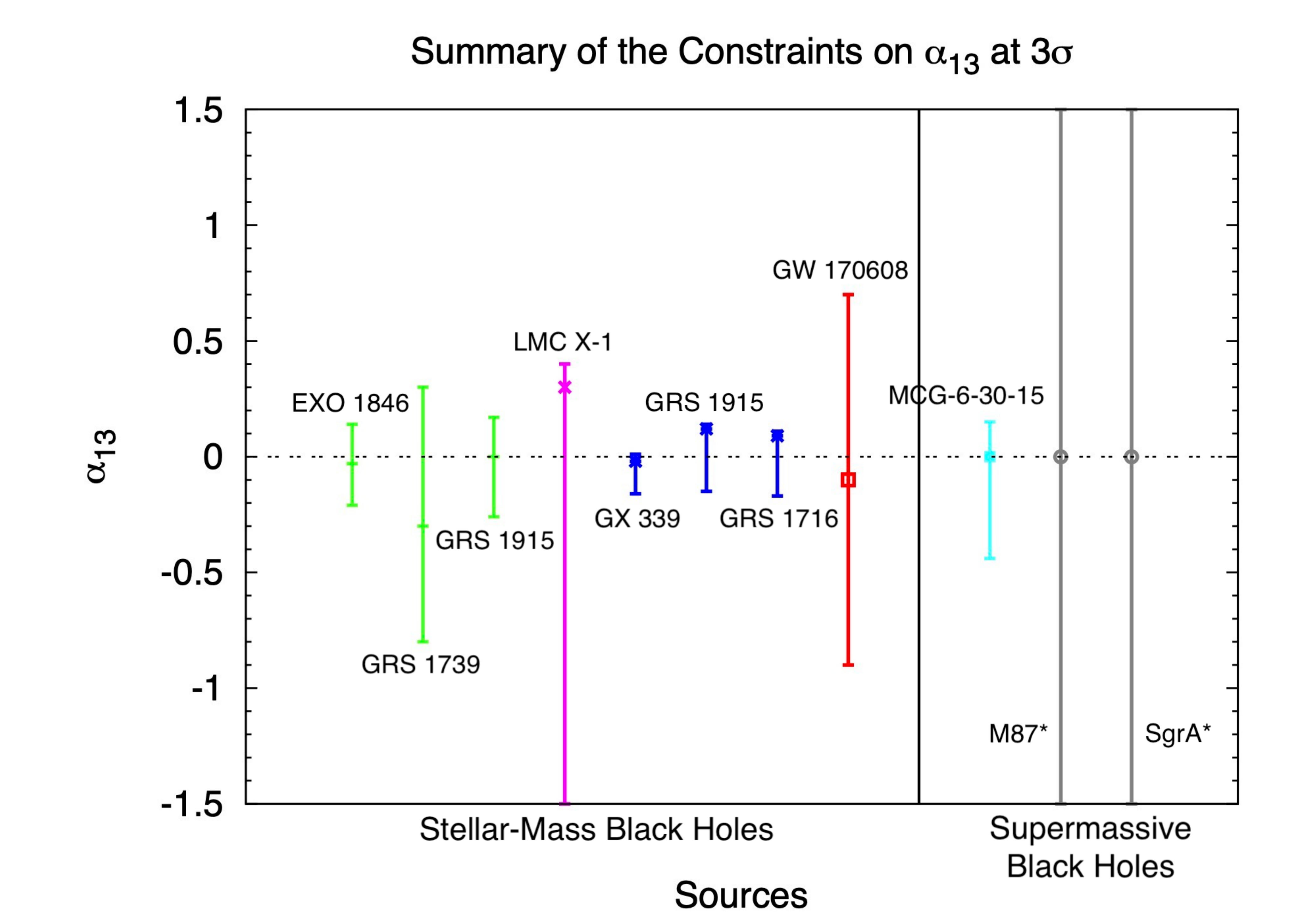}
\caption{Summary of the current constraints on the Johannsen deformation parameter $\alpha_{13}$ from X-ray reflection spectroscopy (green bars for stellar-mass black holes and cyan bar for supermassive black holes), continuum-fitting method (magenta bar), combination of X-ray reflection spectroscopy and continuum-fitting method (blue bars), gravitational waves (red bar), and black hole imaging (gray bars). All constraints are at 3-$\sigma$. The horizontal dotted line at $\alpha_{13}=0$ marks the Kerr solution of General Relativity. The magenta and gray bars extend beyond the y-axis range. Figure from Ref.~\cite{cb22}.}
\label{f-summary}
\end{figure}

Fig.~\ref{f-summary} summarizes the current 3-$\sigma$ constraints on the Johannsen deformation parameter $\alpha_{13}$~\cite{Johannsen:2013szh} with X-ray observations, gravitational wave data, and black hole imaging. The label of every measurement (e.g., EXO~1846, GRS~1739, GRS~1915, etc.) refers to the name of the source. The dotted horizontal line at $\alpha_{13} = 0$ corresponds to the Kerr solution and any deviations from $\alpha_{13} = 0$ at a significant confidence level can be a potential signature of new physics. Since the Johannsen metric is a deformed Kerr metric for agnostic tests of the Kerr hypothesis, Fig.~\ref{f-summary} distinguishes the observational constraints on stellar-mass and supermassive black holes, as they may belong to two different object classes. Let us note that X-ray reflection spectroscopy is currently the only technique that can test both stellar-mass and supermassive black holes.

{\bf Stellar-Mass Black Holes ---} The constraints in green are the most precise and accurate measurements of $\alpha_{13}$ from the analysis of reflection spectra of stellar-mass black holes in X-ray binary systems (see the original papers for the details of the data analysis~\cite{Zhang:2019ldz,Tripathi:2020yts}). The constraint in magenta is obtained from the analysis of the thermal spectrum of the stellar-mass black hole in LMC~X-1~\cite{Tripathi:2020qco}: the constraint is weak because the estimates of the spin parameter and $\alpha_{13}$ are strongly correlated and it is not possible to constrain well both parameters simultaneously from the sole analysis of the thermal spectrum of a source. The constraints in blue are obtained from sources for which it is possible to combine the analysis of the reflection features with that of the thermal spectrum~\cite{Tripathi:2020dni,Tripathi:2021rqs,Zhang:2021ymo}: even if the analysis of the thermal spectrum provides weak constraints on $\alpha_{13}$, it can still help to improve the constraints from X-ray reflection spectroscopy if we use the two techniques together. The constraint in red is the best constraints on $\alpha_{13}$ from gravitational waves~\cite{Shashank:2021giy}\footnote{The gravitational wave constraint is inferred by assuming that the emission of gravitational waves is the same as in General Relativity and by analyzing the inspiral phase of the gravitational wave emission of binary black holes. GW~170608 provides the strongest constraint because of its long detected inspiral signal.}.

{\bf Supermassive Black Holes ---} In the case of supermassive black holes in active galactic nuclei, the thermal spectrum of the disk peaks in the UV band, where dust absorption limits the possibility of accurate measurements of the thermal component. X-ray tests of supermassive black holes can thus rely only on the analysis of the reflection features of a source. The constraint in cyan is the most precise and accurate measurements of $\alpha_{13}$ for supermassive black holes: MCG--6--30--15 is quite a bright source, presenting often a spectrum with a very strong iron line, and the measurement in Fig.~\ref{f-summary} is inferred from the analysis of high-quality data of \textsl{NuSTAR} and \textsl{XMM-Newton}~\cite{Tripathi:2018lhx}. The constraints in gray are inferred from black hole imaging by the Event Horizon Telescope Collaboration and, at least for the moment, are clearly weaker than the constraints we can infer from X-ray observations and gravitational wave data~\cite{EventHorizonTelescope:2020qrl,EventHorizonTelescope:2022xqj}.

Current tests of the Kerr hypothesis with X-ray observations can be improved in the future. The X-ray mission \textsl{eXTP}~\cite{eXTP:2016rzs} is currently scheduled to be launched in 2030 and will be able to measure even the polarization of the X-ray radiation of the source: polarization properties of the radiation from the inner part of the accretion disk can also be used to test the spacetime metric around a source~\cite{Krawczynski:2012ac,Liu:2015ibq}. The X-ray mission \textsl{Athena}~\cite{Nandra:2013jka} is currently scheduled to be launched in 2037: with an energy resolution of about 2.5~eV and a large effective area at the iron line, it promises to provide unprecedented high-quality data of accreting black holes~\cite{Liu:2024xim}.

\bibliographystyle{utphys}

\providecommand{\href}[2]{#2}
\begingroup\raggedright

\endgroup

\title{How does a horizonless compact object ringdown?}
\author{Elisa Maggio}
\institute{\textit{Max Planck Institute for Gravitational Physics (Albert Einstein Institute), \\Am Mühlenberg
1, Potsdam 14476, Germany}}

\maketitle 

\begin{abstract}
Black holes are the most compact objects in the universe. General relativity predicts that black holes possess a singularity hidden by an event horizon, representing a breakdown of Einstein's theory.  Recent gravitational-wave detections provide a crucial opportunity to observationally examine these horizons. This is particularly significant as certain quantum gravity models propose the existence of singularity-free compact objects lacking horizons. Here, we review the signatures of horizonless compact objects in the ringdown of compact binary coalescences.
\end{abstract}

The ground-based detectors LIGO and Virgo have detected ninety gravitational-wave (GW) events from the coalescence of compact binaries~\cite{LIGOScientific:2018mvr,LIGOScientific:2020ibl,KAGRA:2021vkt}.
In 2023, the LIGO-Virgo-KAGRA (LVK) collaboration started its fourth observing run, reporting about two hundred significant detection candidates~\cite{KAGRA:2013rdx}.

Three main stages characterise the GW signal emitted by a compact binary coalescence: the inspiral, when the two bodies orbit around each other and the emission of GWs shrinks the orbit; the merger, when a common apparent horizon forms; and the ringdown, when the final remnant settles down to a stationary configuration. GWs provide a unique channel for probing general relativity (GR)~\cite{LIGOScientific:2019fpa,LIGOScientific:2020tif,LIGOScientific:2021sio}. Parametrised tests introduce deviations in the stages of the GR waveform to constrain the degree to which the GW data agree with GR.

The ringdown stage is dominated by the characteristic oscillation frequencies of the remnant, the so-called quasinormal modes (QNMs), $\omega_{\ell m n} = \omega_{R, \ell m n} + i \omega_{I, \ell m n}$, which depend on the angular number $\ell$, the azimuthal number $m$ and the overtone number $n$ of the perturbation. The ringdown is modeled as a sum of exponentially damped sinusoids whose frequencies and damping times are related to the QNMs of the remnant as
\begin{equation}
f_{\ell m n} = \frac{\omega_{R, \ell m n}}{2 \pi} \,, \qquad
\tau_{\ell m n} = -\frac{1}{\omega_{I, \ell m n}} \,.
\end{equation}

Parametrised test of GR introduce fractional deviations to the frequency and the decay time of the fundamental QNMs~\cite{Ghosh:2021mrv,Brito:2018rfr,Maggio:2022hre},
\begin{equation}
\omega_{\ell m 0} = \omega_{\ell m 0}^{\text{GR}} (1+\delta \omega_{\ell m 0}) \,, \qquad
\tau_{\ell m 0} = \tau_{\ell m 0}^{\text{GR}} (1+\delta \tau_{\ell m 0}) \,,
\end{equation}
where the GR values are predicted from estimates of the mass and spin of the remnant and fits with numerical relativity.                                                      
The least-damped dominant QNM ($\ell=m=2$, $n=0$) has been observed in the ringdown of twelve GW events~\cite{LIGOScientific:2020tif,LIGOScientific:2021sio}. The observations are compatible with Kerr black hole (BH) remnants with an accuracy of 10\% and 20\% on the QNM frequency and damping time, respectively. 

The BH spacetime has been probed approximately until the location of the light ring, which is the innermost circular orbit of photons. 
Indeed, the damping time of the  ringdown is
associated with the instability timescale of the photon orbits at the light ring.
BHs are characterised by horizons, where the curvature changes the causal structure, and a curvature singularity, where Einstein's field equations break down.
New physics can prevent the formation of the horizon and the singularity in quantum-gravity extensions of GR and in GR with dark matter or exotic fields. These ideas have inspired a plethora of models including boson stars~\cite{Liebling:2012fv}, fuzzballs~\cite{Mathur:2005zp} and gravastars~\cite{Mazur:2004fk}.

Exotic compact objects (ECOs) can mimic black holes and quantify the existence of horizons~\cite{Giudice:2017dde,Cardoso:2019rvt}. A horizonless compact object deviates from a BH for its~\cite{Maggio:2021ans}: compactness, since it radius is located at $r_{\text{ECO}} = r_{\text{BH}} (1+\epsilon)$ where $r_{\text{BH}}$ is the BH horizon; and reflectivity, $\mathcal{R}(\omega)$, which differs from the totally absorbing BH case with $\mathcal{R}=0$.

\section*{Signatures of horizonless compact objects in the ringdown}

We analyse a horizonless compact object, which is described by the Schwarzschild metric and perturbed by a gravitational perturbation.
The radial component of the perturbation is governed by a Schrödinger-like equation~\cite{Regge:1957td,Zerilli:1970se}
\begin{equation}
\frac{d^2 \psi(r)}{dr_*^2} + \left[ \omega^2 - V(r) \right] \psi(r) = 0 \,,
\end{equation}
where $r_*$ is the tortoise coordinate, and $V(r)$ is the effective potential to which the perturbation is subjected. To derive the QNMs of the system, we add two boundary conditions: at infinity, we impose purely outgoing waves; and at the object's radius, we impose a superposition of ingoing and outgoing waves with reflectivity $\mathcal{R}(\omega)$.

\begin{figure}[t]
\centering
\includegraphics[width=0.49\textwidth]{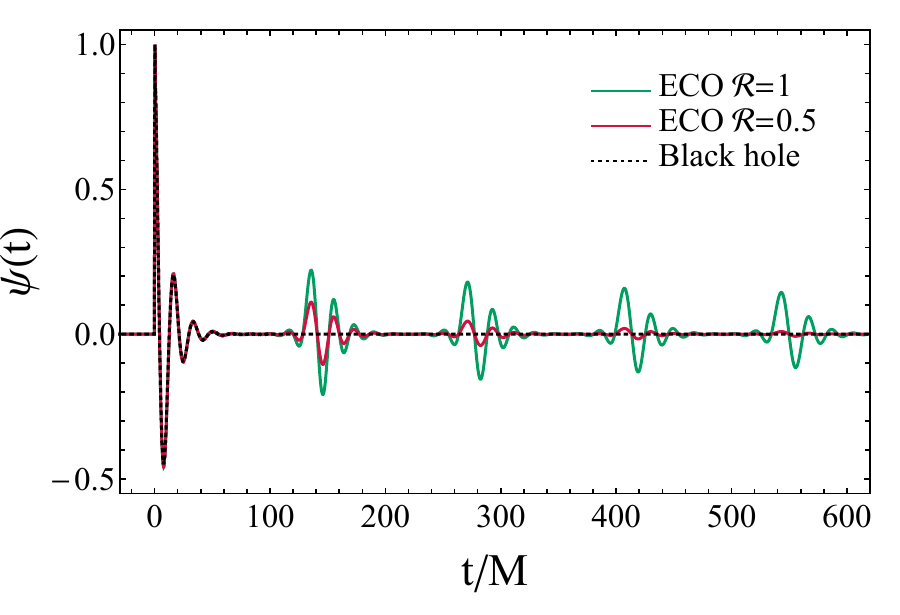}
\includegraphics[width=0.49\textwidth]{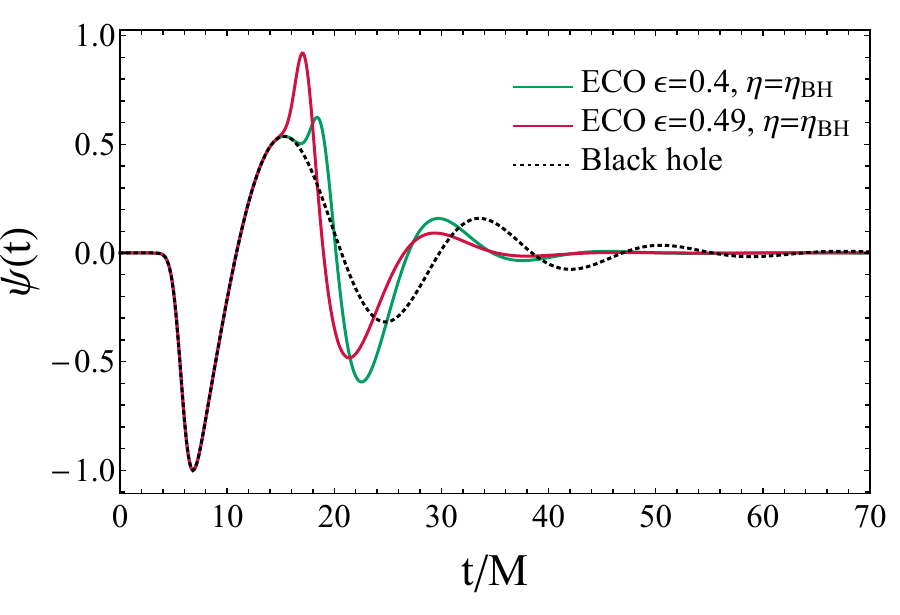}
\caption{Left panel: GW echoes emitted by a static ultracompact horizonless object with a given compactness and several values of the  reflectivity ($\mathcal{R}=1, 0.5$). Right panel: Ringdown of an ECO with radius $r_{\rm{ECO}} = 2M (1+\epsilon)$ and the same reflective properties of a BH~\cite{Maggio:2019zyv,Maggio:2020jml,Maggio:2021ans}.
} 
\label{fig}
\end{figure}

For ultracompact ($\epsilon \ll 1$) and reflective objects, axial and polar modes are not isospectral differently from the BH case. Moreover, for $\epsilon \to 0$ the QNMs are low-frequencies and long-lived~\cite{Cardoso:2016rao}. In the time domain, the waveform has the  same BH ringdown due to the excitation of the light ring, and GW echoes due to long-lived modes, as shown in the left panel of Fig.~\ref{fig}.

For horizonless objects with diﬀerent compactnesses and
interior solutions, we make use of the membrane paradigm to derive the boundary condition at the object's radius~\cite{1982mgm..conf..587D,Thorne:1986iy}. Indeed, the interior of the object can be replaced by a fictitious membrane located at the object's radius, which is a viscous fluid with viscosity related to the object's reflectivity~\cite{Maggio:2020jml}. The right panel of Fig.~\ref{fig} shows the modifications to the ringdown when the remnant is horizonless and has the same reflective properties of a BH. In this case, the ringdown is modified due to the interference with the first GW echo, and no subsequent echoes are emitted since there are no long-lived modes.

Horizonless compact objects with $\epsilon \lesssim 0.1$ are compatible with the measurement accuracy of the least-damped dominant QNM in the event GW150914~\cite{LIGOScientific:2016lio}. 

The membrane paradigm was generalised to spinning horizonless compact objects described by the Kerr metric to linear order in spin~\cite{Saketh:2024ojw}.
From a comparison with the BH case, this description is valid until the maximum value of the spin $\chi=0.2$.
The spin makes the constraints on the compactness of objects more stringent. Horizonless compact objects with $\epsilon \lesssim 0.04$ are compatible with the measurement accuracy of the fundamental quasinormal mode of GW150914~\cite{Saketh:2024ojw}.

Spinning compact objects with an ergoregion but without an event horizon might be unstable due to the ergoregion instability~\cite{Friedman:1978ygc,Brito:2015oca} above a critical value of the spin~\cite{Maggio:2018ivz}.
If the compact objects we observe are horizonless, they would be subject to the ergoregion instability and spin down until they reach the critical value of the spin.
Using spin measurements from 104 compact objects from the third GW catalogue, we infer that the population of compact binary coalescences cannot be composed by more than 70\% of reflective ECOs (28\% for ultracompact ECOs)~\cite{Mastrogiovanni:2025ixe}.

Partial absorption at the surface can make spinning horizonless compact objects stable. The minimum absorption rate to quench the instability is the maximum amplification factor of superradiance of BHs~\cite{Maggio:2018ivz}.

\section*{Conclusions and future prospects}

We can test the physics at the horizon scale with GWs.
Horizonless compact objects are not excluded by current GW observations, however next-generation detectors will allow us to perform unprecedented tests of the BH paradigm.
As future prospects, we need to model horizonless compact objects numerically to extract information on the waveform for different binary configurations.
Finally, we need to develop strategies to constrain the properties of the remnants with Bayesian frameworks.

\section*{Acknowledgments}
E.M. is supported by the European Union’s Horizon Europe research and innovation programme under the Marie Skłodowska-Curie grant agreement No. 101107586. E.M. acknowledges funding from the Deutsche Forschungsgemeinschaft (DFG) - project number:
386119226.

\bibliographystyle{utphys}
\providecommand{\href}[2]{#2}\begingroup\raggedright\endgroup

\title{Topological stars and solitons from supergravity}
\author{Pierre Heidmann}
\institute{\textit{Department of Physics and Center for Cosmology and AstroParticle Physics (CCAPP),\\
		The Ohio State University, Columbus, OH 43210, USA}}

\maketitle 

\begin{abstract}
	Understanding the fundamental structure of black holes is a central challenge in quantum gravity. Quantum information arguments and supersymmetric constructions in string theory suggest that horizon-scale physics must significantly deviate from its classical description in General Relativity (GR) and could be captured by coherent,  horizonless microstates of black holes in supergravity.  These \emph{topological solitons} resemble black holes but replace the horizon with a smooth topological structure.
 
    We present recent advances in constructing such solitons beyond supersymmetry, particularly the topological star, which remain regular while closely resembling nonextremal black holes,  including the Schwarzschild black hole. These solutions provide semi-classical prototypes for microstates of realistic black holes, evading the no-hair theorem through well-motivated string-theoretic mechanisms. Their stability and gravitational signatures offer new insights into black hole microstructure and horizon-scale physics beyond GR,  providing a pathway to probe quantum gravity through astrophysical black holes.
\end{abstract}

In quantum gravity, the key question is not merely what could serve as an alternative to black holes or a ``black hole mimicker,'' but rather what black holes truly are and what they are made of. These two perspectives share a fundamental dissatisfaction with the description of black holes in General Relativity (GR), which features a singularity and a horizon, necessitating significant corrections to this picture.

From Penrose's singularity theorem \cite{Penrose:1969pc}, it has long been understood that black holes in GR arise from the breakdown of effective field theory in the strong-gravity regime. Initially, it was believed that resolving this issue required only Planck-scale modifications confined near the singularity. However, the Hawking information paradox \cite{Hawking:1974sw,Hawking:1976ra}, later sharpened by the small corrections theorem \cite{Mathur:2009hf}, demonstrates that resolving the paradox necessitates large modifications at the horizon scale,  unless one is willing to abandon either quantum unitarity or locality in quantum field theory. A more fundamental quantum theory of gravity is therefore needed to describe black hole microstructure and explain how it modifies horizon-scale physics.

String theory has successfully provided a microscopic description of black hole microstates in terms of fundamental ingredients like strings, branes, or antibranes when gravity is ``turned off'' \cite{Strominger:1996sh,Heidmann:2023kry}.  However, a complete description of these microstates at finite Newton's constant remains elusive. A priori, generic states should be quantum horizon-scale entities where the classical notions of spacetime and fields break down. However, if new physics emerges at the horizon, far exceeding the string and Planck scales above the singularity, black hole microstructure could be captured by the massless sector of string theory: supergravity. A promising approach to constructing black hole microstates at finite Newton’s constant is to focus on \emph{coherent microstates} known as \emph{microstate geometries} \cite{Bena:2022rna,Bena:2025pcy}. These are \emph{gravitational solitons}: smooth, horizonless spacetimes admitting a semi-classical description within supergravity.

Over the years, extensive families of supersymmetric microstate geometries have been constructed in supergravity \cite{Lunin:2001fv,Bena:2007kg,Bena:2015bea,Heidmann:2019xrd}. These ``black hole mimickers'' closely resemble extremal black holes but replace the horizon with a smooth, horizonless structure. They evade GR's no-hair theorem by using the only viable classical support mechanism at the horizon: topological spacetime deformations involving compact dimensions and electromagnetic fluxes \cite{Gibbons:2013tqa}. Remarkably, this result applies broadly to any classical gravity theory aiming to construct smooth, horizonless geometries with genuine black-hole features, extending its implications well beyond string theory.

The existence of supersymmetric black hole microstates in supergravity challenges the conventional GR picture and suggests a new correspondence principle: just as GR dictates that any exotic quantum system must resemble a classical black hole \emph{above} the horizon, supergravity might capture collective behavior of physics \emph{at} the horizon scale.  However, most results to date apply to supersymmetric black holes, which are idealized models.  The black holes we observe through gravitational wave detections and direct imaging — best exemplified by the Kerr metric — require to extend beyond supersymmetric configurations. 

To probe the microstructure of nonextremal black holes, explicit gravitational solitons must be constructed by solving Einstein’s equations in supergravity without supersymmetric shortcuts. These equations are notoriously difficult to solve analytically. However, recent advances \cite{Bah:2020pdz,Bah:2021owp,Bah:2021rki,Heidmann:2021cms} have built techniques that impose sufficient symmetry to reduce these equations to manageable two-dimensional problems while preserving the key ingredients: extra compact dimensions and electromagnetic fluxes. These methods generalize the electrostatic Ernst formalism of four-dimensional Einstein-Maxwell theory \cite{Ernst:1967by,Ernst:1967wx} to supergravity theories \cite{Heidmann:2021cms}.

The \emph{topological star} was the first solution constructed using this approach \cite{Bah:2020ogh,Bah:2020pdz}. It is a simple, static, nonextremal gravitational soliton with the same conserved charges as a nonextremal black hole, but replacing the horizon with a smooth topological structure supported by magnetic flux. This solution captures the essential features of black hole microstructure while remaining conceptually simple, making it an excellent prototype for nonextremal black hole microstates. The topological star is a solution of five-dimensional Einstein-Maxwell theory, a consistent truncation of five-dimensional supergravity, with metric and magnetic field strength:
\begin{equation}
\begin{split}
ds_5^2 &= - \left(1-\frac{r_\text{S}}{r} \right) \,dt^2 +\left(1-\frac{r_\text{B}}{r} \right) \,dy^2 + \frac{dr^2}{\left(1-\frac{r_\text{S}}{r} \right)\left(1-\frac{r_\text{B}}{r} \right)} + r^2 \left(d\theta^2 +\sin^2\theta \,d\phi^2 \right)\,,\\
F & = \sqrt{3 \,r_\text{S}r_\text{B}} \,\sin \theta \,d\theta \wedge d\phi\,,
\label{eq:TSSol}
\end{split}
\end{equation}
where $y$ is a compact dimension with periodicity $y=y+2\pi R_y$ above an asymptotically flat four-dimensional spacetime, $(t,r,\theta,\phi)$. Depending on $(r_\text{S},r_\text{B})$, the solutions describe either a black hole ($r_\text{S} \geq r_\text{B}$), including Schwarzschild at $r_\text{B}=0$, or a smooth, horizonless geometry ($r_\text{B} > r_\text{S}$) where the spacetime ends smoothly at $r=r_\text{B}$  as the coordinate degeneracy of the $y$-circle with a S$^2$ Kaluza-Klein bubble of radius $r_\text{B}$.  

Unlike many beyond-GR black hole mimickers, the topological star does not require exotic matter but is a pure solution of the theory, exploiting gravitational and topological deformations to sustain gravitational contraction and replace the horizon by a smooth structure. For further details, see \cite{Bah:2020ogh,Bah:2020pdz} for the construction, \cite{Bah:2021irr,Bianchi:2023sfs,Bena:2024hoh,Dima:2025zot} for stability analysis, and \cite{Heidmann:2022ehn,Heidmann:2023ojf,Bianchi:2024rod} for gravitational signatures.

Moving forward, constructing nonextremal gravitational solitons with net zero charge is crucial for comparisons with realistic black holes. The generalization of the Ernst formalism in supergravity allows for the construction of regular ``bound states of topological stars'' \cite{Bah:2021rki,Heidmann:2021cms}, including neutral geometries directly comparable to the Schwarzschild black hole \cite{Bah:2022yji,Bah:2023ows,Heidmann:2023kry}. These were the first explicit examples of Schwarzschild microstructure in string theory, replacing the Schwarzschild horizon with a smooth topological structure supported by electromagnetic flux. While significantly more intricate than the topological star \eqref{eq:TSSol}, these solutions confine their microstructure to a deep interior region near the would-be horizon.  Consequently, they are indistinguishable from singular deformations of the Schwarzschild black hole with a scalar field, called  \emph{Schwarzschild Scalarwalls}:
\begin{align}
ds^2_\text{SW} &= - f^\alpha\, dt^2 + \frac{1}{f^{\alpha-1}} \left[\left(1+\frac{M^2\sin^2\theta}{\alpha^2 r^2 f} \right)^{1-\alpha^2-\beta^2} \left( \frac{dr^2}{f}+r^2 d\theta^2\right) +r^2 \sin^2 \theta \,d\phi^2 \right],  \nonumber\\
e^\Phi &= f^{\beta},\qquad f = 1- \frac{2M}{\alpha \,r},
\end{align}
where $\alpha$ and $\beta$ are constants. The scalarwalls have a singular horizon at $r=2M/\alpha$ but this singularity is resolved by the smooth, horizonless solitons in higher dimensions.  Very close to $r\sim 2M/\alpha$, electromagnetic fluxes and deformations along compact dimensions emerge, smoothly terminating the spacetime \cite{Bah:2022yji,Bah:2023ows}.

 \begin{figure}[t]
 \begin{center}
\hspace{-0.5cm} \includegraphics[width=\textwidth]{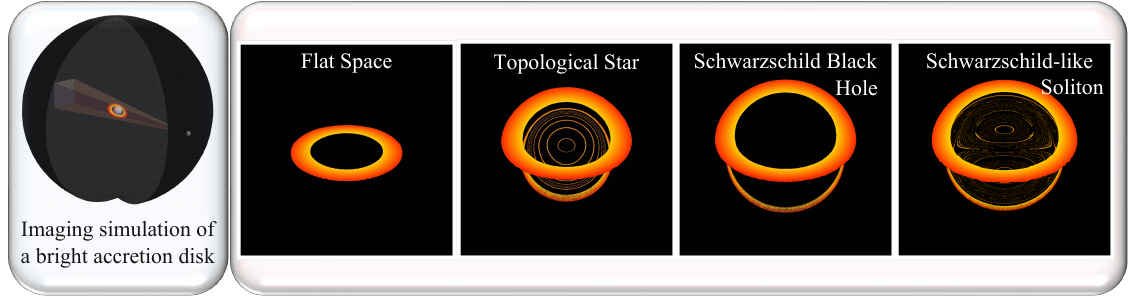}
\caption{Imaging simulation of a bright accretion disk around different backgrounds. From left to right: a schematic representation of the setup, showing the observer, the accretion disk, and the gravitational object; simulation results for flat space, the topological star, the Schwarzschild black hole, and the Schwarzschild-like solitons. For further details, see \cite{Heidmann:2022ehn}.\label{fig:ImagingTS}}
\end{center}
\vspace{-0.7cm}
\end{figure}

In \cite{Heidmann:2022ehn}, a ray-tracing code for geometries with compact dimensions was developed to simulate the scattering properties of various nonextremal solitons in comparison to black holes. Figure \ref{fig:ImagingTS} summarizes the main results. Since these solitons are not sourced by any matter, light passes through them and is nontrivially scattered by their topological structure. The topological star, being a highly coherent object, behaves like a spacetime mirror with black hole-like properties, coherently reflecting light from an accretion disk. The Schwarzschild-like solitons, despite resembling a Schwarzschild scalarwall, produce an image nearly identical to that of a Schwarzschild black hole. However, a key difference is that light passing through the photon ring undergoes chaotic internal scattering before eventually being deflected away.

This series of work represents the first explicit construction of horizon-scale microstructure for astrophysical black holes within supergravity, bridging the black hole microstate paradigm in string theory with black hole astrophysics. These top-down string-theoretic solutions, which resemble black holes in GR but replace the vacuum region near the horizon with a new string-theoretic phase of matter, may provide insights into the fundamental structure within black holes and exhibit gravitational signatures that deviate from the classical black hole picture.

\bibliographystyle{utphys}      

\providecommand{\href}[2]{#2}\begingroup\raggedright\endgroup

\end{document}